\documentclass[journal]{IEEEtran}
\usepackage{epsbox}
\usepackage[dvips]{graphicx,color}
\usepackage{latexsym}
\usepackage{amssymb}

\usepackage{array}
\begin{document}
\arraycolsep 0.5mm

\newcommand{\bfig}{\begin{figure}[t]}
\newcommand{\efig}{\end{figure}}
\setcounter{page}{1}
\newenvironment{indention}[1]{\par
\addtolength{\leftskip}{#1}\begingroup}{\endgroup\par}
%
\newcommand{\namelistlabel}[1]{\mbox{#1}\hfill} 
\newenvironment{namelist}[1]{%
\begin{list}{}
{\let\makelabel\namelistlabel
\settowidth{\labelwidth}{#1}
\setlength{\leftmargin}{1.1\labelwidth}}
}{%
\end{list}}
%
%
%
%
\newcommand{\bc}{\begin{center}}  %
\newcommand{\ec}{\end{center}}
\newcommand{\befi}{\begin{figure}[h]}  %
\newcommand{\enfi}{\end{figure}}
\newcommand{\bsb}{\begin{shadebox}\begin{center}}   %
\newcommand{\esb}{\end{center}\end{shadebox}}
\newcommand{\bs}{\begin{screen}}     %
\newcommand{\es}{\end{screen}}
\newcommand{\bib}{\begin{itembox}}   %
\newcommand{\eib}{\end{itembox}}
\newcommand{\bit}{\begin{itemize}}   %
\newcommand{\eit}{\end{itemize}}
\newcommand{\defeq}{\stackrel{\triangle}{=}}
\newcommand{\qed}{\hbox{\rule[-2pt]{3pt}{6pt}}}
\newcommand{\beq}{\begin{equation}}
\newcommand{\eeq}{\end{equation}}
\newcommand{\beqa}{\begin{eqnarray}}
\newcommand{\eeqa}{\end{eqnarray}}
\newcommand{\beqno}{\begin{eqnarray*}}
\newcommand{\eeqno}{\end{eqnarray*}}
\newcommand{\ba}{\begin{array}}
\newcommand{\ea}{\end{array}}
\newcommand{\vc}[1]{\mbox{\boldmath $#1$}}
\newcommand{\lvc}[1]{\mbox{\scriptsize \boldmath $#1$}}
\newcommand{\svc}[1]{\mbox{\scriptsize\boldmath $#1$}}

\newcommand{\wh}{\widehat}
\newcommand{\wt}{\widetilde}
\newcommand{\ts}{\textstyle}
\newcommand{\ds}{\displaystyle}
\newcommand{\scs}{\scriptstyle}
\newcommand{\vep}{\varepsilon}
\newcommand{\rhp}{\rightharpoonup}
\newcommand{\cl}{\circ\!\!\!\!\!-}
\newcommand{\bcs}{\dot{\,}.\dot{\,}}
\newcommand{\eqv}{\Leftrightarrow}
\newcommand{\leqv}{\Longleftrightarrow}
\newtheorem{co}{Corollary} 
\newtheorem{lm}{Lemma} 
\newtheorem{Ex}{Example} 
\newtheorem{Th}{Theorem}
\newtheorem{df}{Definition} 
\newtheorem{pr}{Property} 
\newtheorem{pro}{Proposition} 
\newtheorem{rem}{Remark} 

\newcommand{\lcv}{convex } 

\newcommand{\hugel}{{\arraycolsep 0mm
                    \left\{\ba{l}{\,}\\{\,}\ea\right.\!\!}}
\newcommand{\Hugel}{{\arraycolsep 0mm
                    \left\{\ba{l}{\,}\\{\,}\\{\,}\ea\right.\!\!}}
\newcommand{\HUgel}{{\arraycolsep 0mm
                    \left\{\ba{l}{\,}\\{\,}\\{\,}\vspace{-1mm}
                    \\{\,}\ea\right.\!\!}}
\newcommand{\huger}{{\arraycolsep 0mm
                    \left.\ba{l}{\,}\\{\,}\ea\!\!\right\}}}
\newcommand{\Huger}{{\arraycolsep 0mm
                    \left.\ba{l}{\,}\\{\,}\\{\,}\ea\!\!\right\}}}
\newcommand{\HUger}{{\arraycolsep 0mm
                    \left.\ba{l}{\,}\\{\,}\\{\,}\vspace{-1mm}
                    \\{\,}\ea\!\!\right\}}}

\newcommand{\hugebl}{{\arraycolsep 0mm
                    \left[\ba{l}{\,}\\{\,}\ea\right.\!\!}}
\newcommand{\Hugebl}{{\arraycolsep 0mm
                    \left[\ba{l}{\,}\\{\,}\\{\,}\ea\right.\!\!}}
\newcommand{\HUgebl}{{\arraycolsep 0mm
                    \left[\ba{l}{\,}\\{\,}\\{\,}\vspace{-1mm}
                    \\{\,}\ea\right.\!\!}}
\newcommand{\hugebr}{{\arraycolsep 0mm
                    \left.\ba{l}{\,}\\{\,}\ea\!\!\right]}}
\newcommand{\Hugebr}{{\arraycolsep 0mm
                    \left.\ba{l}{\,}\\{\,}\\{\,}\ea\!\!\right]}}
\newcommand{\HUgebr}{{\arraycolsep 0mm
                    \left.\ba{l}{\,}\\{\,}\\{\,}\vspace{-1mm}
                    \\{\,}\ea\!\!\right]}}

\newcommand{\hugecl}{{\arraycolsep 0mm
                    \left(\ba{l}{\,}\\{\,}\ea\right.\!\!}}
\newcommand{\Hugecl}{{\arraycolsep 0mm
                    \left(\ba{l}{\,}\\{\,}\\{\,}\ea\right.\!\!}}
\newcommand{\hugecr}{{\arraycolsep 0mm
                    \left.\ba{l}{\,}\\{\,}\ea\!\!\right)}}
\newcommand{\Hugecr}{{\arraycolsep 0mm
                    \left.\ba{l}{\,}\\{\,}\\{\,}\ea\!\!\right)}}

\newcommand{\hugepl}{{\arraycolsep 0mm
                    \left|\ba{l}{\,}\\{\,}\ea\right.\!\!}}
\newcommand{\Hugepl}{{\arraycolsep 0mm
                    \left|\ba{l}{\,}\\{\,}\\{\,}\ea\right.\!\!}}
\newcommand{\hugepr}{{\arraycolsep 0mm
                    \left.\ba{l}{\,}\\{\,}\ea\!\!\right|}}
\newcommand{\Hugepr}{{\arraycolsep 0mm
                    \left.\ba{l}{\,}\\{\,}\\{\,}\ea\!\!\right|}}

\newenvironment{jenumerate}
	{\begin{enumerate}\itemsep=-0.25em \parindent=1zw}{\end{enumerate}}
\newenvironment{jdescription}
	{\begin{description}\itemsep=-0.25em \parindent=1zw}{\end{description}}
\newenvironment{jitemize}
	{\begin{itemize}\itemsep=-0.25em \parindent=1zw}{\end{itemize}}
\renewcommand{\labelitemii}{$\cdot$}

\newcommand{\iro}[2]{{\color[named]{#1}#2\normalcolor}}
\newcommand{\irr}[1]{{\color[named]{Red}#1\normalcolor}}
\newcommand{\irg}[1]{{\color[named]{Green}#1\normalcolor}}
\newcommand{\irb}[1]{{\color[named]{Blue}#1\normalcolor}}
\newcommand{\irBl}[1]{{\color[named]{Black}#1\normalcolor}}
\newcommand{\irWh}[1]{{\color[named]{White}#1\normalcolor}}

\newcommand{\irY}[1]{{\color[named]{Yellow}#1\normalcolor}}
\newcommand{\irO}[1]{{\color[named]{Orange}#1\normalcolor}}
\newcommand{\irBr}[1]{{\color[named]{Purple}#1\normalcolor}}
\newcommand{\IrBr}[1]{{\color[named]{Purple}#1\normalcolor}}
\newcommand{\irBw}[1]{{\color[named]{Brown}#1\normalcolor}}
\newcommand{\irPk}[1]{{\color[named]{Magenta}#1\normalcolor}}
\newcommand{\irCb}[1]{{\color[named]{CadetBlue}#1\normalcolor}}
%
%
\newcommand{\lcov}{
                     convex }

\newcommand{\cars}{s}
\newcommand{\sigNi}{\sigma_{N_i}^2}
\newcommand{\sigN}{\sigma_{N}^2}

\newcommand{\cef}{c_i}
\newcommand{\cefsq}{c_i^2}

\newcommand{\Iset}{L}

\newcommand{\IsetA}{{\Lambda_L}}
\newcommand{\IsetB}{{\Lambda_K}}

\newcommand{\Ntn}{\mbox{\boldmath $N$}}
\newcommand{\lNtn}{\mbox{\scriptsize\boldmath $N$}}
\newcommand{\Nitn}{\mbox{\boldmath $N$}_i}
\newcommand{\lNitn}{\mbox{\scriptsize\boldmath $N$}_i}
\newcommand{\Nost}{N_{0,t}}
\newcommand{\Nast}{N_{1,t}}
\newcommand{\Nlst}{N_{L,t}}

\newcommand{\tinbNs}{\tilde{N}
                    }
\newcommand{\tiNs}{\tilde{\mbox{\boldmath $N$}}
                  }
\newcommand{\tilNs}{\tilde{\mbox{\scriptsize\boldmath$N$}}
                   }
\newcommand{\Xotn}{\mbox{\boldmath $X$}_0}
\newcommand{\lXotn}{\mbox{\scriptsize\boldmath $X$}_0}
\newcommand{\Xost}{X_{0,t}}

\newcommand{\Xatn}{\mbox{\boldmath $X$}_1}
\newcommand{\Xast}{X_{1}(t)}

\newcommand{\Xbtn}{\mbox{\boldmath $X$}_2}
\newcommand{\Xbst}{X_{2}(t)}

\newcommand{\Xltn}{\mbox{\boldmath $X$}_L}
\newcommand{\Xlst}{X_{L}(t)}
\newcommand{\Xlsn}{X_{L}(n)}

\newcommand{\Xitn}{\mbox{\boldmath $X$}_i}
\newcommand{\lXitn}{\mbox{\scriptsize\boldmath $X$}_i}
\newcommand{\Xisa}{X_{i}(1)}
\newcommand{\Xisb}{X_{i}(2)}
\newcommand{\Xist}{X_{i}(t)}
\newcommand{\Xisn}{X_{i}(n)}
\newcommand{\Xtn}{\mbox{\boldmath $X$}}

\newcommand{\hatXotn}{\hat{\mbox{\boldmath $X$}}_0}
\newcommand{\hatXost}{\hat{X}_{0,t}}

\newcommand{\tiXs}{
                  {\mbox{\boldmath $Y$}}
                  }
\newcommand{\tilXs}{
                   {\mbox{\scriptsize\boldmath $Y$}}
                   }

\newcommand{\Zatn}{\mbox{\boldmath $Z$}_1}
\newcommand{\lZatn}{\mbox{\scriptsize\boldmath $Z$}_1}
\newcommand{\Zbtn}{\mbox{\boldmath $Z$}_2}
\newcommand{\lZbtn}{\mbox{\scriptsize\boldmath $Z$}_2}

\newcommand{\rdf}{J}  

\newcommand{\Dff}{-}  
\newcommand{\coS}{S^{\rm c}} 

\newcommand{\piso}{\pi(k_1)} 
\newcommand{\pisi}{\pi(k_i)} 
\newcommand{\pisiadd}{\pi(k_{i+1})} 
\newcommand{\pisimo}{\pi(k_{i-1})} 
\newcommand{\pisj}{\pi(k_j)} 
\newcommand{\pisjmo}{\pi(k_{j-1})} 
\newcommand{\pisjadd}{\pi(k_{j+1})} 
\newcommand{\piss}{\pi(k_s)} 
\newcommand{\pissmo}{\pi(k_{s-1})} 
\newcommand{\pios}{\pi(S)} 

\newcommand{\pibi}{\pi(B_i)} 
\newcommand{\picobi}{\pi(S\Dff B_{i})}
\newcommand{\pibimo}{\pi(B_{i-1})} 
\newcommand{\picobimo}{\pi(S\Dff B_{i-1})}
\newcommand{\pibj}{\pi(B_j)} 
\newcommand{\picobj}{\pi(S\Dff B_{j})}
\newcommand{\pibs}{\pi(B_s)} 
\newcommand{\pibsmo}{\pi(B_{s-1})} 

\newcommand{\maho}{many-help-one } 
\newcommand{\Mho}{Many-help-one } 
\newcommand{\oho}{one-helps-one } 
\newcommand{\RateDist}{source coding } 

\newcommand{\D}{\mbox{\rm d}} %
\newcommand{\E}{\mbox{\rm E}} %

\newcommand{\conv}{\mbox{\rm conv}} %
\newcommand{\rsub}{\empty}

\newcommand{\DisT}{\Sigma_d}
\newcommand{\DisTi}{\Sigma_{\tilde{d}}}
\newcommand{\Npre}{\preceq{\!\!\!\!\!|\:\:}}
\newcommand{\EP}[1]
{
\ts \frac{1}{2\pi{\rm e}}{\rm e}^{\scriptstyle \frac{2}{n}h(#1)}
}
\newcommand{\CdEP}[2]
{
\ts \frac{1}{2\pi{\rm e}}{\rm e}^{\frac{2}{n}
h(\scriptstyle #1|\scriptstyle #2)}
}
\newcommand{\MEq}[1]{\stackrel{
{\rm (#1)}}{=}}

\newcommand{\MLeq}[1]{\stackrel{
{\rm (#1)}}{\leq}}

\newcommand{\ML}[1]{\stackrel{
{\rm (#1)}}{<}}

\newcommand{\MGeq}[1]{\stackrel{
{\rm (#1)}}{\geq}}

\newcommand{\MG}[1]{\stackrel{
{\rm (#1)}}{>}}

\newcommand{\MPreq}[1]{\stackrel{
{\rm (#1)}}{\preceq}}

\newcommand{\MSueq}[1]{\stackrel{
{\rm (#1)}}{\succeq}}
%
\title{
Distributed Source Coding 
of Correlated 
Gaussian Sources
}
%
\author{Yasutada~Oohama
\thanks{Manuscript received xxx, 20XX; revised xxx, 20XX.}%
\thanks{Y. Oohama is with the Department of Information Science 
        and Intelligent Systems, 
        University of Tokushima,
        2-1 Minami Josanjima-Cho, Tokushima 
        770-8506, Japan.}
}
\markboth{
}
{Oohama: 
}
%
%
%
%
\maketitle

\begin{abstract}
We consider the distributed source coding system of $L$ correlated
Gaussian sources $Y_l,l=1,2,\cdots,L$ which are noisy observations of
correlated Gaussian remote sources $X_k, k=1,2,\cdots,K$. We assume that
$Y^{L}={}^{\rm t}(Y_1,Y_2,$ $\cdots, Y_L)$ is an observation of 
the source vector $X^K={}^{\rm t}(X_1,X_2,\cdots, X_K)$, having the form
$Y^L=AX^K+N^L$, where $A$ is a $L\times K$ matrix and 
$N^L={}^{\rm t}(N_1,N_2,\cdots,N_L)$ 
is a vector of $L$ independent Gaussian
random variables also independent of $X^K$. In this system $L$
correlated Gaussian observations are separately compressed by $L$
encoders and sent to the information processing center. We study the
remote source coding problem where the decoder at the center attempts
to reconstruct the remote source $X^K$.
We consider three distortion criteria based on the covariance
matrix of the estimation error on $X^K$. For each of those three
criteria we derive explicit inner and outer bounds of the rate
distortion region. Next, in the case of $K=L$ and $A=I_L$, we
study the multiterminal \RateDist problem where the decoder
wishes to reconstruct the observation $Y^L=X^L+N^L$.  To investigate
this problem we shall establish a result which provides a strong
connection between the remote source coding problem and the 
multiterminal \RateDist problem. Using this result, we drive 
several new partial solutions to the 
multiterminal \RateDist problem.
\end{abstract}


\IEEEpeerreviewmaketitle

\section{Introduction}

Distributed source coding systems of correlated information sources
are a form of communication system which is significant from both
theoretical and practical points of view in multi-user source
networks. The first fundamental theory in those coding systems was
established by Slepian and Wolf \cite{sw}. They considered a
distributed source coding system of two correlated information
sources.  Those two sources are separately encoded and sent to a
single destination, where the decoder wishes to decode the original
sources.  In the above distributed source coding systems we can
consider a situation where the source outputs should be reconstructed
with average distortions smaller than prescribed levels. This
situation yields a kind of multiterminal rate distortion theory in the
framework of distributed source coding.  The rate distortion region is
defined by the set of all rate vectors for which the source outputs
are reconstructed with average distortions smaller than prescribed
levels. The determination problem of the rate distortion region is
often called the multiterminal source coding problem.

The multiterminal source coding problem was intensively studied by
\cite{wz}-\cite{wa}. 
Wagner and Anantharam \cite{wg} gave a new method to evaluate an outer bound of 
the rate distortion region. 
Wagner {\it et al.} \cite{wg3} gave a complete
solution to this problem in the case of Gaussian information sources
and quadratic distortion by proving that the sum rate part of the
inner bound of Berger \cite{bt} and Tung \cite{syt} is optimal. Wang
{\it et al.} \cite{wa} gave a new alternative proof of the sum rate
part optimality.  In spite of a recent progress made by those three 
works, the multiterminal source coding problem still largely remains
open.

As a practical situation of the distributed source coding system, we can
consider a case where the distributed encoders can not directly access
the source outputs but can access their noisy observations. This
situation was first studied by Yamamoto and Ito \cite{yam0}. They call
the investigated coding system the communication system with a remote
source. Subsequently, a similar distributed source coding 
system was studied by Flynn and Gray \cite{fg}.

In this paper we consider the distributed source coding system of $L$
correlated Gaussian sources $Y_l,l=1,2,\cdots,L$ which are noisy
observations of $X_k,k=1,2,\cdots,K$. We assume that 
$Y^{L}={}^{\rm t}(Y_1,Y_2,$ $\cdots, Y_L)$ is an observation of 
the source vector
$X^K={}^{\rm t}(X_1,X_2,\cdots, X_K)$, having the form
$Y^L=AX^K+N^L\,,$ where $A$ is a $L\times K$ matrix and $N^L={}^{\rm
  t}(N_1,N_2,\cdots,N_L)$ is a vector of $L$ independent Gaussian
random variables also independent of $X^K$.  In this system $L$
correlated Gaussian observations are separately compressed by $L$
encoders and sent to the information processing center. We study the
remote source coding problem where the decoder at the center attempts
to reconstruct the remote source $X^K$.

We consider three distortion criteria based on the covariance matrix of 
the average estimation error on $X^K$. The first criterion is called 
the distortion matrix criterion, where the estimation error must not 
exceed an arbitrary prescribed covariance matrix in the meaning of 
positive semi definite. The second criterion is called the vector 
distortion criterion, where for a fixed positive vector $D^K=(D_1,D_2,$ 
$\cdots, D_K)$ and for each $k=1,2,\cdots,K$, the diagonal $(k,k)$ element 
of the covariance matrix is upper bounded by $D_k$. The third criterion 
is called the sum distortion criterion, where the trace of the 
covariance matrix must not exceed a prescribed positive level $D$. 
For each distortion criterion the rate distortion region is defined 
by a set of all rates vectors for which the estimation error 
does not exceed an arbitrary prescribed distortion level. 

For the first distortion criterion, i.e., the distortion matrix 
criterion we derive explicit inner and outer bounds of the rate 
distortion region. Those two bounds have a form of positive semi 
definite programming with respect to covariance matrices. Using this 
results, for each of the second and third distortion criteria we 
derive explicit inner and outer bounds of the rate distortion 
region. In the case of vector distortion criterion our 
outer bound includes that of Oohama \cite{oh8} 
as a special case by letting $K=L$ and $A=I_L$.   
In the case of sum distortion criterion 
we derive more explicit outer bound of the rate distortion region 
having a form of water filling solution. In this case 
we further show that if the prescribed distortion level 
$D$ does not exceed a certain threshold, 
the inner and outer bounds match and derive 
two different thresholds. The first threshold improves 
the threshold obtained by Oohama \cite{oh7b},\cite{oh9a} in the case of 
$K=L,A=I_L$. The second threshold improves the first one 
for some cases but neither subsumes the other.   

When $K=1$, the distributed source coding system treated in this paper
becomes the quadratic Gaussian CEO problem investigated by \cite{wa},
\cite{vb}-\cite{ptr}. The system in the case of $K=L$ and sum
distortion criterion was studied by Pandya {\it et al.}
\cite{pdya}. They derived lower and upper bounds of the minimum sum
rate in the rate distortion region. Several partial solutions in the
case of $K=L$, $A=I_L$, and sum distortion criterion were obtained by
\cite{oh5}-\cite{oh9a}. The case of $K=L$, $A=I_L$,  and vector
distortion criterion was studied by \cite{oh8}.

Recently, Yang and Xiong \cite{yx} have studied the same problem. They
have derived two outer bounds of the rate distortion
region in the case of sum rate distortion criterion. 
When $K=L,A=I_L$, the first outer bound does not
coincide with the outer bound obtained by Oohama
\cite{oh7}-\cite{oh9a}.   
When ${}^{\rm t}AA=I_K$, they have obtained the second 
outer bound tighter than the first one. This bound 
is the same as that of our result of this paper.
When ${}^{\rm t}AA = I_K$, Yang {\it et al.} \cite{yx2} have 
derived a threshold on the distortion level $D$ such that for $D$
below this threshold their second outer bound is tight. Their
threshold also improves that of Oohama \cite{oh7b},\cite{oh9a} in the
case of $K=L,A=I_L$. Comparing the formula of our first threshold with
that of and Yang {\it et al.} \cite{yx2}, we can see that we have no obvious
superiority of either to the other. On the other hand, our second
threshold is better than their threshold for some nontrivial cases.

In this paper, in the case of $K=L$ and $A=I_L$, we study the
multiterminal \RateDist problem where the decoder wishes to
reconstruct the observation $Y^L=X^L+N^L$.  Similarly to the case of
remote source coding problem, we consider three types of distortion
criteria based on the covariance matrix of the estimation error on
$Y^L$. Based on the above three criteria, three rate distortion
regions are defined.

The remote source coding problem is often referred to as the indirect
distributed source coding problem.  On the other hand, the
multiterminal \RateDist problem in the frame work of distributed
source coding is often called the direct distributed source coding
problem. As shown in the paper of Wagner {\it et al.}  \cite{wg3} and
in the recent work by Wang {\it et al.} \cite{wa}, we have a strong
connection between the direct and indirect distributed source coding
problems. To investigate the determination problem of the three rate
distortion regions for the multiterminal \RateDist problem we shall
establish a result which provides a strong connection between the
remote source coding problem and the multiterminal \RateDist
problem. This result states that all results on the rate distortion
region of the remote source coding problem can be converted into those
on the rate distortion region of the multiterminal source coding
problem. Using this relation and our results on the remote source
coding problem, we drive new three outer bounds of the rate distortion
regions for each of three distortion criteria.

In the case of vector distortion criterion, we can obtain a lower
bound of the sum rate part of the rate distortion region by using the
established outer bound in this case. This bound has a form of
positive semidefinite programming. By some analytical computation we
can show that this lower bound is equal to the lower bound obtained by
Wang {\it et al.} \cite{wa} and tight when $L=2$.  Our method to
derive this result essentially differs from the method of Wang {\it et
al.}  \cite{wa}. It is also quite different from that of Wagner {\it
et al.}  \cite{wg3}.  Hence in the case of two terminal Gaussian
sources there exists three different proofs of the optimality of the
sum rate part of the inner bound of Berger \cite{bt} and Tung
\cite{syt}.

In the case of sum distortion criterion we derive an explicit threshold 
such that for the distortion level $D$ below 
this threshold the outer bound coincides with 
the inner bound. An important feature of the multiterminal rate 
distortion problem is that the rate distortion region remains the same 
for any choice of covariance matrix $\Sigma_{X^L}$ and diagonal 
covariance matrix $\Sigma_{N^L}$ satisfying $\Sigma_{Y^L}$ 
$=\Sigma_{X^L}+\Sigma_{N^L}$. Using this feature, we find a pair 
$(\Sigma_{X^L},$ $\Sigma_{N^L})$ which maximizes the threshold 
subject to $\Sigma_{Y^L}$ $=\Sigma_{X^L}+\Sigma_{N^L}$. 

Let 
$\tau(Y^L)\defeq (Y_2,Y_3,\cdots,Y_L,Y_1)$  
be a cyclic shift of the source $Y^L=(Y_1,Y_2,Y_3,$ $\cdots,Y_L)$. 
We say that the source $Y^L$ has the cyclic shift invariant
property if the covariance matrix $\Sigma_{\tau({Y^L})}$ of $\tau(Y^L)$
is the same as the covariance matrix $\Sigma_{Y^L}$ of $Y^L$.  When
$Y^L$ has the cyclic shift invariant property, 
we investigate the sum rate part of the rate distortion region. We derive an explicit
upper bound of the sum rate part from the inner bounds 
of the rate distortion region. 
On a lower bound of the sum rate part we derive a
new explicit bound by making full use of the cyclic shift invariance
property of $\Sigma_{Y^L}$. We further derive an explicit sufficient 
condition for the lower bound to coincide with the
upper bound.  We show that the lower and upper bounds match if 
the distortion does not exceed a threshold which is a function
of $\Sigma_{Y^L}$ and find an explicit form of this threshold.
As a corollary of this result, in the case of 
vector distortion criterion we obtain the optimal sum rate when
${Y^L}$ is cyclic shift invariant and $D^L$ has $L$ components with an
identical value $D$ below a certain threshold depending 
only on $\Sigma_{Y^L}$

%
%
%

\section{Problem Statement and Previous Results}

\subsection{
Formal Statement of Problem
}

\newcommand{\baseN}{\rm e}

In this subsection we present a formal statement 
of problem. Throughout this paper all logarithms 
are taken to the base natural. Let 
$
\IsetB\defeq \{1,2,\cdots,K\}
$
and 
$
\IsetA\defeq \{1,2,\cdots,L\}.
$ 
Let $X_k, k\in\IsetB$ be correlated zero 
mean Gaussian random variables. 
For each $k\in\IsetB$, $X_k$ takes values 
in the real line 
$\mathbb{R}$. We write a $K$ dimensional random vector as 
$X^K=$ ${}^{\rm t}(X_1,X_2,$ $\cdots, X_K)$. 
We denote the covariance matrix of $X^K$ by $\Sigma_{X^K}$. 
Let $Y^{L}\defeq {}^{\rm t}(Y_1,Y_2,$ $\cdots, Y_L)$ 
be an observation of the source vector $X^K$, 
having the form $Y^L=AX^K+N^L$, where $A$ is 
a $L\times K$ matrix and $N^L={}^{\rm t}(N_1,N_2,\cdots,N_L)$ 
is a vector of $L$ independent zero mean Gaussian random 
variables also independent of $X^K$. For $l\in\IsetA$, $\sigma_{N_l}^2$ 
stands for the variance of $N_l$. 
Let $\{(\Xast,$ $\Xbst, \cdots, X_K(t))\}_{t=1}^{\infty}$
be a stationary memoryless multiple Gaussian source. 
For each $t=1,$$2,\cdots,$ $X^K(t)\defeq $ 
${}^{\rm t}(X_{1}(t),X_{2}(t),\cdots,$ $\!X_{k}(t))\,$ 
has the same distribution as $X^K$. A random vector 
consisting of $n$ independent copies of 
the random variable $X_k$ is denoted by 
$$
{\vc X}_k\defeq (X_{k}(1),X_{k}(2),\cdots, X_{k}(n)).
$$
For each $t=1,2,\cdots$, 
$
Y^L(t)\defeq {}^{\rm t}(Y_1(t),$ $Y_2(t),\cdots,Y_L(t))
$
is a vector of $L$ correlated observations of 
$X^K(t)$, having the form 
$
Y^L(t)=AX^K(t)+N^L(t),
$
where $N^L(t),t=1,2,\cdots$, are independent 
identically distributed (i.i.d.) Gaussian random 
vector having the same distribution as $N^L$. 
We have no assumption on the number of observations $L$, 
which may be $L\geq K$ or $L<K$.   


\bfig
\setlength{\unitlength}{1.00mm}
\begin{picture}(80,48)(2,0)

\put(5,33){\framebox(6,6){$X_1$}}
\put(11,36){\vector(1,0){7.5}}

\put(5,19){\framebox(6,6){$X_2$}}
\put(11,22){\vector(1,0){7.5}}

\put(7,12){$\vdots$}

\put(5,2){\framebox(6,6){$X_K$}}
\put(11,5){\vector(1,0){7.5}}

\put(18.5,1.5){\framebox(4,38){$A$}}

\put(12.5,38.6){${\vc X}_1$}
\put(30,40.6){${\vc Y}_1$}
\put(26,45){${\vc N}_1$}

\put(12.5,24.6){${\vc X}_2$}
\put(30,25.6){${\vc Y}_2$}
\put(26,30){${\vc N}_2$}

\put(12.5,7.6){${\vc X}_K$}
\put(30,5.6){${\vc Y}_L$}
\put(26,10){${\vc N}_L$}

\put(22.5,38){\vector(1,0){4}}
\put(29.5,38){\vector(1,0){7.5}}
\put(28,44){\vector(0,-1){4.5}}
\put(28,37){\line(0,1){2}}
\put(27,38){\line(1,0){2}}
\put(28,38){\circle{3.0}}

\put(22.5,23){\vector(1,0){4}}
\put(29.5,23){\vector(1,0){7.5}}
\put(28,29){\vector(0,-1){4.5}}
\put(28,22){\line(0,1){2}}
\put(27,23){\line(1,0){2}}
\put(28,23){\circle{3.0}}

\put(22.5,3){\vector(1,0){4}}
\put(29.5,3){\vector(1,0){7.5}}
\put(28,9){\vector(0,-1){4.5}}
\put(28,2){\line(0,1){2}}
\put(27,3){\line(1,0){2}}
\put(28,3){\circle{3.0}}

\put(37,34.5){\framebox(7,7){$\varphi_1^{(n)}$}}
\put(45,40.6){$\varphi_1^{(n)}({\vc Y}_1)$}

\put(37,19.5){\framebox(7,7){$\varphi_2^{(n)}$}}
\put(45,25.6){$\varphi_2^{(n)}({\vc Y}_2)$}
\put(39.5,11){$\vdots$}
\put(37,-0.5){\framebox(7,7){$\varphi_L^{(n)}$}}
\put(45,5.6){$\varphi_L^{(n)}({\vc Y}_L)$}

\put(44,38){\line(1,0){15}}
\put(59,38){\vector(1,-3){5}}

\put(44,23){\line(1,0){15}}
\put(59,23){\vector(1,0){5}}

\put(44,3){\line(1,0){15}}
\put(59,3){\vector(1,4){5}}

\put(64,19.5){\framebox(7,7){$\psi^{(n)}$}}
\put(71,23){\vector(1,0){3}}
\put(74,22){$\left[
             \ba[c]{c}
              \hat{\vc X}_1\\
              \hat{\vc X}_2\\
              \vdots\\
             \hat{\vc X}_K
             \ea\right]
             $}
\end{picture}
\caption{
Distributed source coding system for $L$ correlated 
Gaussian observations
}
\label{fig:Fig1}
\efig

The distributed source coding system for $L$ 
correlated Gaussian observations treated in this 
paper is shown in Fig. \ref{fig:Fig1}. In this coding system 
the distributed encoder functions $\varphi_l, l\in\IsetA$ 
are defined by
$
\varphi_l^{(n)}: \mathbb{R}^n 
\mapsto {\cal M}_l 
\defeq \left\{1,2,\cdots, M_l\right\}.  
\label{eqn:enc}
$
For each $l\in\IsetA$, set 
$
R_l^{(n)}\defeq \frac{1}{n}\log M_l\,, 
$
which stands for the transmission rate of 
the encoder function $\varphi_l^{(n)}$.
The joint decoder function $\psi^{(n)}=$ 
$(\psi_1^{(n)},$ $\psi_2^{(n)},$ $\cdots,\psi_K^{(n)})$ 
is defined by 
\beqno
& &\psi^{(n)}\defeq (\psi_1^{(n)},\psi_2^{(n)},\cdots,\psi_K^{(n)}),
\\
& &\psi_k^{(n)}: {\cal M}_1 \times \cdots \times {\cal M}_L 
\mapsto \mathbb{R}^n, k\in \IsetB.
\eeqno
For ${\vc X}^K$ $=({\vc X}_1,$ ${\vc X}_2,$ $\cdots,$ 
${\vc X}_K)$, set
\beqno
\varphi^{(n)}({\vc Y}^{L}) & \defeq &
            (\varphi_1^{(n)}({\vc Y}_1),
             \varphi_2^{(n)}({\vc Y}_2),
      \cdots,\varphi_L^{(n)}({\vc Y}_L)),
\\
   \hat{\vc X}^{K}
   &=&\left[
\ba{c}
\hat{\vc X}_1\\
\hat{\vc X}_2\\
      \vdots\\
\hat{\vc X}_K\\
\ea
\right]
\defeq 
\left[
\ba{c}
\psi_1^{(n)}(\varphi^{(n)}({\vc Y}^L))\\
\psi_2^{(n)}(\varphi^{(n)}({\vc Y}^L))\\
\vdots\\ 
\psi_K^{(n)}(\varphi^{(n)}({\vc Y}^L))\\
\ea
\right]\,,
\\
d_{kk}
& \defeq & {\rm E}||{\vc X}_k-\hat{\vc X}_k||^2, 
1\leq k\leq K,
\nonumber\\
d_{kk^{\prime}}
& \defeq &
{\rm E} \langle {\vc X}_k-\hat{\vc X}_k,
        {\vc X}_{k^{\prime}}-\hat{\vc X}_{k^{\prime}}\rangle, 
1\leq k\ne {k^{\prime}}\leq K,
\nonumber
\eeqno
where $||{\vc a}||$ stands for the Euclid norm of $n$ dimensional
vector ${\vc a}$ and $\langle {\vc a},{\vc b}\rangle$ stands for the
inner product between ${\vc a}$ and ${\vc b}$. Let 
$\Sigma_{{\lvc X}^K-\hat{\lvc X}^K}$ be a covariance matrix with 
$d_{k k^{\prime}}$ 
in its $(k,{k^{\prime}})$ element. Let $\DisT$ be a given
$K\times K$ covariance matrix which serves as a distortion
criterion. We call this matrix a distortion matrix.

For a given distortion matrix $\DisT$, the rate 
vector $(R_1,$ $R_2,\cdots, R_L)$ is $\DisT$-{\it admissible} 
if there exists a sequence 
$\{(\varphi_1^{(n)},$
   $\varphi_2^{(n)}, \cdots,$ 
   $\varphi_L^{(n)},$ 
   $\psi^{(n)})\}_{n=1}^{\infty}$ 
such that
\beqno
& &\limsup_{n\to\infty}R_l^{(n)}\leq R_l, 
   \mbox{ for }l\in\IsetA\,,
\\
& &\limsup_{n\to\infty}{\ts \frac{1}{n}}
   \Sigma_{{\lvc X}^K-\hat{\lvc X}^K} \preceq \DisT \,, 
\eeqno
where $A_1\preceq A_2$ means that $A_2-A_1$ 
is a positive semi-definite matrix. 
Let ${\cal R}_{\Iset}(\DisT|\Sigma_{X^KY^L})$ 
denote the set of all $\DisT$-admissible 
rate vectors. We often have a particular interest 
in the minimum sum rate part of the rate distortion 
region. To examine this quantity, we set  
$$ 
R_{{\rm sum}, L}(\DisT|\Sigma_{X^KY^L})
\defeq \min_{\scs (R_1,R_2,\cdots,R_L)
\atop{\scs \in {\cal R}_{\Iset}(\Gamma,D^K|\Sigma_{X^KY^L})}}
\left\{\sum_{l=1}^{L}R_l\right\}.
$$
We consider two types of distortion criterion. For each 
distortion criterion we define the determination problem 
of the rate distortion region. 

{\it Problem 1. Vector Distortion Criterion: }  
Fix $K\times K$ invertible matrix $\Gamma$ and 
positive vector ${D}^K=$ $(D_1,$ $D_2,\cdots$ $, D_K)$. 
For given $\Gamma$ and $D^K$, the rate vector 
$(R_1,R_2,\cdots, R_L)$ is $(\Gamma,D^K)$-{\it admissible} 
if there exists a sequence 
$\{(\varphi_1^{(n)},$ 
   $\varphi_2^{(n)}, \cdots,$ 
   $\varphi_L^{(n)},$ $\psi^{(n)})\}_{n=1}^{\infty}$ 
such that
\beqno
& &\limsup_{n\to\infty}R^{(n)}\leq R_l,\mbox{ for }l\in\IsetA,  
\\
& &\limsup_{n\to\infty}
\left[\Gamma\left({\ts \frac{1}{n}}
\Sigma_{{\lvc X}^K-\hat{\lvc X}^K}\right){}^{\rm t}\Gamma\right]_{kk} 
\leq D_k\,,\mbox{ for }k\in\IsetB, 
\eeqno
where $[C]_{ij}$ stands for the $(i,j)$ element of the matrix $C$. 
Let ${\cal R}_{\Iset}(\Gamma,D^K|\Sigma_{X^KY^L})$ denote the set 
of all $(\Gamma,D^K)$-admissible rate vectors. When $\Gamma$ 
is equal to the $K\times K$ identity matrix $I_K$, 
we omit $\Gamma$ in 
${\cal R}_{\Iset}(\Gamma,D|\Sigma_{X^KY^L})$ 
to simply write ${\cal R}_{\Iset}(D|\Sigma_{X^KY^L})$. 
Similar notations are used for other sets or quantities. 
The sum rate part of 
${\cal R}_{\Iset}(\Gamma,D^K|\Sigma_{X^KY^L})$ is defined by 
$$ 
R_{{\rm sum}, L}(\Gamma,D^K|\Sigma_{X^KY^L})
\defeq \min_{\scs (R_1,R_2,\cdots,R_L)
\atop{\scs \in {\cal R}_{\Iset}(\Gamma,D^K|\Sigma_{X^KY^L})}}
\left\{\sum_{l=1}^{L}R_l\right\}.
$$

{\it Problem 2. Sum Distortion Criterion:}
Fix $K\times K$ positive definite invertible matrix $\Gamma$ and 
positive $D$. For given $\Gamma$ and $D$, 
the rate vector $(R_1,R_2,\cdots, R_L)$ is 
$(\Gamma,D)$-{\it admissible} 
if there exists a sequence 
$\{(\varphi_1^{(n)},$ 
   $\varphi_2^{(n)}, \cdots,$ 
   $\varphi_L^{(n)},$ $\psi^{(n)})\}_{n=1}^{\infty}$ 
such that
\beqno
&&\limsup_{n\to\infty}R^{(n)}\leq R_l,
\mbox{ for }l\in\IsetA, 
\\
&&\limsup_{n\to\infty}
{\rm tr}\left[
\Gamma\left({\ts \frac{1}{n}}\Sigma_{{\lvc X}^K-\hat{\lvc X}^K}\right)
{}^{\rm t}\Gamma\right] 
\leq D. 
\eeqno
The sum rate part of 
${\cal R}_{\Iset}(\Gamma,D|\Sigma_{X^KY^L})$ is defined by
$$ 
R_{{\rm sum},L}(\Gamma,D|\Sigma_{X^KY^L})
\defeq \min_{\scs (R_1,R_2,\cdots,R_L)
\atop{\scs \in{\cal R}_{\Iset}(\Gamma,D|\Sigma_{X^KY^L})}}
\left\{\sum_{l=1}^{L}R_l\right\}.
$$
Let ${\cal S}_K(D^K)$ be a set of all $K\times K$ 
covariance matrices whose $(k,k)$ element do not 
exceed $D_k$ for $k\in\IsetB$. Then we have
\beqa
& &{\cal R}_L(\Gamma,D^K|\Sigma_{X^KY^L})
   =\bigcup_{\Gamma \DisT {}^{\rm t}\Gamma \in {\cal S}_K(D^K)}
   \hspace*{-2mm}{\cal R}_L({\DisT}|\Sigma_{X^KY^L}),
\label{eqn:char1z}
\\
& &{\cal R}_L(\Gamma,D|\Sigma_{X^KY^L})
  =\bigcup_{{\rm tr}[\Gamma\DisT {}^{\rm t}\Gamma] \leq D}
   \hspace*{-2mm}{\cal R}_L({\DisT}|\Sigma_{X^KY^L}).
\label{eqn:char2z}
\eeqa
Furthermore, we have 
\beq
{\cal R}_L(\Gamma,D|\Sigma_{X^KY^L})
=\bigcup_{\sum_{k=1}^K D_k\leq D}{\cal R}_L(\Gamma,D^K|\Sigma_{X^KY^L}).
\eeq
In this paper we establish explicit inner and outer bounds 
of ${\cal R}_L({\DisT}|\Sigma_{X^KY^L})$. 
Using the above bounds and equations (\ref{eqn:char1z}) 
and (\ref{eqn:char2z}), we give new outer 
bounds of ${\cal R}_L(\Gamma,D|\Sigma_{X^KY^L})$ and 
${\cal R}_L(\Gamma,D^K|\Sigma_{X^KY^L})$. 

\subsection{
Inner Bounds and Previous Results
}

In this subsection we present inner bounds of 
${\cal R}_L(\DisT$ $|\Sigma_{X^KY^L})$,
${\cal R}_L(\Gamma,D^L$ $|\Sigma_{X^KY^L})$, and
${\cal R}_L(\Gamma,D$ $|\Sigma_{X^KY^L})$.
Those inner bounds can be obtained by a standard technique 
developed in the field of multiterminal source coding.

For $l\in \IsetA$, let ${U}_l$ 
be a random variable taking values in 
the real line $\mathbb{R}$. 
For any subset 
$S\subseteq \IsetA$, 
we introduce the notation 
$U_S=(U_l)_{l\in S}$. In particular $U_\IsetA=$ $U^L=$ 
$(U_1,$ $U_2,$ $\cdots, U_L)$.   
Define 
\beqno
{\cal G}(\DisT)
&\defeq& \ba[t]{l}
 \left\{U^L \right.:
  \ba[t]{l} 
  U^L\mbox{ is a Gaussian }
  \vspace{1mm}\\
  \mbox{random vector that satisfies}
  \vspace{1mm}\\
  U_S\to Y_S \to X^K \to Y_{S^{\rm c}} \to U_{S^{\rm c}}\,, 
  \vspace{1mm}\\
  U^L \to Y^L \to X^K
  \vspace{1mm}\\
  \mbox{for any $S\subseteq \IsetA$ and }\\ 
  \Sigma_{X^K-{\psi}(U^L)} \preceq \DisT
  \vspace{1mm}\\
  \mbox{for some linear mapping }
  \vspace{1mm}\\
  {\psi}: \mathbb{R}^L\to \mathbb{R}^K. 
  \left. \right\}
  \ea
\ea
\eeqno
and set 
\beqno
& &
\hat{\cal R}_{L}^{({\rm in})}(\DisT|\Sigma_{X^KY^L})
\nonumber\\
&\defeq&{\rm conv}\ba[t]{l}
\left\{R^L \right. : 
  \ba[t]{l}
  \mbox{There exists a random vector}
  \vspace{1mm}\\ 
  U^L\in {\cal G}(\DisT) \mbox{ such that }
  \vspace{1mm}\\
  \ds \sum_{l \in S} R_l \geq I(U_S;Y_S|U_{S^{\rm c}})
  \vspace{1mm}\\
  \mbox{ for any } S\subseteq \IsetA.
  \left. \right\}\,,
  \ea
\ea
\eeqno 
where $\conv\{A\}$ stands for the convex hull of the set $A$. 
Set 
\beqno
& &\hat{\cal R}_L^{\rm (in)}(\Gamma,D^K|\Sigma_{X^KY^L})
\\
&\defeq& \conv\left\{
    \bigcup_{\Gamma \DisT {}^{\rm t}\Gamma \in {\cal S}_K(D^K)}
    {\cal R}_L({\DisT}|\Sigma_{X^KY^L})
    \right\},
\\
& &\hat{\cal R}_L^{\rm (in)}(\Gamma,D|\Sigma_{X^KY^L})
\\
&\defeq&\conv\left\{
    \bigcup_{{\rm tr}[\Gamma\DisT {}^{\rm t}\Gamma] \leq D}
   {\cal R}_L({\DisT}|\Sigma_{X^KY^L})
   \right\}.
\eeqno
Define 
\beqno
\Sigma_{X^K|Y^L}
\defeq (\Sigma_{X^K}^{-1}+{}^{\rm t}A\Sigma_{N^L}^{-1}A)^{-1}
\eeqno
and set
\beqno
d^K(\Gamma\Sigma_{X^K|Y^L}{}^{\rm t}\Gamma)
&\defeq&\left([\Gamma\Sigma_{X^K|Y^L}{}^{\rm t}\Gamma]_{11},
         [\Gamma\Sigma_{X^K|Y^L}{}^{\rm t}\Gamma]_{22},
\right.\\
& &\left.
\:\cdots,[\Gamma\Sigma_{X^K|Y^L}{}^{\rm t}\Gamma]_{KK}\right).
\eeqno
We can show that  
$\hat{\cal R}_{L}^{({\rm in})}(\DisT|\Sigma_{X^KY^L})$,
$\hat{\cal R}_{L}^{({\rm in})}(\Gamma,D^L|\Sigma_{X^KY^L})$,
and $\hat{\cal R}_{L}^{({\rm in})}(\Gamma,D|\Sigma_{X^KY^L})$
satisfy the following property.
\begin{pr}
\label{pr:prz001z} 

\noindent
\begin{itemize}
\item[{\rm a)}]The set 
$\hat{\cal R}_{L}^{({\rm in})}(\DisT|\Sigma_{X^KY^L})$ is not void 
if and only if $\DisT \succ \Sigma_{X^K|Y^L}$.
\item[{\rm b)}]The set 
$\hat{\cal R}_{L}^{({\rm in})}(\Gamma,D^K|\Sigma_{X^KY^L})$ is not void 
if and only if $D^K > d^K(\Gamma$ 
$\Sigma_{X^K|Y^L}{}^{\rm t}\Gamma)$.
\item[{\rm c)}]The set 
$\hat{\cal R}_{L}^{({\rm in})}(\Gamma,D|\Sigma_{X^KY^L})$ is not void 
if and only if $D > {\rm tr }[\Gamma\Sigma_{X^K|Y^L}{}^{\rm t}\Gamma]$.
\end{itemize}
\end{pr}

On inner bounds of 
${\cal R}_{L}(\DisT|\Sigma_{X^KY^L})$, 
${\cal R}_{L}(\Gamma,D^L|\Sigma_{X^KY^L}$ $)$,
and $\hat{\cal R}_{L}(\Gamma,D|\Sigma_{X^KY^L})$,  
we have the following result.
\begin{Th}[Berger \cite{bt} and Tung \cite{syt}]\label{th:direct}
For any $\DisT$ $\succ$ \\$\Sigma_{X^K|Y^L}$, we have
\beqno
& &\hat{\cal R}_{L}^{({\rm in})}(\DisT|\Sigma_{X^KY^L})
   \subseteq {\cal R}_{L}(\DisT|\Sigma_{X^KY^L}).
\eeqno
For any $\Gamma$ and any $D^K$ $>$ 
$d^K(\Gamma\Sigma_{X^K|Y^L}{}^{\rm t}\Gamma)$, 
we have 
\beqno
& &\hat{\cal R}_{L}^{({\rm in})}(\Gamma,D^K|\Sigma_{X^KY^L})
   \subseteq {\cal R}_{L}(\Gamma,D^K|\Sigma_{X^KY^L}).
\eeqno
For any $\Gamma$ and any 
$D$ $>{\rm tr}[\Gamma\Sigma_{X^K|Y^L}{}^{\rm t}\Gamma]$, 
we have 
\beqno
& &\hat{\cal R}_{L}^{({\rm in})}(\Gamma,D|\Sigma_{X^KY^L})
   \subseteq {\cal R}_{L}(\Gamma,D|\Sigma_{X^KY^L}).
\eeqno
\end{Th}

The above three inner bounds can be regarded as variants 
of the inner bound which is well known as that of 
Berger \cite{bt} and Tung \cite{syt}.

When $K=1$ and $L\times 1$ column vector $A$ has the form 
$
A={}^{\rm t}[{11\cdots 1}],$ 
the system considered here becomes the quadratic Gaussian CEO problem. 
This problem was first posed and investigated by Viswanathan and Berger 
\cite{vb}. They further assumed $\Sigma_{N^L}=\sigma^2I_L$. 
Set $\sigma_X^2\defeq\Sigma_X$ and 
$$ 
R_{\rm sum}(D|\sigma_X^2,\sigma^2)
\defeq \liminf_{L\to\infty} R_{{\rm sum},L}(D|\Sigma_{XY^L}).
$$ 
Viswanathan and Berger \cite{vb} studied an asymptotic form 
of $R_{\rm sum}(D|\sigma_X^2,\sigma^2)$ for small $D$. 
Subsequently, Oohama \cite{oh2} determined an exact form 
of $R_{\rm sum}(D|\sigma_X^2,\sigma^2)$. 
The region ${\cal R}_L(D|\Sigma_{XY^L})$ 
was determined independently by Oohama \cite{oh4} 
and Prabhakaram {\it et al.} \cite{ptr}.
Wang {\it et al.}\cite{wa} obtained the same characterization 
of $R_{{\rm sum},L}(D|\Sigma_{XY^L})$ as that of Oohama \cite{oh4} 
in a new alternative method. Their method is based on 
the order of the variances associated
with the minimum mean square error (MMSE) estimation.
Unlike the method of Oohama \cite{oh4}, the method 
of Wang {\it et al.} \cite{wa} 
is not directly applicable to the characterization of 
the entire rate distortion region 
${\cal R}_{L}(D|\Sigma_{XY^L})$.  

In the case where $K=L=2$ and $\Gamma=A=I_2$, 
Wagner {\it et al.} \cite{wg3}
determined ${\cal R}_2(D^2|$ $\Sigma_{X^2Y^2})$. 
Their result is as follows. 
\begin{Th}[Wagner {\it et al.} \cite{wg3}]
For any $D^2>d^2([\Sigma_{X^2|}$ ${}_{Y^2}])$, we have  
$$
{\cal R}_{2}(D^2|\Sigma_{X^2Y^2})
=\hat{\cal R}_{2}^{({\rm in})}(D^2|\Sigma_{X^2Y^2}).
$$
\end{Th}

Their method for the proof depends heavily 
on the specific property of $L=2$. It is hard 
to generalize it to the case of $L\geq 3$.

In the case where $K=L$ and $\Gamma=A=I_L$, 
Oohama \cite{oh5}-\cite{oh9a} derived inner and outer 
bounds of ${\cal R}_{L}(D|\Sigma_{X^LY^L})$. 
Oohama \cite{oh7}, \cite{oh7b}, \cite{oh9a} also derived explicit 
sufficient conditions for inner and outer bounds to 
match. In \cite{oh8}, Oohama derived explicit outer bounds of 
${\cal R}_{L}(\DisT$ $|\Sigma_{X^LY^L}),$
${\cal R}_{L}(D^L$ $|\Sigma_{X^LY^L}),$ and 
${\cal R}_{L}(D$ $|\Sigma_{X^LY^L}).$

The determination problem of ${\cal R}_{L}(D|\Sigma_{X^KY^L})$ 
in the case where $A$ is a general $K\times L$ matrix 
and $\Gamma=I_K$ was studied by Yang and Xiong \cite{yx} 
and Yang {\it et al.} \cite{yx2}. Relations between 
their results and our results of the present paper 
will be discussed in the next section.

\section{Main Results}

\subsection{Inner and Outer Bounds of the Rate Distortion Region}

In this subsection we state our result on the 
characterizations of 
${\cal R}_L(\DisT$ $|\Sigma_{X^KY^L})$, 
${\cal R}_L(\Gamma, D^K$ $|\Sigma_{X^KY^L})$, and 
${\cal R}_L(\Gamma, D$ $|\Sigma_{X^KY^L})$. 
To describe those results we define several functions 
and sets. For each $l\in \IsetA$ and for $r_l\geq 0$, 
let $N_{l}(r_l)$ be a Gaussian random variable 
with mean 0 and variance 
$\sigma_{N_l}^2/(1-{\baseN}^{-2r_l})$. 
We assume that $N_l(r_l),l\in \IsetA$ are independent. 
When $r_l=0$, we formally think that the inverse 
value $\sigma_{N(0)}^{-1}$ of the variance of $N_l(0)$ 
is zero. Let $\Sigma_{N^L(r^L)}$ be a covariance 
matrix of the random vector 
$$N^L(r^L)=N_{\IsetA}(r_{\IsetA})
          =\{N_l(r_l)\}_{l\in \Lambda}.$$ 
When $r_S={\vc 0}$, we formally define 
\beqno
\Sigma_{N_{S^{\rm c}}(r_{S^{\rm c}})}^{-1} 
&\defeq& 
\left. \Sigma_{N^L(r^L)}^{-1} \right|_{r_{S}={\lvc 0}}\,.
\eeqno
Fix nonnegative vector $r^L$. For $\theta >0$ and 
for $S \subseteq  \IsetA$, define
\beqno
\underline{J}_{S}(\theta, r_S|r_{\coS})
&\defeq &\frac{1}{2}\log^{+}
   \left[\ts 
   \frac{\ds \prod_{l\in S} {\baseN}^{2r_l} }
        {\ds  \theta \left|\Sigma_{X^K}^{-1}
                 +{}^{\rm t}A\Sigma_{N_{S^{\rm c}}(r_{S^{\rm c}})}^{-1}A
                           \right|
        }
  \right],
\\
{J}_{S}\left(r_S|r_{\coS}\right)
&\defeq &\frac{1}{2}\log
   \left[\ts 
   \frac{\ds \left|\Sigma_{X^K}^{-1}
    +{}^{\rm t}A\Sigma_{N^L(r^L)}^{-1}A\right|
             \prod_{l\in S} {\baseN}^{2r_l}
        }
        {\ds  \left|\Sigma_{X^K}^{-1}
                 +{}^{\rm t}A
                 \Sigma_{N_{S^{\rm c}}(r_{S^{\rm c}})}^{-1}
                            A\right|
        }
  \right],
\eeqno
where $S^{\rm c}=\IsetA-S$ 
and $\log^{+}[x]\defeq\max\{\log x,0\}.$
Set
$$
{\cal A}_L(\DisT)
\defeq 
\left\{ r^L\geq 0:
\left[\Sigma_{X^K}^{-1}+
{}^{\rm t}A\Sigma_{N^L(r^L)}^{-1}A\right]^{-1}
             \preceq \DisT\right\}.
$$
We can show that for $S\subseteq \IsetA$, 
$\underline{J}_S(|\DisT|,$ $r_S|r_{\coS})$ and $J_S(r_S|r_{\coS})$ 
satisfy the following two properties.
\begin{pr}{
\label{pr:prz01z}
$\quad$
\begin{itemize}
\item[{\rm a)}] If $r^L\in {\cal A}_L(\DisT)$, then for any 
$S\subseteq \IsetA$, 
$$
\underline{J}_S(|\DisT|,r_S|r_{\coS})\leq J_S(r_S|r_{\coS}).
$$
\item[{\rm b)}] Suppose that $r^L\in {\cal A}_L(\DisT)$. 
If $\left. r^L\right|_{r_S={\lvc 0}}$ still belongs to 
${\cal A}_L(\DisT)$, then 
\beqno
& &\left. \underline{J}_S(|\DisT|, r_S|r_{\coS})
 \right|_{r_S={\lvc 0}}
 =\left. J_S(r_S|r_{\coS})\right|_{r_S={\lvc 0}}
\\
& &=0.
\eeqno
\end{itemize}
}\end{pr}

\begin{pr}\label{pr:matroid}{\rm 
Fix $r^L\in {\cal A}_L(\DisT)$. 
For $S \subseteq \IsetA$, set 
\beqno
{f}_S&=&{f}_S(r_S|r_{\coS})
\defeq \underline{J}_S(|\DisT|,r_S|r_{\coS}).
\eeqno
By definition, it is obvious that ${f}_S, S \subseteq \IsetA$ 
are nonnegative. We can show that
$f\defeq \{{f}_S\}_{S \subseteq \IsetA}$ 
satisfies the followings:
\begin{itemize}
\item[{\rm a)}] ${f}_{\emptyset}=0$. 
\item[{\rm b)}] 
${f}_A\leq {f}_B$ 
for $A\subseteq B\subseteq \IsetA$.  
\item[{\rm c)}] ${f}_A+{f}_B \leq {f}_{A \cap B}+{f}_{A\cup B}.$
\end{itemize}
In general $(\IsetA,f)$ is called a {\it co-polymatroid} 
if the nonnegative function $\rho$ on $2^{\IsetA}$ satisfies 
the above three properties. Similarly, we set
\beqno
\tilde{f}_S&=&\tilde{f}_S(r_S|r_{\coS})\defeq J_S(r_S|r_{\coS})\,,
\quad \tilde{f}=\left\{\tilde{f}_S\right\}_{S \subseteq \IsetA}.
\eeqno
Then $(\IsetA,\tilde{f})$ also has the same three properties 
as those of $(\IsetA,f)$ and becomes a co-polymatroid. 
}\end{pr}

To describe our result 
on ${\cal R}_L(\DisT|\Sigma_{X^KY^L})$, set 
\beqno
& &
{\cal R}_L^{({\rm out})}(\theta,r^L|\Sigma_{X^KY^L})
\\
&\defeq&
\ba[t]{l}
  \left\{R^L \right. : 
  \ba[t]{l}
  \ds \sum_{i \in S} R_l 
  \geq \underline{J}_{S}\left(\theta,r_S|r_{\coS}\right)
  \vspace{1mm}\\
  \mbox{ for any }S \subseteq \IsetA. 
  \left. \right\}\,,
  \ea
\ea
\nonumber\\
& &{\cal R}_{L}^{({\rm out})}(\DisT|\Sigma_{X^KY^L})
\\
&\defeq& 
\bigcup_{r^L \in {\cal A}_L(\DisT)}
{\cal R}_L^{({\rm out})}(|\DisT|, r^L|\Sigma_{X^KY^L})\,, 
\nonumber\\
& &
{\cal R}_L^{({\rm in})}(r^L)
\\
&\defeq&
\ba[t]{l}
  \left\{R^L \right.: 
  \ba[t]{l}
  \ds \sum_{l \in S} R_l 
  \geq {J}_{S}\left(r_S|r_{\coS}\right)
  \vspace{1mm}\\
  \mbox{ for any }S \subseteq \IsetA. 
  \left. \right\}\,,
  \ea
\ea
\nonumber\\
& &{\cal R}_L^{({\rm in})}(\DisT|\Sigma_{X^KY^L})
\nonumber\\
&\defeq&{\rm conv}
        \left\{
        \bigcup_{r^L \in {\cal A}_L(\DisT)}
        {\cal R}_L^{({\rm in})}(r^L|\Sigma_{X^KY^L})
        \right\}. 
\eeqno
We can show that 
${\cal R}_L^{({\rm in})}(\DisT|\Sigma_{X^KY^L})$
and 
${\cal R}_L^{({\rm out})}(\DisT|\Sigma_{X^KY^L})$
satisfy the following property.
\begin{pr}
The sets 
${\cal R}_{L}^{({\rm in})}(\DisT|\Sigma_{X^KY^L})$ 
and ${\cal R}_{L}^{({\rm out})}(\DisT$ $|\Sigma_{X^KY^L})$ 
are not void if and only if $\DisT \succ \Sigma_{X^K|Y^L}$.
\end{pr}

Our result on inner and outer bounds of 
${\cal R}_{L}(\DisT|\Sigma_{X^KY^L})$ is as follows.
\begin{Th}\label{th:conv2}
For any $\DisT$$\succ$ $\Sigma_{X^K|Y^L}$, 
we have 
\beqno
& &{\cal R}_{L}^{({\rm in})}(\DisT|\Sigma_{X^KY^L})
=\hat{\cal R}_{L}^{({\rm in})}(\DisT|\Sigma_{X^KY^L})
\\
&\subseteq& {\cal R}_{L}(\DisT|\Sigma_{X^KY^L})
 \subseteq {\cal R}_{L}^{({\rm out})}(\DisT|\Sigma_{X^KY^L}).
\eeqno
\end{Th}

Proof of this theorem is given in Section V. This result 
includes the result of Oohama \cite{oh8} as a special 
case by letting $K=L$ and $\Gamma=A=I_L$. From this 
theorem we can derive outer and inner bounds of   
${\cal R}_{L}(\Gamma,D^K|$ $\Sigma_{X^KY^L})$ 
and 
${\cal R}_{L}(\Gamma,$$D|\Sigma_{X^KY^L}).$
To describe those bounds, set
\beqno
& &{\cal R}_L^{\rm (out)}(\Gamma,D^K|\Sigma_{X^KY^L})
\\
&\defeq& 
    \bigcup_{\Gamma \DisT {}^{\rm t}\Gamma \in {\cal S}_K(D^K)}
    {\cal R}_L^{\rm (out)}({\DisT}|\Sigma_{X^KY^L}),
\\
& &{\cal R}_L^{\rm (in)}(\Gamma,D^K|\Sigma_{X^KY^L})
\\
&\defeq& \conv\left\{
    \bigcup_{\Gamma \DisT {}^{\rm t}\Gamma \in {\cal S}_K(D^K)}
    {\cal R}_L^{\rm (in)}({\DisT}|\Sigma_{X^KY^L})
    \right\},
\\
& &{\cal R}_L^{\rm (out)}(\Gamma,D|\Sigma_{X^KY^L})
\\
&\defeq&
    \bigcup_{{\rm tr}[\Gamma\DisT {}^{\rm t}\Gamma] \leq D}
   {\cal R}_L^{\rm (out)}({\DisT}|\Sigma_{X^KY^L}),
\\
& &{\cal R}_L^{\rm (in)}(\Gamma,D|\Sigma_{X^KY^L})
\\
&\defeq&\conv\left\{
    \bigcup_{{\rm tr}[\Gamma\DisT {}^{\rm t}\Gamma] \leq D}
   {\cal R}_L^{\rm (in)}({\DisT}|\Sigma_{X^KY^L})
   \right\}.
\eeqno
Set 
\beqno
{\cal A}(r^L)
&\defeq & 
    \left\{\DisT:
    \DisT 
    \succeq (\Sigma_{X^K}^{-1}
+{}^{\rm t}A\Sigma_{N^L(r^L)}^{-1}A)^{-1}\right\}\,,
\\
\theta(\Gamma,D^K,r^L)
&\defeq &  
\max_{\scs \DisT:\DisT \in {\cal A}_L(r^L), 
      \atop{\scs
      \Gamma\DisT {}^{\rm t}\Gamma\in{\cal S}_K(D^K)}
     }
\left|\DisT \right|\,,
\\
\theta(\Gamma,D,r^L)
&\defeq & 
\max_{\scs \DisT: \DisT \in {\cal A}_L({r^L}),
      \atop{\scs 
       {\rm tr}[\Gamma\DisT {}^{\rm t}\Gamma]\leq D}
      }
\left|\DisT \right|.
\eeqno
Furthermore, set 
\beqno
& &{\cal B}_L(\Gamma,D^K)
\\
&\defeq&
\left\{r^L\geq 0:
\Gamma(\Sigma_{X^K}^{-1}+{}^{\rm t}
A\Sigma_{N^L(r^L)}^{-1}A)^{-1}{}^{\rm t}\Gamma\in{\cal S}_K(D^K)
\right\}\,,
\\
& &{\cal B}_L(\Gamma,D)
\\
&\defeq &
\left\{r^L\geq 0:
{\rm tr}[\Gamma(\Sigma_{X^K}^{-1}
+{}^{\rm t}A\Sigma_{N^L(r^L)}^{-1}A)^{-1}{}^{\rm t}\Gamma]
\leq D 
\right\}.
\eeqno
It can easily be verified that 
${\cal R}_{L}^{({\rm out})}(\Gamma,$ $D^K|\Sigma_{X^KY^L})$,
${\cal R}_{L}^{({\rm in})}($ $\Gamma,$ $D^K|\Sigma_{X^KY^L})$,
${\cal R}_{L}^{({\rm out})}(\Gamma,$ $D|\Sigma_{X^KY^L})$, and 
${\cal R}_{L}^{({\rm in})}(\Gamma,$ $D|$ $\Sigma_{X^KY^L})$ 
satisfies the following property.  
\begin{pr} $\quad$
\begin{itemize}
\item[{\rm a)}] 
The sets ${\cal R}_{L}^{({\rm in})}(\Gamma,D^K|\Sigma_{X^KY^L})$ 
and  ${\cal R}_{L}^{({\rm out})}(\Gamma,D^K|\Sigma_{X^K}$ ${}_{Y^L})$ 
are not void if and only if 
$D^K > d^K(\Gamma\Sigma_{X^K|Y^L}{}^{\rm t}\Gamma)$.
\item[{\rm b)}] 
The sets ${\cal R}_{L}^{({\rm in})}(\Gamma,D|\Sigma_{X^KY^L})$ 
and ${\cal R}_{L}^{({\rm out})}(\Gamma, D |\Sigma_{X^K}$ ${}_{Y^L})$ 
are not void if and only if $D>{\rm tr}[\Gamma \Sigma_{X^K|Y^L}$ 
${}^{\rm t}\Gamma]$.
\item[{\rm c)}] 
\beqno
& &{\cal R}_{L}^{({\rm out})}(\Gamma,D^K|\Sigma_{X^KY^L})
\\
&=& 
\bigcup_{r^L \in {\cal B}_L(\Gamma,D^K)}
{\cal R}_L^{({\rm out})}(\theta(\Gamma,D^K,r^L),r^L|\Sigma_{X^KY^L})\,, 
\nonumber\\
& &  {\cal R}_L^{({\rm in})}(\Gamma,D^K|\Sigma_{X^KY^L})
\\
&=&
        \conv\left\{\bigcup_{r^L \in {\cal B}_L(\Gamma,D^K)}
        {\cal R}_L^{({\rm in})}(r^L|\Sigma_{X^KY^L})\right\}\,,
\\
& &{\cal R}_{L}^{({\rm out})}(\Gamma,D|\Sigma_{X^KY^L})
\\
&=& 
\bigcup_{r^L \in {\cal B}_L(\Gamma,D)}
{\cal R}_L^{({\rm out})}(\theta(\Gamma,D,r^L), r^L|\Sigma_{X^KY^L})\,, 
\\
& &
{\cal R}_L^{({\rm in})}(\Gamma,D|\Sigma_{X^KY^L})
\\
&=&\conv\left\{
        \bigcup_{r^L \in {\cal B}_L(\Gamma,D)}
        {\cal R}_L^{({\rm in})}(r^L)\right\}.
\eeqno
\end{itemize}
\end{pr}

The following result is obtained as a 
simple corollary from Theorem \ref{th:conv2}. 
\begin{co}\label{co:conv2z}
For any $\Gamma$ and any $D^K>$ 
$d^K(\Gamma\Sigma_{X^K|Y^L}{}^{\rm t}\Gamma)$, 
we have 
\beqno
& &{\cal R}_{L}^{({\rm in})}(\Gamma,D^K|\Sigma_{X^KY^L})
= \hat{\cal R}_{L}^{({\rm in})}(\Gamma,D^K|\Sigma_{X^KY^L})
\\
&\subseteq& {\cal R}_{L}(\Gamma,D^K|\Sigma_{X^KY^L})
 \subseteq {\cal R}_{L}^{({\rm out})}(\Gamma,D^K|\Sigma_{X^KY^L}).
\eeqno
For any $\Gamma$ and any 
$D>{\rm tr}[\Gamma\Sigma_{X^K|Y^L}{}^{\rm t}\Gamma]$, 
we have 
\beqno
& &{\cal R}_{L}^{({\rm in})}(\Gamma,D|\Sigma_{X^KY^L})
= \hat{\cal R}_{L}^{({\rm in})}(\Gamma,D|\Sigma_{X^KY^L})
\\
&\subseteq& {\cal R}_{L}(\Gamma,D|\Sigma_{X^KY^L})
 \subseteq  {\cal R}_{L}^{({\rm out})}(\Gamma,D|\Sigma_{X^KY^L}).
\eeqno
\end{co}

Those result includes the result of Oohama \cite{oh8} 
as a special case by letting $K=L$ and $\Gamma=A=I_L$. 
Next we compute $\theta(\Gamma,D,r^L)$ to 
derive a more explicit expression of 
${\cal R}_{L}^{({\rm out})}
(\Gamma$ $,D|\Sigma_{X^KY^L})$. This expression 
will be quite useful for finding a sufficient 
condition for the outer bound 
${\cal R}_{L}^{({\rm out})}
(\Gamma$ $,D|\Sigma_{X^KY^L})$
to be tight. Let $\alpha_k=\alpha_k(r^L), k\in\IsetB$ 
be $K$ eigenvalues of the matrix 
$$
\Gamma^{-1}
\left(\Sigma_{X^K}^{-1}
+{}^{\rm t}A\Sigma_{N^L(r^L)}^{-1}A\right){}^{\rm t}\Gamma^{-1}.
$$
Let $\xi$ be a nonnegative number that satisfy 
$$
\sum_{k=1}^K\left\{[\xi- \alpha_k^{-1}]^{+}
+\alpha_k^{-1}\right\}=D.
$$    
Define  
$$
\omega(\Gamma,D,r^L)\defeq |\Gamma|^{-2}
\prod_{k=1}^K\left\{[\xi-\alpha_k^{-1}]^{+}
+\alpha_k^{-1}\right\}.
$$
The function ${\omega}(\Gamma, D,r^L)$ has 
an expression of the so-called water filling 
solution to the following optimization problem:
\beqa
{\omega}(\Gamma, D,r^L)
=|\Gamma|^{-2}
  \max_{\scs \xi_k\alpha_{k}\geq 1,k\in\IsetB\,, 
  \atop{\scs
          \sum_{k=1}^K\xi_k\leq D
       }   
      }\prod_{k=1}^K\xi_{k}. 
\eeqa
Then we have the following theorem.
\begin{Th}\label{th:conv2a}
For any ${\Gamma}$ and any positive $D$, we have 
$$
\theta(\Gamma,D,r^L)=\omega(\Gamma,D,r^L).
$$
A more explicit expression of 
${\cal R}_{L}^{({\rm out})}(\Gamma,D|\Sigma_{X^KY^L})$ 
using $\omega(\Gamma,D,r^L)$ is given by
\beqno
& &{\cal R}_{L}^{({\rm out})}(\Gamma,D|\Sigma_{X^KY^L})
\\
&\defeq& 
\bigcup_{r^L \in {\cal B}_L(\Gamma, D)}
{\cal R}_L^{({\rm out})}(\omega
(\Gamma,D,r^L),r^L|\Sigma_{X^KY^L}). 
\eeqno
\end{Th}

Proof of this theorem will be given in Section V. The above
expression of the outer bound includes the result of Oohama \cite{oh8}
as a special case by letting $K=L$ and $\Gamma=A=I_L$. In the next
subsection we derive a matching condition for 
${\cal R}_{L}^{({\rm out})}(\Gamma, D|\Sigma_{X^KY^L})$ 
to coincide with ${\cal R}_{L}(\Gamma, D|\Sigma_{X^KY^L})$.

Two other outer bounds of ${\cal R}_{L}(D|\Sigma_{X^KY^L})$
were obtained by Yang and Xiong \cite{yx}. 
They derived the first outer bound for general $L\times K$ 
matrix $A$. This outer bound denoted by
$\check{\cal R}_{L}^{({\rm out})}($ $D|\Sigma_{X^KY^L})$ does not
coincide with ${\cal R}_{L}^{({\rm out})}($ $D|\Sigma_{X^KY^L})$ when
$K=L$ and $A=I_L$. 
%
%
When $A$ is semi orthogonal, i.e., ${}^{\rm t}AA=I_K$, Yang and Xiong
\cite{yx} derived the second outer bound $\tilde{\cal R}_{L}^{\rm
(out)}(D|\Sigma_{X^KY^L})$ tighter than $\check{\cal R}_{L}^{\rm
(out)}(D|\Sigma_{X^KY^L})$. The outer bound 
$\tilde{\cal R}_{L}^{\rm (out)}(D|\Sigma_{X^KY^L})$ 
is the same as our outer bound 
${\cal R}_{L}^{({\rm out})}($ $D|\Sigma_{X^KY^L})$ although it
has a form different from that of our outer bound. They further
derived a matching condition for $\tilde{\cal R}_{L}^{({\rm
    out})}(D|\Sigma_{X^KY^L})$ to coincide with ${\cal
  R}_{L}(D|\Sigma_{X^KY^L})$. Their matching condition and its
relation to our matching condition will be presented in the next
subsection.

\subsection{Matching Condition Analysis}

For $L\geq 3$, we present a sufficient condition for 
${\cal R}^{{(\rm out)}}_L(\Gamma,$ $D|$ $\Sigma_{X^KY^L})$ 
$\subseteq$ ${\cal R}_L^{({\rm in})}($$D|\Sigma_{X^KY^L}).$ 
We consider the following condition on 
$\theta(\Gamma, D,r^L)$.

{\it Condition: } For any $l\in \IsetA$, 
${\baseN}^{-2r_l}\theta(\Gamma, D,r^L)$ is a monotone 
decreasing function of $r_l\geq 0$.

We call this condition the MD condition. The following 
is a key lemma to derive the matching condition. 
This lemma is due to 
Oohama \cite{oh7}, \cite{oh7b}.
\begin{lm}[Oohama \cite{oh7},\cite{oh7b}] 
\label{lm:lem1}
If $\theta(\Gamma,D,r^L)$ satisfies the 
MD condition on ${\cal B}_L($ $\Gamma,D)$, then 
\beqno
 {\cal R}_L^{({\rm in})}(\Gamma,D|\Sigma_{X^KY^L})
&=&{\cal R}_L(\Gamma,D|\Sigma_{X^KY^L})
\\
&=&{\cal R}_L^{({\rm out})}(\Gamma,D|\Sigma_{X^KY^L}).
\eeqno 
\end{lm}

Based on Lemma \ref{lm:lem1}, we derive a sufficient 
condition for $\theta(\Gamma,D,r^L)$ to satisfy 
the MD condition. 

Let ${a}_{lk}$ be the $(l,k)$ element of $A$. Set 
$
{\vc a}_l\defeq 
[{a}_{l1}{a}_{l2} \cdots {a}_{lK}]
$
and $\hat{\vc a}_l\defeq {\vc a}_l\Gamma^{-1}$. 
Let ${\cal O}_K$ be the set of all $K\times K$ 
orthogonal matrices.   
For $(l,k)\in \Lambda_L \times \Lambda_K$, 
let ${\cal O}_K(\hat{\vc a}_l,k)$ be 
a set of all $T\in {\cal O}_K$ that 
satisfy
\beqno
[\hat{\vc a}_lT]_j=
\left\{ 
\ba{cl}
||\hat{\vc a}_l||, & \mbox{ if } j=k,     \\
0                , & \mbox{ if }j \neq k. \\
\ea
\right.
\eeqno
For $T\in {\cal O}_K(\hat{\vc a}_l,k)$, 
we consider the following matrix:
\beqno
& &C(\Gamma^{-1}T,r^L)
\defeq
{}^{\rm t}T{}^{\rm t}\Gamma^{-1}
    (\Sigma_{X^K}^{-1}+{}^{\rm t}A\Sigma_{N^L(r^L)}^{-1}A)\Gamma^{-1}T
\\
&=& {}^{\rm t}T{}^{\rm t}\Gamma^{-1}\Sigma_{X^K}^{-1}\Gamma^{-1}T
    +\sum_{l=1}^L
\ts\frac{1}{\sigma_{N_l}^2}(1-{\baseN}^{-2r_l})
{}^{\rm t}(\hat{\vc a}_lT)(\hat{\vc a}_lT).
\eeqno
Let 
$
r^L_{[l]} \defeq r_{1}\cdots r_{l-1} r_{l+1}\cdots r_L 
$
and set
\beqno
\lefteqn{\eta_{k}(\Gamma^{-1}T,r^L_{[l]})}
\\
&\defeq& 
\left[
{}^{\rm t}T{}^{\rm t}\Gamma^{-1}\Sigma_{X^K}^{-1}\Gamma^{-1}T
\right]_{kk}
\\
&&\qquad
+\sum_{i\ne l}\ts\frac{1}{\sigma_{N_i}^2}(1-{\baseN}^{-2r_i})
  \left[{}^{\rm t}(\hat{\vc a}_iT)(\hat{\vc a}_iT)\right]_{kk},  
\\
\lefteqn{\chi_{lk}(\Gamma^{-1}T,r^L_{[l]})
\defeq||\hat{\vc a}_l||^2\ts\frac{1}{\sigma_{N_l}^2}
+\eta_{k}(\Gamma^{-1}T,r^L_{[l]}).}
\eeqno
Then we have 
\beqa
[C(\Gamma^{-1}T,r^L)]_{kk}
&=&
||\hat{\vc a}_l||^2\ts\frac{1}{\sigma_{N_l}^2}(1-{\baseN}^{-2r_l})
+\eta_i(\Gamma^{-1}T,r^L_{[l]})
\nonumber\\
&=&\chi_{lk}(\Gamma^{-1}T,r^L_{[l]})
-||\hat{\vc a}_l||^2\ts\frac{1}{\sigma_{N_l}^2}{\baseN}^{-2r_l}.
\label{eqn:match0}
\eeqa
If $(i^{\prime},i^{\prime \prime})\ne (k,k)$, then the value of 
\beqno
&  &
[C(\Gamma^{-1}T,r^L)]_{i^{\prime}i^{\prime \prime}}
\\
&=&[{}^{\rm t}T{}^{\rm t}\Gamma^{-1}\Sigma_{X^K}^{-1}
   \Gamma^{-1}T]_{i^{\prime}i^{\prime \prime}}
\\
& &\qquad
   +\sum_{j=1}^L\ts\frac{1}{\sigma_{N_j}^2}(1-{\baseN}^{-2r_j})
   \left[
   {}^{\rm t}(\hat{\vc a}_jT)(\hat{\vc a}_jT)
   \right]_{i^{\prime}i^{\prime \prime}}  
\eeqno
does not depend on $r_l$. Note that the matrix 
$C(\Gamma^{-1}T,r^L)$ has the same eigenvalue set 
as that of 
$$
C(\Gamma^{-1},r^L)=
{}^{\rm t}\Gamma^{-1}
(\Sigma_{X^K}^{-1}+{}^{\rm t}A\Sigma_{N^L(r^L)}^{-1}A)\Gamma^{-1}.
$$
We recall here that $\alpha_k=\alpha_k(r^L), k\in \Lambda_K$ are $K$
eigenvalues of the above two matrices. Let
$\alpha_{\min}=\alpha_{\min}(r^L)$ and
$\alpha_{\max}=\alpha_{\max}(r^L)$ be the minimum and maximum
eigenvalues among $\alpha_k,k\in \Lambda_K$.  The matrix
$C(\Gamma^{-1}T,r^L)$ for $T\in {\cal O}_K(\hat{\vc a}_l,k)$, has a
structure that the $(k,k)$ element of this matrix is only one element
which depends on $r_l$ and this element is a monotone increasing
function of $r_l\geq 0$.  Properties on eigenvalues of matrices having
the above structure were studied in detail by Oohama
\cite{oh7},\cite{oh7b}.  The following lemma is a variant of his
result.
\begin{lm}[Oohama \cite{oh7},\cite{oh7b}] 
\label{lm:Egn1}$\:$
For each $(l,k)\in \IsetA \times \IsetB$ and 
each $T\in {\cal O}_K(\hat{\vc a}_l,k)$, we have the followings.
\beqno
& &\ba[t]{rcl}
   &    & \alpha_{\min}(r^L)
   \\ 
   &\leq& 
    ||\hat{\vc a}_l||^2\ts\frac{1}{\sigma_{N_l}^2}(1-{\baseN}^{-2r_l})
    +\eta_{lk}(\Gamma^{-1}T,r_{[l]}^L)
    \leq\alpha_{\max}(r^L),
   \ea
\\
& &\frac{\partial \alpha_j}{\partial r_l}
   \geq 0, \mbox{ for }j\in \Lambda_K,\quad 
\sum_{j=1}^K\frac{\partial \alpha_j}{\partial r_l}
=\frac{2||\hat{\vc a}_l||^2}
    {{\baseN}^{2r_l}\sigma_{N_l}^2}.
\eeqno
\end{lm}

The following is a key lemma to derive a sufficient 
condition for the MD condition to hold. 
\begin{lm}\label{lm:pro3}
If 
$\alpha_{\min}(r^L)$ and 
$\alpha_{\max}(r^L)$ satisfy 
\beqa
& &
\left(
\frac{1}{\alpha_{\min}(r^L)}-\frac{1}{\alpha_{\max}(r^L)}
\right)\cdot\frac{\alpha_{\max}(r^L)}{\alpha_{\min}(r^L)}
\leq \frac{{\baseN}^{2r_l}\sigma_{N_l}^2}{||\hat{\vc a}_l||^2}
\label{eqn:match600}\\
& &\mbox{ for }l\in \IsetA,
\nonumber
\eeqa
on ${\cal B}_L(\Gamma,D)$, then $\theta(\Gamma, D,r^L)$
satisfies the MD condition on ${\cal B}_L(\Gamma,D)$.
\end{lm}

Proof of Lemma \ref{lm:pro3} will be stated in Section V. 
Set
\beqno
\lefteqn{
C^*(\Gamma^{-1}T,r_l)\defeq\lim_{r_{[l]}^L\to\infty}
C(\Gamma^{-1}T,r^{L}),}
\\
\lefteqn{
\chi^*_k(\Gamma^{-1}T)
\defeq\lim_{r_{[l]}^L\to\infty}\chi_{lk}(\Gamma^{-1}T,r_{[l]}^L)}
\\
&=& [{}^{\rm t}T{}^{\rm t}\Gamma^{-1}\left(\Sigma_{X^K}^{-1}
       +{}^{\rm t}A\Sigma_{N^L}^{-1}A
       \right)\Gamma^{-1}T]_{kk}.
\eeqno 
For $k\in \Lambda_K$, we denote the $(k,k)$ element 
of $C^*(\Gamma^{-1}T,r_l)$ by 
$c_{kk}^*=c_{kk}^*(\Gamma^{-1}T,r_l)$. 
When $(j,j^{\prime})\in \Lambda_K^2$ and 
$(j,j^{\prime})\neq (k,k)$, the $(j,j^{\prime})$ 
element of $C^*(\Gamma^{-1}T,r_l)$ 
does not depend on $r_l$. We denote it by 
$c_{jj^{\prime}}^*=c_{jj^{\prime}}^*(\Gamma^{-1}T)$.   
Furthermore, set
$$
{\vc c}^*_{k[k]}=
{\vc c}^*_{k[k]}(\Gamma^{-1}T)\defeq
[c_{k1}^* \cdots c_{kk-1}^* c_{kk+1}^* \cdots c^{*}_{kK}].
$$
By definition we have
$$
c_{kk}^*(\Gamma^{-1}T,r_l)=\chi_{k}^*(\Gamma^{-1}T)
-\frac{||\hat{\vc a}_l||^2}
    {{\baseN}^{2r_l}\sigma_{N_l}^2}.
$$
Define
\beqno
\alpha_{\max}^{*}&\defeq&\lim_{r^L\to\infty}\alpha_{\max}(r^L),
\alpha_{\min}^{*}\defeq\lim_{r^L\to\infty}\alpha_{\min}(r^L),
\\
\alpha_{\max}^{*}(r_i)&\defeq&\lim_{r_{[l]}^L\to\infty}\alpha_{\max}(r^L)
\mbox{ for }l\in \Lambda_L.
\eeqno
By definition, $\alpha_{\max}^{*}$ and $\alpha_{\min}^{*}$ are 
the maximum and minimum eigenvalues of 
$
{}^{\rm t}\Gamma^{-1}
(\Sigma_{X^K}^{-1}+{}^{\rm t}A\Sigma_{N^L}^{-1}A$ $)\Gamma^{-1}, 
$
respectively. By Lemma \ref{lm:Egn1}, we have
\beqa
&&\alpha_{\min}(r^L)\leq \alpha_{\min}^*(r_l) \leq \alpha_{\min}^*,
\mbox{ for }l\in \IsetA,
\label{eqn:match97b} 
\\
&&\chi_{lk}(\Gamma^{-1}T,r_{[l]}^L)
\leq\chi_k^*(\Gamma^{-1}T)\leq \alpha_{\max}^*,
\mbox{ for }l\in \IsetA.
\label{eqn:match98} 
\eeqa
The following lemma provides an effective lower bound of 
$
{{\baseN}^{2r_l}\sigma_{N_l}^2}/{||\hat{\vc a}_l||^2}.
$
\begin{lm}\label{lm:pro4} 
For any $(l,k)\in \Lambda_L\times \Lambda_K$
and $T\in {\cal O}_Y(\hat{\vc a}_l,k)$, we have
\beqno
&     &c_{kk}^*(\Gamma^{-1}T,r_l)=\chi_{k}^*(\Gamma^{-1}T)
      -\frac{||\hat{\vc a}_l||^2}{{\baseN}^{2r_l}\sigma_{N_l}^2}
\\
&\geq & 
\alpha_{\min}^*(r_l)
+\frac{||{\vc c}^*_{k[k]}(\Gamma^{-1}T)||^2}
{\alpha_{\max}^*(r_l)-\alpha_{\min}^*(r_l)}
\\
&\geq & 
\alpha_{\min}(r^L)
+\frac{||{\vc c}^*_{k[k]}(\Gamma^{-1}T)||^2}
{\alpha_{\max}^*-\alpha_{\min}(r^L)}.
\eeqno
\end{lm}

Proof of this lemma will be given in Section V.  
Set
\beqno
\Upsilon_l(\Gamma^{-1})
&\defeq& 
\max_{\scs k\in\IsetB
\atop{ 
     \scs T\in {\cal O}_K({\lvc a}_l\Gamma^{-1},k)
     }
     }
\frac{1+\frac{||{\vc c}^*_{k[k]}(\Gamma^{-1}T)||^2}{(\alpha^*_{\max})^2}}
    {\chi_k^*(\Gamma^{-1}T)
   -\frac{||{\vc c}^*_{k[k]}(\Gamma^{-1}T)||^2}{{\alpha_{\max}^*}}}.
\eeqno
When $\Gamma=I_K$, we simply write 
$\Upsilon_l(I_K)=\Upsilon_l$.
From Lemmas \ref{lm:lem1}-\ref{lm:pro4} 
and an elementary computation we obtain the following. 
\begin{Th} \label{th:matchTh}
If we have 
\beqa
{\rm tr}[\Gamma\Sigma_{X^K|Y^L}{}^{\rm t}\Gamma] 
&<&D \leq {\frac{K}{\alpha_{\max}^{*}}}
   +\min_{l\in\IsetA}\Upsilon_l(\Gamma^{-1})
\label{eqn:match999z}
\eeqa
then 
\beqno
  & &{\cal R}_L^{({\rm in})}(\Gamma, D|\Sigma_{X^KY^L})
=\hat{\cal R}_L^{({\rm in})}(\Gamma, D|\Sigma_{X^KY^L})
\\
&=&{\cal R}_L(\Gamma,D|\Sigma_{X^KY^L})
 ={\cal R}_L^{({\rm out})}(\Gamma,D|\Sigma_{X^KY^L}).
\eeqno
Using (\ref{eqn:match98}), we obtain 
$\Upsilon_l(\Gamma^{-1})\geq 1/\alpha_{\max}^{*}$. Hence we have 
the following matching condition simpler than (\ref{eqn:match999z}):    
\beq
{\rm tr}[\Gamma\Sigma_{X^K|Y^L}{}^{\rm t}\Gamma] 
< D \leq \frac{K+1}{\alpha_{\max}^{*}}.
\label{eqn:match102}  
\eeq
\end{Th}

Proof of Theorem \ref{th:matchTh} will be stated in Section V. 
When $K=L,A=I_L$, the matching condition (\ref{eqn:match102}) is 
the same as that of Oohama \cite{oh7b},\cite{oh9a}. 
It is obvious that in the case of $K=L,A=I_L$, 
the matching condition (\ref{eqn:match999z}) improves that
of Oohama \cite{oh7b},\cite{oh9a}. Yang {\it et al.} \cite{yx2} 
have obtained a matching condition on ${\cal R}_L(D|\Sigma_{X^KY^L})$ 
by an argument quite similar to that of Oohama \cite{oh7b}. 
The matching condition by Yang {\it et al.} \cite{yx2} is as follows:
\beqa
{\rm tr}[\Sigma_{X^K|Y^L}] 
&<&D \leq {\frac{K}{\alpha_{\max}^{*}}}+
\min_{l\in\IsetA}\tilde{\Upsilon}_l,
\label{eqn:match99xx}
\eeqa
where 
\beqno
\tilde{\Upsilon}_l&\defeq&
\max_{T\in {\cal O}_K} 
\max_{k\in \IsetB}
\left\{
\frac{1}{\chi_k^*(T)}
\frac{[{\vc a}_{l}T]_k^2}{||{\vc a}_lT||^2}
\right\}.
\eeqno
The matching condition (\ref{eqn:match99xx}) by 
Yang {\it et al.} \cite{yx2} also improves that of 
Oohama \cite{oh7b},\cite{oh9a} in the case of $K=L,A=I_L$.
When $\Gamma=I_K$, for $l\in \Lambda_L$, we have 
\beqa
\Upsilon_l
&=& 
\max_{\scs k\in\IsetB
\atop{ 
     \scs T\in {\cal O}_K({\lvc a}_l,k)
     }
     }
\frac{1+\frac{||{\vc c}^*_{k[k]}(T)||^2}{(\alpha^*_{\max})^2}}
    {\chi_{k}^*(T)
    -\frac{||{\vc c}^*_{k[k]}(T)||^2}{{\alpha_{\max}^*}}}
\nonumber\\
&\geq&
\max_{\scs k\in\IsetB
\atop{ 
     \scs T\in {\cal O}_K({\lvc a}_l,k)
     }
     }
\frac{1}{\chi_{k}^*(T)}\defeq \underline{\Upsilon}_l. 
\label{eqn:match700}
\eeqa
On the other hand, for $i\in \Lambda_L$, we have
\beqa
\tilde{\Upsilon}_l&= &
\max_{
     \scs T\in {\cal O}_K
     }
\max_{\scs k\in\IsetB}
\left\{ 
\frac{1}{\chi_k^*(T)}
\frac{[{\vc a}_{l}T]_k^2}{||{\vc a}_lT||^2}
\right\}
\nonumber\\
&=&
 \max_{\scs k\in\IsetB} 
 \max_{
\scs T\in {\cal O}_K
     }
\left\{
\frac{1}{\chi_k^*(T)}
\frac{[{\vc a}_{l}T]_k^2}{||{\vc a}_lT||^2}
\right\}
\nonumber\\ 
&\geq &\max_{\scs k\in\IsetB} 
\max_{
      \scs T\in {\cal O}_K({\lvc a}_l,k)
     }
\left\{
\frac{1}{\chi_k^*(T)}
\frac{[{\vc a}_{l}T]_k^2}{||{\vc a}_lT||^2}
\right\}
\label{eqn:match311}\\ 
&= &\max_{\scs k\in\IsetB} 
\max_{
     \scs T\in {\cal O}_K({\lvc a}_l,k)
     }
\frac{1}{\chi_k^*(T)}=
\underline{\Upsilon}_l.
\nonumber 
\eeqa
Thus, we have ${\Upsilon}_l$$\geq\underline{\Upsilon}_l$
and $\tilde{\Upsilon}_l$$\geq\underline{\Upsilon}_l.$
Comparing the two inequalities (\ref{eqn:match700}) 
and (\ref{eqn:match311}), we can see that 
the improvement of ${\Upsilon}_l$ from $\underline{\Upsilon}_l$
is quite differnt from that of $\tilde{\Upsilon}_l$ 
from $\underline{\Upsilon}_l$.
Hence we have no obvious superiority 
of ${\Upsilon}_l$ or $\tilde{\Upsilon}_l$ to the other. 

Next we derive another matching condition, which   
is better than the second matching condition (\ref{eqn:match102}) 
in Theorem \ref{th:matchTh} and the matching condition 
(\ref{eqn:match99xx}) of Yang {\it et al.} \cite{yx2} for 
some nontrivial cases. Set
$$
\tau_l\defeq\frac{\sigma_{N_l}^2}{||\hat{\vc a}_l||^2},
\tau^*\defeq\min_{ l\in \IsetA}\tau_l. 
$$
From Lemmas \ref{lm:lem1}-\ref{lm:pro3} and an elementary 
computation we obtain the following. 
\begin{Th}\label{th:matchTh2z}
If we have 
\beqa
\lefteqn{{\rm tr}[\Gamma\Sigma_{X^K|Y^L}{}^{\rm t}\Gamma]} 
\nonumber\\
<D &\leq &  {\frac{K}{\alpha_{\max}^{*}}}
  +\frac{1}{2\alpha_{\max}^{*}}
 \left\{\sqrt{1+4\alpha_{\max}^{*}\tau^*}-1\right\}, 
 \label{eqn:match109}
 \eeqa
then 
\beqno
  & &{\cal R}_L^{({\rm in})}(\Gamma, D|\Sigma_{X^KY^L})
=\hat{\cal R}_L^{({\rm in})}(\Gamma, D|\Sigma_{X^KY^L})
\\
&=&{\cal R}_L(\Gamma,D|\Sigma_{X^KY^L})
 ={\cal R}_L^{({\rm out})}(\Gamma,D|\Sigma_{X^KY^L}).
\eeqno
\end{Th}

Proof of Theorem \ref{th:matchTh2z} will 
be stated in Section V. When 
$\tau^*$ becomes large, $\alpha_{\max}^*$ and $\alpha_{\min}^*$ 
approach to the maximum and minimum eigenvalues of 
$\Sigma_{X^L}^{-1}$, respectively. 
Hence we have 
\beq
\lim_{\tau^{*}\to+\infty}
\frac{1}{2\alpha_{\max}^{*}}
\left\{\sqrt{1+4\alpha_{\max}^{*}\tau^*}-1\right\}=+\infty,
\label{eqn:MatchCd001} 
\eeq
which implies that there exists a sufficiently large $\tau^{*}$
such that
\beq
\frac{1}{\alpha_{\max}^{*}}\leq 
\frac{1}{\alpha_{\min}^{*}}
< \frac{1}{2\alpha_{\max}^{*}}
\left\{\sqrt{1+4\alpha_{\max}^{*}\tau^*}-1\right\}.
\label{eqn:MatchCd002} 
\eeq
On the other hand, it follows from the definition of 
$\tilde{\Upsilon}_l$ that we have  for $l\in\IsetA$, 
\beqa
\tilde{\Upsilon}_l&\leq&
\max_{
     \scs T\in {\cal O}_K
     }
\max_{\scs k\in\IsetB}
\frac{1}{\chi_k^*(T)}
\leq \frac{1}{\alpha_{\min}^*}. 
\label{eqn:MatchCd003} 
\eeqa
Thus we can see from (\ref{eqn:MatchCd002}) and (\ref{eqn:MatchCd003})
that for sufficiently large $\tau^*$, 
the matching condition (\ref{eqn:match109}) in 
Theorem \ref{th:matchTh2z}
is better than the second matching condition (\ref{eqn:match102}) 
in Theorem \ref{th:matchTh} and the matching condition 
(\ref{eqn:match99xx}) of Yang {\it et al.} \cite{yx2}. 

\section{
Application to the Multiterminal 
\RateDist problem
}

\bfig
\setlength{\unitlength}{1.00mm}
\begin{picture}(79,58)(0,0)
\put(2,-1){\dashbox{0.5}(15,50)}
\put(3,35){\framebox(6,6){$X_1$}}
\put(19,35){\framebox(6,6){$Y_1$}}

\put(3,20){\framebox(6,6){$X_2$}}
\put(19,20){\framebox(6,6){$Y_2$}}

\put(5,11){$\vdots$}
\put(3,0){\framebox(6,6){$X_L$}}
\put(19,0){\framebox(6,6){$Y_L$}}

\put(26,40.6){${\vc Y}_1$}
\put(12,45){${N}_1$}

\put(26,25.6){${\vc Y}_2$}
\put(12,30){${N}_2$}

\put(26,5.6){${\vc Y}_L$}
\put(12,10){${N}_L$}

\put(9,38){\vector(1,0){3.5}}
\put(14,44){\vector(0,-1){4.5}}
\put(14,37){\line(0,1){2}}
\put(13,38){\line(1,0){2}}
\put(14,38){\circle{3.0}}

\put(9,23){\vector(1,0){3.5}}
\put(14,29){\vector(0,-1){4.5}}
\put(14,22){\line(0,1){2}}
\put(13,23){\line(1,0){2}}
\put(14,23){\circle{3.0}}

\put(9,3){\vector(1,0){3.5}}
\put(14,9){\vector(0,-1){4.5}}
\put(14,2){\line(0,1){2}}
\put(13,3){\line(1,0){2}}
\put(14,3){\circle{3.0}}

\put(32,34.5){\framebox(7,7){$\varphi_1^{(n)}$}}
\put(40,40.6){$\varphi_1^{(n)}({\vc Y}_1)$}

\put(32,19.5){\framebox(7,7){$\varphi_2^{(n)}$}}
\put(40,25.6){$\varphi_2^{(n)}({\vc Y}_2)$}

\put(34.5,11){$\vdots$}

\put(32,-0.5){\framebox(7,7){$\varphi_L^{(n)}$}}
\put(40,5.6){$\varphi_L^{(n)}({\vc Y}_L)$}

\put(15.5,38){\vector(1,0){3.5}}
\put(25,38){\vector(1,0){7}}
\put(39,38){\line(1,0){15}}
\put(54,38){\vector(2,-3){10}}

\put(15.5,23){\vector(1,0){3.5}}
\put(25,23){\vector(1,0){7}}
\put(39,23){\line(1,0){15}}
\put(54,23){\vector(1,0){10}}

\put(15.5,3){\vector(1,0){3.5}}
\put(25,3){\vector(1,0){7}}
\put(39,3){\line(1,0){15}}
\put(54,3){\vector(1,2){10}}

\put(64,19.5){\framebox(7,7){$\phi^{(n)}$}}
\put(71,23){\vector(1,0){3}}
\put(74,22){$\left[
             \ba{c}
             \hat{\vc Y}_1\\
             \hat{\vc Y}_2\\
             \vdots \\ 
             \hat{\vc Y}_L
             \ea
             \right]$}
\end{picture}
\caption{
Distributed source coding system for $L$ correlated 
Gaussian sources}
\label{fig:Fig2}
\efig

In this section we consider the case where $K=L$ and $A=I_L$. 
In this case we have $Y^L=X^L+N^L$; Gaussian random variables 
$Y_l$, $l\in\IsetA$ are $L$-noisy components of the Gaussian 
random vector $X^L$. We study the multiterminal \RateDist 
problem for the Gaussian observations $Y_l,l\in \Lambda$. 
The random vector $X^L$ can be regarded as a ``hidden'' 
information source of $Y^L$. Note that
$(X^L,Y^L)$ satisfies $Y_S \to X^L \to Y_{S^{\rm c}} \mbox{ for any }S
\subseteq \IsetA. 
$

\subsection{
Problem Formulation and Previous Results
}

The distributed source coding system for $L$ 
correlated Gaussian source treated here is 
shown in Fig. \ref{fig:Fig2}. Definitions of encoder 
functions $\varphi_l, l\in \IsetA$ 
are the same as the previous definitions.
The decoder function $\phi^{(n)}$
is defined by  
\beqno
&&\phi^{(n)}=(\phi_1^{(n)},\phi_2^{(n)},\cdots,\phi_L^{(n)}) 
\\
& &\phi_l^{(n)}: {\cal M}_1 \times \cdots \times {\cal M}_L 
\mapsto \mathbb{R}^n,l\in\IsetA.
\eeqno
For ${\vc Y}^L$ $=({\vc Y}_1,$ ${\vc Y}_2,$ $\cdots,$ 
${\vc Y}_L)$, set
\beqno
   \hat{\vc Y}^{L}
   &=&\left[
\ba{c}
\hat{\vc Y}_1\\
\hat{\vc Y}_2\\
      \vdots\\
\hat{\vc Y}_L\\
\ea
\right]
\defeq 
\left[
\ba{c}
\phi_1^{(n)}(\varphi^{(n)}({\vc Y}^L))\\
\phi_2^{(n)}(\varphi^{(n)}({\vc Y}^L))\\
\vdots\\ 
\phi_L^{(n)}(\varphi^{(n)}({\vc Y}^L))\\
\ea
\right]\,,
\\
\tilde{d}_{ll}
& \defeq & {\rm E}||{\vc Y}_l-\hat{\vc Y}_l||^2, 
1 \leq l \leq L,
\nonumber\\
\tilde{d}_{ll^{\prime}}
& \defeq &
{\rm E} \langle {\vc Y}_l-\hat{\vc Y}_l,
        {\vc Y}_{l^{\prime}}-\hat{\vc Y}_{l^{\prime}}\rangle\,, 
        1 \leq l \ne l^{\prime} \leq L.
\nonumber
\eeqno
Let $\Sigma_{{\lvc Y}^L-\hat{\lvc Y}^L}$ 
be a covariance matrix with $\tilde{d}_{ll^{\prime}}$ 
in its $(l,l^{\prime})$ element. 

For a given $\DisT$, the rate vector $(R_1,R_2,\cdots, R_L)$ 
is $\DisT$-{\it admissible} if there exists a sequence 
$\{(\varphi_1^{(n)},$
   $\varphi_2^{(n)}, \cdots,$ 
   $\varphi_L^{(n)},$ 
   $\psi^{(n)})\}_{n=1}^{\infty}$ 
such that
\beqno
& &\limsup_{n\to\infty}R_l^{(n)}\leq R_l, 
   \mbox{ for }l\in\IsetA\,,
\\
& &\limsup_{n\to\infty}{\ts \frac{1}{n}}
   \Sigma_{{\lvc Y}^L-\hat{\lvc Y}^L} \preceq \DisT. 
\eeqno
Let ${\cal R}_{\Iset}(\DisT|\Sigma_{Y^L})$ 
denote the set of all $\DisT$-admissible 
rate vectors. 
We consider two types of distortion criterion. For each distortion 
criterion we define the determination 
problem of the rate distortion region. 

{\it Problem 3. Vector Distortion Criterion:}  
For given $L\times L$ invertible matrix $\Gamma$ and 
$D^L>0$, the rate vector $(R_1,R_2,$ $\cdots,R_L)$ is 
$(\Gamma,D^L)$-{\it admissible} if there exists a sequence 
$\{(\varphi_1^{(n)},$ 
   $\varphi_2^{(n)}, \cdots,$ 
   $\varphi_L^{(n)},$ $\phi^{(n)})\}_{n=1}^{\infty}$ 
such that
\beqno
& &\limsup_{n\to\infty}R^{(n)}\leq R_l\,,
\mbox{ for }l\in\IsetA\,,  
\\
& &\limsup_{n\to\infty}
\left[\Gamma\left(
{\ts \frac{1}{n}}
\Sigma_{{\lvc Y}^L-\hat{\lvc Y}^L}
\right){}^{\rm t}\Gamma\right]_{ll} 
\leq D_l\,,
\mbox{ for }l\in\IsetA. 
\eeqno
Let ${\cal R}_{\Iset}(\Gamma,D^L|\Sigma_{Y^L})$ denote the set 
of all $(\Gamma,D^L)$-admissible rate vectors. The sum rate 
part of the rate distortion region is defined by
$$ 
R_{{\rm sum}, L}(\Gamma,D^L|\Sigma_{Y^L})
\defeq \min_{\scs (R_1,R_2,\cdots,R_L)
\atop{\scs \in {\cal R}_{\Iset}(\Gamma,D^L|\Sigma_{Y^L})}}
\left\{\sum_{l=1}^{L}R_l\right\}.
$$

{\it Problem 4. Sum Distortion Criterion:}
For given $L\times L$ invertible matrix $\Gamma$ 
and $D>0$, the rate vector 
$(R_1,R_2,\cdots, R_L)$ is 
$(\Gamma,D)$-{\it admissible} 
if there exists a sequence 
$\{(\varphi_1^{(n)},$ 
   $\varphi_2^{(n)}, \cdots,$ 
   $\varphi_L^{(n)},$ $\phi^{(n)})\}_{n=1}^{\infty}$ 
such that
\beqno
&&\limsup_{n\to\infty}R^{(n)}\leq R_l,
\mbox{ for }l\in\IsetA, 
\\
&&\limsup_{n\to\infty}
{\rm tr}\left[
\Gamma
\left({\ts \frac{1}{n}}\Sigma_{{\lvc Y}^L-\hat{\lvc Y}^L}
\right){}^{\rm t}\Gamma\right] 
\leq D. 
\eeqno
Let ${\cal R}_{\Iset}(\Gamma,D|\Sigma_{Y^L})$ denote 
the set of all admissible rate vectors. The sum 
rate part of the rate distortion region 
is defined by
$$ 
R_{{\rm sum}, L}(\Gamma,D|\Sigma_{Y^L})
\defeq \min_{\scs (R_1,R_2,\cdots,R_L)
\atop{\scs \in {\cal R}_{\Iset}(\Gamma,D|\Sigma_{Y^L})}}
\left\{\sum_{l=1}^{L}R_l\right\}.
$$

Relations between 
${\cal R}_{\Iset}(\DisT|\Sigma_{Y^L}),$ 
${\cal R}_{\Iset}(\Gamma,D^L|\Sigma_{Y^L}),$ and  
${\cal R}_{\Iset}(\Gamma,$ $D|\Sigma_{Y^L})$ 
are as follows.
\beqa
& &{\cal R}_L(\Gamma,D^L|\Sigma_{Y^L})
  =\bigcup_{\Gamma \DisT {}^{\rm t}\Gamma\in {\cal S}_L(D^L)}
   {\cal R}_L({\DisT}|\Sigma_{Y^L}),
\label{eqn:char1}
\\
& &{\cal R}_L(\Gamma,D|\Sigma_{Y^L})
   =\bigcup_{{\rm tr}[\Gamma\DisT{}^{\rm t}\Gamma ] \leq D}
   {\cal R}_L({\DisT}|\Sigma_{Y^L}).
\label{eqn:char2}
\eeqa
Furthermore, we have 
\beq
{\cal R}_L(\Gamma,D|\Sigma_{Y^L})
=\bigcup_{\sum_{l=1}^L D_l\leq D}{\cal R}_L(\Gamma,D^L|\Sigma_{Y^L}).
\eeq

We first present inner bounds of 
${\cal R}_L(\DisT$ $|\Sigma_{Y^L})$,
${\cal R}_L(\Gamma,D^L$ $|\Sigma_{Y^L})$, and 
${\cal R}_L(\Gamma,$ $D|\Sigma_{Y^L})$.
Those inner bounds can be obtained by a standard technique 
of multiterminal source coding. Define 
\beqno
\tilde{\cal G}(\DisT)
&\defeq& \ba[t]{l}
 \left\{U^L \right.:
  \ba[t]{l} 
  U^L\mbox{ is a Gaussian }
  \vspace{1mm}\\
  \mbox{random vector that satisfies}
  \vspace{1mm}\\
  U_S\to Y_S \to X^L \to 
  Y_{S^{\rm c}} \to U_{S^{\rm c}} 
  \vspace{1mm}\\
  U^L \to Y^L \to X^L
  \vspace{1mm}\\
  \mbox{for any $S\subset \IsetA$ and }\\ 
\Sigma_{Y^L-{\phi}(U^L)} \preceq \DisT
  \vspace{1mm}\\
  \mbox{for some linear mapping }
  \vspace{1mm}\\
  {\phi}: \mathbb{R}^L\to \mathbb{R}^L. 
  \left. \right\}
  \ea
\ea
\eeqno
and set 
\beqno
& &
\hat{\cal R}_{L}^{({\rm in})}(\DisT|\Sigma_{Y^L})
\nonumber\\
&\defeq&{\rm conv}\ba[t]{l}
\left\{R^L \right. : 
  \ba[t]{l}
  \mbox{There exists a random vector}
  \vspace{1mm}\\ 
  U^L\in \tilde{\cal G}(\DisT) \mbox{ such that }
  \vspace{1mm}\\
  \ds \sum_{i \in S} R_l \geq I(U_S;Y_S|U_{S^{\rm c}})
  \vspace{1mm}\\
  \mbox{ for any } S\subseteq \IsetA.
  \left. \right\}\,,
  \ea
\ea
\eeqno
\beqno
& &\hat{\cal R}_{L}^{({\rm in})}(\Gamma,D^L|\Sigma_{Y^L}) 
\\
&\defeq& 
\conv 
\left\{
\bigcup_{\Gamma\DisT{}^{\rm t}\Gamma \in {\cal S}_L(D^L)}
\hat{\cal R}_{L}^{({\rm in})}(\DisT|\Sigma_{Y^L})
\right\}\,,
\\
& &\hat{\cal R}_{L}^{({\rm in})}(\Gamma,D|\Sigma_{Y^L}) 
\\
&\defeq &\conv 
\left\{
\bigcup_{{\rm tr}[\Gamma\DisT{}^{\rm t}\Gamma]\leq D}
\hat{\cal R}_{L}^{({\rm in})}(\DisT|\Sigma_{Y^L}) 
\right\}.
\eeqno
Then we have the following result.
\begin{Th}[Berger \cite{bt} and Tung \cite{syt}]\label{th:direct2}
For any positive \\ definite $\DisT$, we have
\beqno
& &\hat{\cal R}_{L}^{({\rm in})}(\DisT|\Sigma_{Y^L})
\subseteq {\cal R}_{L}(\DisT|\Sigma_{Y^L}).
\eeqno
For any invertible $\Gamma$ and any $D^L>0$, we have
\beqno
& &\hat{\cal R}_{L}^{({\rm in})}(\Gamma,D^L|\Sigma_{Y^L})
  \subseteq {\cal R}_{L}(\Gamma,D^L|\Sigma_{Y^L}).
\eeqno
For any invertible $\Gamma$ and any $D>0$, 
we have
\beqno
& &\hat{\cal R}_{L}^{({\rm in})}(\Gamma,D|\Sigma_{Y^L})
\subseteq {\cal R}_{L}(\Gamma,D|\Sigma_{Y^L}).
\eeqno
\end{Th}

The inner bound $\hat{\cal R}_{L}^{({\rm in})}(D^L|\Sigma_{Y^L})$
for $\Gamma=I_L$ is well known as the inner bound of 
Berger \cite{bt} and Tung \cite{syt}. The above 
three inner bounds are variants of this inner bound. 

Optimality of $\hat{\cal R}_{2}^{({\rm in})}(D^2|\Sigma_{Y^2})$
was first studied by Oohama \cite{oh1}. 
Let 
$$
{\arraycolsep 1mm
\Sigma_{Y^2}=
\left[
\ba{cc}
\sigma_1^2 &\rho{\sigma_1\sigma_2}\\
\rho{\sigma_1\sigma_2}& \sigma^2_2\\
\ea
\right],\quad \rho\in [0,1).
}
$$
For $l=1,2$, set
$$
{\cal R}_{l,2}(D_l|\Sigma_{Y^2})
\defeq \bigcup_{D_{3-l}>0}{\cal R}_{2}(D^2|\Sigma_{Y^2}).
$$
Oohama \cite{oh1} obtained the following result.
\begin{Th}[Oohama \cite{oh1}] \label{th:Oh97} 
For $l=1,2$, we have 
$$
{\cal R}_{l,2}(D_l|\Sigma_{Y^2})={\cal R}_{l,2}^{*}(D_l|\Sigma_{Y^2}),
$$
where 
$$
\ba{l} 
{\cal R}_{l,2}^{*}(D_l|\Sigma_{Y^2})\defeq
\vspace{2mm}\\
\hspace*{-1.5mm}
        \ba[t]{rl}
        \Bigl\{(R_1,R_2):
        &\,\hspace*{-1mm}R_l\geq\ts\frac{1}{2}\log^{+}
                       \left[
                       (1-\rho^2)\frac{\sigma^2_l}{D_l}
                       \left(
                       1+\frac{\rho^2}{1-\rho^2}\cdot s
                       \right)
                       \right],
\vspace{2mm}\\
         &\,\hspace*{-1mm}R_{3-l}\geq\ts
                        \frac{1}{2}\log \left[\frac{1}{s}\right]
 \vspace{2mm}\\
         &\,\mbox{ for some }0<s\leq 1\:\Bigl.\Bigr\}.
\ea
\ea
$$
\end{Th}

Since ${\cal R}^{*}_{l,2}(D_l|\Sigma_{Y^2}),$ $l=1,2$ serve
as outer bounds of ${\cal R}_{2}(D^2|\Sigma_{Y^2})$, we have 
\beq
{\cal R}_{2}(D^2|\Sigma_{Y^2})
\subseteq {\cal R}_{1,2}^*(D_1|\Sigma_{Y^2})
\cap{\cal R}_{2,2}^*(D_2|\Sigma_{Y^2}).
\label{eqn:outer0}
\eeq
Wagner {\it et al.} \cite{wg3} derived the 
condition where the outer bound in the right hand side of 
(\ref{eqn:outer0}) is tight. To describe their result 
set
\beqno
\lefteqn{{\cal D}\defeq \Bigl\{(D_1,D_2):D_1,D_2>0,
}\\
& &
\ba[t]{l}
\left.
\max\left\{\frac{D_1}{\sigma_1^2},\frac{D_2}{\sigma_2^2}\right\}
\leq 
\min
\left\{1,\rho^2
\min\left\{
\frac{D_1}{\sigma_1^2},\frac{D_2}{\sigma^2_2}
\right\}+1-\rho^2
\right\}
\right\}.
\ea
\eeqno
Wagner {\it et al.} \cite{wg3} showed that 
if $D^2\notin {\cal D}$, 
we have 
$$
{\cal R}_{2}(D^2|\Sigma_{Y^2})
={\cal R}_{1,2}^*(D_1|\Sigma_{Y^2})
\cap{\cal R}_{2,2}^*(D_2|\Sigma_{Y^2}).
$$
Next we consider the case of $D^2\in {\cal D}$. 
In this case by an elementary computation 
we can show that 
$\hat{\cal R}_2^{\rm(in)}(D^2|\Sigma_{Y^2})$ 
has the following form:
\beqno
& &\hat{\cal R}_2^{\rm(in)}(D^2|\Sigma_{Y^2})
\\
&=& {\cal R}_{1,2}^{*}(D_1|\Sigma_{Y^2})
\cap{\cal R}_{2,2}^{*}(D_2|\Sigma_{Y^2})
\cap{\cal R}^{*}_{3,2}(D^2|\Sigma_{Y^2})\,,
\eeqno
where
\beqno
& &{\cal R}_{3,2}^*(D^2|\Sigma_{Y^2})
\\
&\defeq&\Bigl\{(R_1,R_2):
       R_1+R_2 \geq R_{\rm sum,2}^{\rm (u)}(D^2|\Sigma_{Y^2})
       \Bigr\},
\\
& &R_{\rm sum,2}^{\rm (u)}(D^2|\Sigma_{Y^2})
\\
&\defeq &\min_{(R_1,R_2)\in 
         \hat{\cal R}_2^{\rm (in)}(D^2|\Sigma_{Y^2})}
     \left\{R_1+R_2\right\}
\\
&=&
\ts\frac{1}{2}\log
           \left[
           \frac{1-\rho^2}{2}\cdot
           \left\{
  \frac{\sigma_1^2\sigma_2^2}{D_1D_2}
 +\sqrt{
  \left(\frac{\sigma_1^2\sigma_2^2}{D_1D_2}\right)^2
  +\frac{4\rho^2}{(1-\rho^2)^2}}
 \right\}
\right].
\eeqno
The boundary of $\hat{\cal R}_2^{\rm (in)}(D^2|\Sigma_{Y^2})$ 
consists of one straight line segment defined 
by the boundary of ${\cal R}_{3,2}^*(D^2|\Sigma_{Y^2})$ 
and two curved portions defined by the boundaries of 
${\cal R}_{1,2}^{*}(D_1|\Sigma_{Y^2})$ and
${\cal R}_{2,2}^{*}(D_2|\Sigma_{Y^2})$.
Accordingly, the inner bound established 
by Berger \cite{bt} and Tung \cite{syt} partially
coincides with ${\cal R}_2(D^2|\Sigma_{Y^2})$ 
at two curved portions of its boundary.

Wagner {\it et al.} \cite{wg3} have completed the proof of the 
optimality of $\hat{\cal R}_2^{\rm (in)}(D^2|\Sigma_{Y^2})$ 
by determining the sum rate part 
${R}_{{\rm sum}, 2}(D^2|\Sigma_{Y^2})$. 
Their result is as follows.
\begin{Th}[Wagner {\it et al.} \cite{wg3}]\label{th:SumRateOptL2}
For any $D^2\in {\cal D}$, we have 
\beqno
& &
 {R}_{{\rm sum}, 2}(D^2|\Sigma_{Y^2})
={R}_{{\rm sum}, 2}^{\rm (u)}(D^2|\Sigma_{Y^2})
\\ 
&=&
\ts\frac{1}{2}\log
           \left[
           \frac{1-\rho^2}{2}\cdot
           \left\{
  \frac{\sigma_1^2\sigma_2^2}{D_1D_2}
 +\sqrt{
  \left(\frac{\sigma_1^2\sigma_2^2}{D_1D_2}\right)^2
  +\frac{4\rho^2}{(1-\rho^2)^2}}
 \right\}
\right].
\eeqno
\end{Th}

According to Wagner {\it et al.} \cite{wg3}, the results of Oohama
\cite{oh2}, \cite{oh4} play an essential role in deriving their
result. Their method for the proof depends heavily on the specific
property of $L=2$.  It is hard to generalize it to the case of $L\geq
3$. Recently, Wang {\it et al.} \cite{wa} have given an alternative
proof of Theorem \ref{th:SumRateOptL2}. Their method of the proof is
quite different from the previous method employed by Oohama \cite{oh2}, 
\cite{oh4} and Wagner {\it et al.} \cite{wg3} and also has
a great advantage that it is also applicable to the characterization
of ${R}_{{\rm sum}, L}(D^L|\Sigma_{Y^2})$ for $L\geq3$. Their result
and its relation to our result in the present paper will be discussed
in the next subsection.

\subsection{New Outer Bounds of Positive Semidefinite Programming} 

In this subsection we state our results on 
the characterizations of 
${\cal R}_L(\DisT|\Sigma_{Y^L})$,  
${\cal R}_L(\Gamma,D^L|\Sigma_{Y^L})$, and 
${\cal R}_L(\Gamma,$ $D|\Sigma_{Y^L})$. 
Before describing those results we derive an important 
relation between remote source coding problem and 
multiterminal \RateDist problem. We first observe that 
by an elementary computation we have 
\beq
X^L=\tilde{A}Y^L+\tilde{N}^L\,,
\label{eqn:assad}
\eeq
where 
$\tilde{A}=(\Sigma_{X^L}^{-1}$
$+\Sigma_{N^L}^{-1})^{-1}\Sigma_{N^L}^{-1}$
and $\tilde{N}^L$ is a zero mean Gaussian random vector 
with covariance matrix 
$\Sigma_{\tilde{N}^L}$
$ 
=(\Sigma_{X^L}^{-1}$ $+\Sigma_{N^L}^{-1})^{-1}.
$
The random vector $\tilde{N}^L$ is independent of $Y^L$. 
Set
\beqno
B&\defeq&\tilde{A}^{-1}
  \Sigma_{\tilde{N}^L}{}^{\rm t}\tilde{A}^{-1}
 =\Sigma_{N^L}+\Sigma_{N^L}\Sigma_{X^L}^{-1}\Sigma_{N^L}\,,
\\
b^L&\defeq& {}^{\rm t}([B]_{11},[B]_{22},\cdots,[B]_{LL})\,,
\\
\tilde{B}&\defeq&\Gamma B {}^{\rm t}\Gamma\,,
\\
\tilde{b}^L&\defeq&{}^{\rm t}
([\tilde{B}]_{11},[\tilde{B}]_{22},\cdots,[\tilde{B}]_{LL}).
\eeqno
From (\ref{eqn:assad}), we have the following 
relation between ${\vc X}^L$ and ${\vc Y}^L$:   
\beq
{\vc X}^L=\tilde{A}{\vc Y}^L+\tilde{\vc N}^L,
\label{eqn:relation1}
\eeq
where $\tilde{\vc N}^L$ is a sequence of $n$ 
independent copies of $\tilde{N}^L$ and is 
independent of ${\vc Y}^L$. Now, we fix  
$\{(\varphi_1^{(n)},$
   $\varphi_2^{(n)}, \cdots,$ 
   $\varphi_L^{(n)},$ 
   $\psi^{(n)})\}_{n=1}^{\infty}$, arbitrarily. 
For each $n=1,2,\cdots$, the estimation $\hat{\vc X}^{L}$ 
of ${\vc X}^L$ is given by    
\beqno
   \hat{\vc X}^{L}
   &=&
\left[
\ba{c}
        \psi_1^{(n)}(\varphi^{(n)}({\vc Y}^L))\\
        \psi_2^{(n)}(\varphi^{(n)}({\vc Y}^L))\\
        \vdots\\ 
        \psi_L^{(n)}(\varphi^{(n)}({\vc Y}^L))\\
\ea
\right].
\eeqno
Using this estimation, we construct an estimation 
$\hat{\vc Y}^{L}$ of ${\vc Y}^L$ by 
$
\hat{\vc Y}^L=\tilde{A}^{-1}\hat{\vc X}^L\,,
$
which is equivalent to 
\beq
\hat{\vc X}^L=\tilde{A}\hat{\vc Y}^L.
\label{eqn:relation3}
\eeq
From (\ref{eqn:relation1}) 
 and (\ref{eqn:relation3}), we have
\beq
           {\vc X}^L-\hat{\vc X}^L
=\tilde{A}({\vc Y}^L-\hat{\vc Y}^L) + \tilde{{\vc N}^L}.
\label{eqn:relation4}
\eeq
Since $\hat{\vc Y}^L$ is a function of ${\vc Y}^L$, 
$\hat{\vc Y}^L-{\vc Y}^L$ is independent 
of $\tilde{\vc N}^L$. Based on (\ref{eqn:relation4}), 
we compute
${\ts \frac{1}{n}}\Sigma_{{\lvc X}^L-\hat{\lvc X}^L}$ 
to obtain
\beq
{\ts \frac{1}{n}}\Sigma_{{\lvc X}^L-\hat{\lvc X}^L}=
\tilde{A}
\left({\ts \frac{1}{n}}\Sigma_{{\lvc Y}^L-\hat{\lvc Y}^L}
\right){}^{\rm t}\tilde{A}+\Sigma_{\tilde{N}^L}.
\label{eqn:zaa}
\eeq
From (\ref{eqn:zaa}), we have 
\beqa
{\ts \frac{1}{n}}\Sigma_{{\lvc Y}^L-\hat{\lvc Y}^L}
&=& \tilde{A}^{-1}
\left({\ts \frac{1}{n}}\Sigma_{{\lvc X}^L-\hat{\lvc X}^L}
-\Sigma_{\tilde{N}^L}\right){}^{\rm t}\tilde{A}^{-1}
\nonumber\\
&=&\tilde{A}^{-1}
\left({\ts \frac{1}{n}}\Sigma_{{\lvc X}^L-\hat{\lvc X}^L}\right)
{}^{\rm t}\tilde{A}^{-1}-B.
\label{eqn:relation5}
\eeqa
Conversely, we fix  
$\{(\varphi_1^{(n)},$
   $\varphi_2^{(n)}, \cdots,$ 
   $\varphi_L^{(n)},$ 
   $\phi^{(n)})\}_{n=1}^{\infty}$, arbitrarily. 
For each $n=1,2,\cdots$, using the estimation 
$\hat{\vc Y}^{L}$ of ${\vc Y}^L$ 
given by    
\beqno
   \hat{\vc Y}^{L}
   &=&
\left[
\ba{c}
\phi_1^{(n)}(\varphi^{(n)}({\vc Y}^L))\\
\phi_2^{(n)}(\varphi^{(n)}({\vc Y}^L))\\
\vdots\\ 
\phi_L^{(n)}(\varphi^{(n)}({\vc Y}^L))\\
\ea
\right]\,,
\eeqno
we construct an estimation $\hat{\vc X}^L$ of ${\vc X}^L$ by  
(\ref{eqn:relation3}). Then using (\ref{eqn:relation1}) 
  and (\ref{eqn:relation3}),
we obtain (\ref{eqn:relation4}). Hence we have the 
relation (\ref{eqn:zaa}).

The following proposition provides an important strong 
connection between remote source coding problem and 
multiterminal \RateDist problem.

\begin{pro}\label{pro:MainTh1} For any positive 
definite $\DisT$, we have 
\beqno
{\cal R}_{L}(\DisT|\Sigma_{Y^L})
&=&
{\cal R}_{L}(\tilde{A}(\DisT+B){}^{\rm t}\tilde{A}|\Sigma_{X^LY^L}).
\eeqno
For any invertible $\Gamma$ and any $D^L>0$, we have
\beqno
{\cal R}_{L}(\Gamma,D^L|\Sigma_{Y^L})
&=&{\cal R}_{L}(\Gamma\tilde{A}^{-1},D^L+\tilde{b}^L|\Sigma_{X^LY^L}).
\eeqno
For any invertible $\Gamma$ and any $D>0$, we have
\beqno
{\cal R}_{L}(\Gamma,D|\Sigma_{Y^L})
&=&{\cal R}_{L}(\Gamma\tilde{A}^{-1},
D+{\rm tr}[\tilde{B}]|\Sigma_{X^LY^L}).
\eeqno
\end{pro}

{\it Proof:}
Suppose that 
$R^L$ $\in {\cal R}_L(\tilde{A}(\DisT+B){}^{\rm t}\tilde{A}
|\Sigma_{X^LY^L})$. 
Then there exists
$\{(\varphi_1^{(n)},$
   $\varphi_2^{(n)}, \cdots,$ 
   $\varphi_L^{(n)},$ 
   $\psi^{(n)})\}_{n=1}^{\infty}$ 
such that
\beqno
& &\limsup_{n\to\infty}R^{(n)}\leq R_l, 
   \mbox{ for }l\in\IsetA\,,
\\
& &\limsup_{n\to\infty}{\ts \frac{1}{n}}
   \Sigma_{{\lvc X}^L-\hat{\lvc X}^L} \preceq 
\tilde{A}(\DisT+B){}^{\rm t}\tilde{A}. 
\eeqno
Using $\hat{\vc X}^L$, we construct an 
estimation $\hat{\vc Y}^L$ of ${\vc Y}^L$ by 
$\hat{\vc Y}^L$$=\tilde{A}^{-1}\hat{\vc X}^L$. 
Then from (\ref{eqn:relation5}), we have
\beqno
& &\limsup_{n\to\infty}{\ts \frac{1}{n}}\Sigma_{{\lvc Y}^L-\hat{\lvc Y}^L}
\\
&=& \limsup_{n\to\infty}
\tilde{A}^{-1}
\left({\ts \frac{1}{n}}\Sigma_{{\lvc X}^L-\hat{\lvc X}^L}\right)
{}^{\rm t}\tilde{A}^{-1}-B
\nonumber\\
&\preceq&
\tilde{A}^{-1}
\tilde{A}(\DisT+B){}^{\rm t}
\tilde{A}{}^{\rm t}\tilde{A}^{-1}-B
=\DisT\,,
\eeqno
which implies that 
$R^L\in$
${\cal R}_{L}(\tilde{A}(\DisT+B){}^{\rm t}\tilde{A}|
 \Sigma_{X^LY^L}).$
Thus
$$
{\cal R}_{L}(\DisT|\Sigma_{Y^L})
\supseteq 
{\cal R}_{L}(\tilde{A}(\DisT+B){}^{\rm t}\tilde{A}|\Sigma_{X^LY^L})
$$
is proved. Next we prove the reverse inclusion. Suppose 
that $R^L$ $\in {\cal R}_L(\DisT|\Sigma_{Y^L})$. Then 
there exists
$\{(\varphi_1^{(n)},$
$\varphi_2^{(n)}, \cdots,$ 
$\varphi_L^{(n)},$ 
$\phi^{(n)})\}_{n=1}^{\infty}$ 
such that
\beqno
& &\limsup_{n\to\infty}R^{(n)}\leq R_l, 
   \mbox{ for }l\in\IsetA\,,
\\
& &\limsup_{n\to\infty}{\ts \frac{1}{n}}
   \Sigma_{{\lvc Y}^L-\hat{\lvc Y}^L} \preceq \DisT. 
\eeqno
Using $\hat{\vc Y}^L$, we construct an 
estimation $\hat{\vc X}^L$ of ${\vc X}^L$ by 
$\hat{\vc X}^L$$=\tilde{A}\hat{\vc Y}^L$. 
Then from (\ref{eqn:zaa}), we have
\beqno
& &\limsup_{n\to\infty}{\ts \frac{1}{n}}\Sigma_{{\lvc X}^L-\hat{\lvc X}^L}
\\
&=& \limsup_{n\to\infty}
\tilde{A}
\left({\ts \frac{1}{n}}\Sigma_{{\lvc Y}^L-\hat{\lvc Y}^L}\right)
{}^{\rm t}\tilde{A}+\Sigma_{\tilde{N}^L}
\nonumber\\
&\preceq&
\tilde{A}\DisT{}^{\rm t}\tilde{A}{}^{\rm t}+\Sigma_{\tilde{N}^L}
=\tilde{A}(\DisT+B){}^{\rm t}\tilde{A}{}^{\rm t}\,,
\eeqno
which implies that
$R^L\in$ 
${\cal R}_{L}(\tilde{A}(\DisT+B){}^{\rm t}\tilde{A}$ 
$|\Sigma_{X^LY^L}).$
Thus,
$$
{\cal R}_{L}(\DisT|\Sigma_{Y^L})
\subseteq 
{\cal R}_{L}(\tilde{A}(\DisT+B){}^{\rm t}\tilde{A}|\Sigma_{X^LY^L})
$$
is proved. Next we prove the second equality. We have the following
chain of equalities:
\beqa
& &{\cal R}_{L}(\Gamma,D^L|\Sigma_{Y^L})
=
\bigcup_{\Gamma\DisT{}^{\rm t}\Gamma \in {\cal S}_L(D^L)}
{\cal R}_L({\DisT}|\Sigma_{Y^L})
\nonumber\\
&=&
\bigcup_{\Gamma\DisT{}^{\rm t}\Gamma\in {\cal S}_L(D^L)}
{\cal R}_L(\Gamma\tilde{A}(\DisT+B){}^{\rm t}\tilde{A}|\Sigma_{X^L Y^L})
\nonumber\\
&=&
\bigcup_{
\scs
\Gamma
\tilde{A}^{-1}
\tilde{A}(\DisT+B){}^{\rm t}\tilde{A}{}^{\rm t}\tilde{A}^{-1}
{}^{\rm t}\Gamma
\atop{\scs -\Gamma B {}^{\rm t}\Gamma \in {\cal S}_L(D^L)}}
{\cal R}_L(\tilde{A}(\DisT+B){}^{\rm t}\tilde{A}|\Sigma_{X^LY^L})
\nonumber\\
&=&
\bigcup_{
\scs
\Gamma\tilde{A}^{-1}\tilde{A}(\DisT+B){}^{\rm t}\tilde{A}
{}^{\rm t}(\Gamma\tilde{A}^{-1})
\atop{\scs \in {\cal S}_L(D^L+\tilde{b}^L)}}
{\cal R}_L(\tilde{A}(\DisT+B){}^{\rm t}\tilde{A}|\Sigma_{X^L Y^L})
\nonumber\\
&=&
\bigcup_{
\scs \hat{\Sigma}_d=\tilde{A}(\DisT+B){}^{\rm t}\tilde{A}
\succ \Sigma_{X^L|Y^L}, 
\atop{\scs
\Gamma\tilde{A}^{-1}\hat{\Sigma}_d
{}^{\rm t}(\Gamma\tilde{A}^{-1})\in {\cal S}_L(D^L+\tilde{b}^L)}}
{\cal R}_L(\hat{\Sigma}_d|\Sigma_{X^L Y^L})
\nonumber\\
&=&{\cal R}_{L}(\Gamma\tilde{A}^{-1},D^L+\tilde{b}^L|\Sigma_{X^LY^L}).
\nonumber
\eeqa
Thus the second equality is proved. 
Finally we prove the third equality. 
We have the following
chain of equalities:
\beqa
& &{\cal R}_{L}(\Gamma,D|\Sigma_{Y^L})
=
\bigcup_{{\rm tr}[\Gamma\DisT{}^{\rm t}\Gamma]\leq D}
{\cal R}_L({\DisT}|\Sigma_{Y^L})
\nonumber\\
&=&
\bigcup_{{\rm tr}[\Gamma\DisT{}^{\rm t}\Gamma]\leq D}
{\cal R}_L(\Gamma\tilde{A}(\DisT+B){}^{\rm t}\tilde{A}|\Sigma_{X^LY^L})
\nonumber\\
&=&
\bigcup_{
\scs
{\rm tr}[\Gamma\tilde{A}^{-1}
\tilde{A}(\DisT+B){}^{\rm t}\tilde{A}{}^{\rm t}
\tilde{A}^{-1}{}^{\rm t}\Gamma]
\atop{\scs -
{\rm tr}[\Gamma B {}^{\rm t}\Gamma]\leq D}}
{\cal R}_L(\tilde{A}(\DisT+B){}^{\rm t}\tilde{A}|\Sigma_{X^LY^L})
\nonumber\\
&=&
\bigcup_{
\scs
{\rm tr}[
\Gamma\tilde{A}^{-1}\tilde{A}(\DisT+B){}^{\rm t}\tilde{A}
{}^{\rm t}(\Gamma\tilde{A}^{-1})]
\atop{\scs \leq D+{\rm tr}[\tilde{B}]}}
{\cal R}_L(\tilde{A}(\DisT+B){}^{\rm t}\tilde{A}|\Sigma_{X^LY^L})
\nonumber\\
&=&
\bigcup_{
\scs \hat{\Sigma}_d=\tilde{A}(\Sigma_d +B){}^{\rm t}\tilde{A} 
\succ \Sigma_{X^L|Y^L} ,
\atop{\scs
{\rm tr}[\Gamma\tilde{A}^{-1}
\hat{\Sigma}_d{}^{\rm t}(\Gamma \tilde{A}^{-1})]
\leq D+{\rm tr}[\tilde{B}]}}
{\cal R}_L(\hat{\Sigma}_d|\Sigma_{X^LY^L})
\nonumber\\
&=&  {\cal R}_{L}(\Gamma\tilde{A}^{-1},
   D+{\rm tr}[\tilde{B}]|\Sigma_{X^LY^L}).
\nonumber
\eeqa
Thus the third equality is proved. 
\hfill\IEEEQED

Proposition \ref{pro:MainTh1} implies that all results on the 
rate distortion regions for the remote source coding 
problems can be converted into those on the multiterminal 
source coding problems. In the following we derive inner 
and outer bounds of ${\cal R}_{L}(\DisT|\Sigma_{Y^L})$, 
${\cal R}_{L}(\Gamma,D^L|\Sigma_{Y^L})$, and 
${\cal R}_{L}(\Gamma,D|\Sigma_{Y^L})$ 
using Proposition \ref{pro:MainTh1}. 
We first derive inner and outer bounds of 
${\cal R}_{L}(\DisT|\Sigma_{Y^L})$.
For each $l\in \IsetA$ and for $r_l\geq 0$, 
let $V_{l}(r_l),$ $l\in \IsetA$ be a Gaussian random variable with 
mean 0 and variance $\sigma_{N_l}^2/({\baseN}^{2r_l}-1)$. 
We assume that $V_l(r_l),l\in \IsetA$ are independent. 
When $r_l=0$, we formally think that the inverse value 
$\sigma_{V_l(0)}^{-1}$ of $V_l(0)$ is zero. 
Let $\Sigma_{V^L(r^L)}$ be a covariance matrix 
of the random vector $V^L(r^L)$. 
When $r_S={\vc 0}$, we formally define
%
%
\beqno
\Sigma_{V_{S^{\rm c}}(r_{S^{\rm c}})}^{-1} 
&\defeq& 
\left. \Sigma_{V^L(r^L)}^{-1} \right|_{r_{S}={\lvc 0}}\,.
\eeqno
Fix nonnegative vector $r^L$. For $\theta >0$ 
and for $S \subseteq  \IsetA$, define
\beqno
\underline{\tilde{J}}_{S}(\theta, r_S|r_{\coS})
&\defeq &\frac{1}{2}\log^{+}
   \left[\ts 
   \frac{\ds |\Sigma_{Y^L}+B|\prod_{l=1}^L {\baseN}^{2r_i} }
        {\ds \theta |\Sigma_{Y^L}|\left|\Sigma_{Y^L}^{-1}
                 +\Sigma_{V_{S^{\rm c}}(r_{S^{\rm c}})}^{-1}\right|
        }
  \right],
\\
\tilde{J}_{S}\left(r_S|r_{\coS}\right)
&\defeq &\frac{1}{2}\log
   \frac{\ds \left|\Sigma_{Y^L}^{-1}
    +\Sigma_{V^L(r^L)}^{-1}\right|
        }
        {\ds  \left|\Sigma_{Y^L}^{-1}
        +\Sigma_{V_{S^{\rm c}}(r_{S^{\rm c}})}^{-1}\right|.
        }
\eeqno
Set
$$
\tilde{\cal A}_L(\DisT)
\defeq 
\left\{ r^L\geq 0:
\left[\Sigma_{Y^L}^{-1}+
\Sigma_{V^L(r^L)}^{-1}\right]^{-1}
\preceq \DisT\right\}.
$$
Define four regions by 
\beqno
{\cal R}_L^{({\rm out})}(\theta,r^L|\Sigma_{Y^L})
&\defeq&
\ba[t]{l}
  \left\{R^L \right. : 
  \ba[t]{l}
  \ds \sum_{l \in S} R_l 
  \geq \underline{\tilde{J}}_{S}\left(\theta,r_S|r_{\coS}\right)
  \vspace{1mm}\\
  \mbox{ for any }S \subseteq \IsetA. 
  \left. \right\}\,,
  \ea
\ea
\nonumber\\
{\cal R}_{L}^{({\rm out})}(\DisT|\Sigma_{Y^L})
&\defeq& 
\bigcup_{r^L \in \tilde{\cal A}_L(\DisT)}
{\cal R}_L^{({\rm out})}(|\DisT+B|,r^L|\Sigma_{Y^L})\,, 
\nonumber\\
{\cal R}_L^{({\rm in})}(r^L|\Sigma_{Y^L})
&\defeq&
\ba[t]{l}
  \left\{R^L \right.: 
  \ba[t]{l}
  \ds \sum_{l \in S} R_l 
  \geq {J}_{S}\left(r_S|r_{\coS}\right)
  \vspace{1mm}\\
  \mbox{ for any }S \subseteq \IsetA. 
  \left. \right\}\,,
  \ea
\ea
\nonumber\\
{\cal R}_L^{({\rm in})}(\DisT|\Sigma_{Y^L})
&\defeq&{\rm conv}
        \left\{
        \bigcup_{r^L \in \tilde{\cal A}_L(\DisT)}
        {\cal R}_L^{({\rm in})}(r^L|\Sigma_{Y^L})
        \right\}. 
\eeqno
The functions and sets defined above have 
properties shown in the following.   
\begin{pr}{
\label{pr:prz01zs}
$\quad$
\begin{itemize}
\item[{\rm a)}]$\:$For any positive definite 
$\DisT$,  
$\tilde{\cal G}(\DisT)=
 {\cal G}(\tilde{A}(\DisT+B){}^{\rm t}\tilde{A})$.
\vspace*{1mm}
\item[{\rm b)}]$\:$For any positive definite 
$\DisT$, we have   
$$
\hat{\cal R}_{L}^{\rm (in)}(\DisT|\Sigma_{Y^L})
=\hat{\cal R}_{L}^{\rm (in)}
(\tilde{A}(\DisT+B){}^{\rm t}\tilde{A}
|\Sigma_{X^LY^L}).
$$
\item[{\rm c)}]$\:$For any positive definite $\DisT$ 
and any $S\subseteq \IsetA$, we have
\beqno
& &
\underline{\tilde{J}}_S(|\DisT+B|,r_S|r_{\coS})
=\underline{J}_S
(|\tilde{A}(\DisT+B){}^{\rm t}\tilde{A}|,r_S|r_{\coS}),
\\
& &\tilde{J}_S(r_S|r_{\coS})=J_S(r_S|r_{\coS}).
\eeqno
\item[{\rm d)}]$\:$For any positive definite $\DisT$, 
$
\tilde{\cal A}_L(\DisT)
={\cal A}_L(\tilde{A}(\DisT+B){}^{\rm t}\tilde{A})
.$
\vspace*{1mm}
\item[{\rm e)}]$\:$For any positive definite $\DisT$, we have 
$$
\ba[t]{l}
{\cal R}_{L}^{\rm (out)}(\DisT|\Sigma_{Y^L})
={\cal R}_{L}^{\rm (out)}
   (\tilde{A}(\DisT+B){}^{\rm t}\tilde{A}|\Sigma_{X^LY^L})\,,
\vspace*{1mm}\\
{\cal R}_{L}^{\rm (in)}(\DisT|\Sigma_{Y^L})
={\cal R}_{L}^{\rm (in)}
(\tilde{A}(\DisT+B){}^{\rm t}\tilde{A}|\Sigma_{X^LY^L}).
\ea
$$
\end{itemize}
}\end{pr}

From Theorem \ref{th:conv2}, Proposition \ref{pro:MainTh1} 
and Property \ref{pr:prz01zs}, 
we have the following.
\begin{Th}\label{tho:MainTh2a}$\:$For any positive 
definite $\DisT$, we have
\beqno
& &{\cal R}_{L}^{({\rm in})}(\DisT|\Sigma_{Y^L})
= \hat{\cal R}_{L}^{({\rm in})}(\DisT|\Sigma_{Y^L}) 
\\
&\subseteq & 
{\cal R}_{L}(\DisT|\Sigma_{Y^L}) 
\subseteq {\cal R}_{L}^{({\rm out})}(\DisT|\Sigma_{Y^L}). 
\eeqno
\end{Th}

Next, we derive inner and outer bounds of 
${\cal R}_{L}(\Gamma,$$D^K|\Sigma_{Y^L})$
and 
${\cal R}_{L}(\Gamma,$$D|\Sigma_{Y^L})$.
Set 
\beqno
\tilde{\cal A}_L(r^L)
&\defeq &\{\DisT:
    \DisT 
    \succeq (\Sigma_{Y^L}^{-1}
+\Sigma_{V^L(r^L)}^{-1})^{-1}\}\,,
\\
\tilde{\theta}(\Gamma,D^L,r^L)
&\defeq &  
\max_{\scs \DisT:\DisT \in \tilde{\cal A}_L(r^L), 
      \atop{\scs
      \Gamma\DisT {}^{\rm t}\Gamma \in{\cal S}_L(D^L)}
     }
\left|\DisT+B\right|\,,
\\
\tilde{\theta}(\Gamma,D,r^L)
&\defeq & 
\max_{\scs \DisT: \DisT \in \tilde{\cal A}_L({r^L}),
      \atop{\scs 
      {\rm tr}[\Gamma\DisT {}^{\rm t}\Gamma]\leq D}
      }
\left|\DisT+B\right|.
\eeqno
Furthermore, set 
\beqno
& &\tilde{\cal B}_L(\Gamma,D^L)
\\
&\defeq&
\left\{r^L\geq 0:
\Gamma
(\Sigma_{Y^L}^{-1}+\Sigma_{V^L(r^L)}^{-1})^{-1}
{}^{\rm t}\Gamma
\in{\cal S}_L(D^L)
\right\}\,,
\\
& &\tilde{\cal B}_L(\Gamma,D)
\\
&\defeq&
\left\{r^L\geq 0:
{\rm tr}\left[
\Gamma
(\Sigma_{Y^L}^{-1}+\Sigma_{V^L(r^L)}^{-1})^{-1}
{}^{\rm t}\Gamma
\right]
\leq D 
\right\}.
\eeqno
Define four regions by
\beqno
& &{\cal R}_{L}^{({\rm out})}(\Gamma,D^L|\Sigma_{Y^L})
\\
&\defeq& 
\bigcup_{r^L \in \tilde{\cal B}_L(\Gamma,D^L)}
{\cal R}_L^{({\rm out})}(\tilde{\theta}
(\Gamma,D^L,r^L),r^L|\Sigma_{Y^L}), 
\\
& &{\cal R}_L^{({\rm in})}(\Gamma,D^L|\Sigma_{Y^L})
\\
&\defeq&
        \conv\left\{\bigcup_{r^L \in \tilde{\cal B}_L(\Gamma,D^L)}
        {\cal R}_L^{({\rm in})}(r^L|\Sigma_{Y^L})\right\},
\\
& &{\cal R}_{L}^{({\rm out})}(\Gamma,D|\Sigma_{Y^L})
\\
&\defeq& 
\bigcup_{r^L \in \tilde{\cal B}_L(\Gamma,D)}
{\cal R}_L^{({\rm out})}(\tilde{\theta}(\Gamma,D,r^L),r^L|\Sigma_{Y^L}), 
\\
& &{\cal R}_L^{({\rm in})}(\Gamma,D|\Sigma_{Y^L})
\\
&\defeq&\conv\left\{
        \bigcup_{r^L \in \tilde{\cal B}_L(\Gamma,D)}
        {\cal R}_L^{({\rm in})}(r^L|\Sigma_{Y^L})\right\}.
\eeqno
It can easily be verified that the functions and sets defined 
above have the properties shown in the following.
\begin{pr}{
\label{pr:PropZ}
$\quad$
\begin{itemize}
\item[{\rm a)}]$\:$For any invertible $\Gamma$ and any $D^L>0$, 
we have  
\beqno
& & \hat{\cal R}_{L}^{\rm (in)}(\Gamma,D^L|\Sigma_{Y^L})
\\
&=&\hat{\cal R}_{L}^{\rm (in)}(\Gamma\tilde{A}^{-1},
    D^L+\tilde{b}^L|\Sigma_{X^LY^L}).
\eeqno
For any invertible $\Gamma$ and any $D>0$, we have    
\beqno
& & \hat{\cal R}_{L}^{\rm (in)}(\Gamma,D|\Sigma_{Y^L})
\\
&=&\hat{\cal R}_{L}^{\rm (in)}(\Gamma\tilde{A}^{-1},
    D+{\rm tr}[\tilde{B}]|\Sigma_{X^LY^L}).
\eeqno
\item[{\rm b)}] For any $r^L\geq 0$, we have   
\beqno
& &\DisT\in\tilde{\cal A}(r^L)
\Leftrightarrow \tilde{A}(\DisT+B){}^{\rm t}\tilde{A}
\in{\cal A}(r^L),
\\
& &\tilde{\theta}(\Gamma,D^L,r^L)
=\left|\tilde{A}\right|^{-2}\theta(\Gamma\tilde{A}^{-1},D^L,r^L), 
\\
& &\tilde{\theta}(\Gamma,D,r^L)
=\left|\tilde{A}\right|^{-2}\theta(\Gamma\tilde{A}^{-1},D,r^L).
\eeqno

\item[{\rm c)}]$\:$For any invertible $\Gamma$ and 
any $D^L>0$, we have    
\beqno
& &{\cal R}_{L}^{\rm (out)}(\Gamma,D^L|\Sigma_{Y^L})
\\
&=&{\cal R}_{L}^{\rm (out)}(\Gamma\tilde{A}^{-1},
    D^L+\tilde{b}^L|\Sigma_{X^LY^L}),
\\
& &{\cal R}_{L}^{\rm (in)}(\Gamma,D^L|\Sigma_{Y^L})
\\
&=&{\cal R}_{L}^{\rm (in)}(\Gamma\tilde{A}^{-1},
    D^L+\tilde{b}^L|\Sigma_{X^LY^L}).
\eeqno
For any invertible $\Gamma$ and any $D>0$, we have   
\beqno
& &{\cal R}_{L}^{\rm (out)}(\Gamma,D|\Sigma_{Y^L})
\\
&=&{\cal R}_{L}^{\rm (out)}(\Gamma\tilde{A}^{-1},
    D+{\rm tr}[\tilde{B}]|\Sigma_{X^LY^L}),
\\
& &{\cal R}_{L}^{\rm (in)}(\Gamma,D|\Sigma_{Y^L})
\\
&=&{\cal R}_{L}^{\rm (in)}(\Gamma\tilde{A}^{-1},
    D+{\rm tr}[\tilde{B}]|\Sigma_{X^LY^L}).
\eeqno
\end{itemize}
}\end{pr}

From Corollary \ref{co:conv2z}, Proposition \ref{pro:MainTh1} 
and Property \ref{pr:PropZ}, we have the following theorem. 
\begin{Th}\label{th:conv4vv}\ For any invertible 
$\Gamma$ and any $D>0$, we have       
\beqno
&         &    {\cal R}_{L}^{({\rm in})}(\Gamma,D^L|\Sigma_{Y^L})
=  \hat{\cal R}_{L}^{({\rm in})}(\Gamma,D^L|\Sigma_{Y^L}) 
\\
&\subseteq&    {\cal R}_{L}(\Gamma,D^L|\Sigma_{Y^L}) 
 \subseteq     {\cal R}_{L}^{({\rm out})}(\Gamma,D^L|\Sigma_{Y^L}). 
\eeqno
For any invertible $\Gamma$ and any $D>0$, we have    
\beqno
 &        &    {\cal R}_{L}^{({\rm in})}(\Gamma,D|\Sigma_{Y^L})
 = \hat{\cal R}_{L}^{({\rm in})}(\Gamma,D|\Sigma_{Y^L})    
\\
&\subseteq&    {\cal R}_{L}(\Gamma,D|\Sigma_{Y^L}) 
 \subseteq     {\cal R}_{L}^{({\rm out})}(\Gamma,D|\Sigma_{Y^L}). 
\eeqno
\end{Th}

The outer bound ${\cal R}_{L}^{({\rm out})}(\Gamma,D^L|\Sigma_{Y^L})$
has a form of positive semidefinite programming.  To find a matching
condition for inner and outer bounds to match, we must examine a
property of the solution to this positive semidefinite programming.  On
the sum rate part of the rate distortion region in the case of vector
distortion criterion we have the following corollary from Theorem
\ref{th:conv4vv}.

\begin{co}\label{co:coZ}
For any $D^L>0$, we have
\beqno
R_{{\rm sum},L}^{(\rm l)}(D^L|\Sigma_{Y^L}) 
&\leq & R_{{\rm sum},L}(D^L|\Sigma_{Y^L}) 
\\
&\leq & R_{{\rm sum},L}^{(\rm u)}(D^L|\Sigma_{Y^L}), 
\eeqno
where
\beqno
&  &
R_{{\rm sum},L}^{(\rm u)}(D^L|\Sigma_{Y^L}) 
\\
&\defeq&
\min_{\scs 
r^L: \scs (\Sigma_{Y^L}^{-1}+\Sigma_{V^L(r^L)}^{-1})^{-1}
\atop{\scs \in{\cal S}_L(D^L)}
}
\frac{1}{2}\log|I+\Sigma_{Y^L}\Sigma_{V^L(r^L)}^{-1}|
\\
&=&
\min_{
\scs 
(r^L,\Sigma_d): 
\atop{
\scs \Sigma_d\in{\cal S}_L(D^L),      
\atop{\scs \Sigma_d=(\Sigma_{Y^L}^{-1}+\Sigma_{V^L(r^L)}^{-1})^{-1}
}}}
\frac{1}{2}\log\frac{|\Sigma_{Y^L}|}{|\Sigma_d|}
\\
&=&
\min_{
\scs 
(r^L,\Sigma_d): 
\atop{ \scs\Sigma_d\in{\cal S}_L(D^L),
\atop{\scs \Sigma_d=(\Sigma_{Y^L}^{-1}+\Sigma_{V^L(r^L)}^{-1})^{-1}
}}}
\left\{\frac{1}{2} 
\log
\frac{|\Sigma_{Y^L}+B|}{|\Sigma_d+B|}
+\sum_{l=1}^Lr_i\right\},
\\
&  &
R_{{\rm sum},L}^{(\rm l)}(D^L|\Sigma_{Y^L}) 
\\
&\defeq&
\min_{\scs 
\scs 
(r^L,\Sigma_d): 
\atop{\scs \Sigma_d\in{\cal S}_L(D^L),
\atop{\scs \Sigma_d\succeq(\Sigma_{Y^L}^{-1}+\Sigma_{V^L(r^L)}^{-1})^{-1}
}}}
\left\{
\frac{1}{2} 
\log
    \frac{|\Sigma_{Y^L}+B|}{|\Sigma_d+B|}
+\sum_{l=1}^Lr_i
\right\}.
\eeqno
\end{co}

A lower bound of $R_{{\rm sum},L}(D^L|\Sigma_{Y^L})$ 
in a form of positive semidefinite programming   
was first obtained by Wang {\it et al.}
\cite{wa}.
Their lower bound denoted by 
$\tilde{R}_{{\rm sum},L}^{\rm (l)}(D^L|\Sigma_{Y^L})$
is as follows. Let 
$\delta^L\defeq(\delta_1,\delta_2,\cdots,\delta_L)$
be a positive vector whose components 
$\delta_l$, $l\in \IsetA$ belong to $(0,\sigma_{N_l}^2]$. 
Let ${\rm Diag.}(\delta^L)$ 
be a diagonal matrix whose 
$(l,l)$ element is $\delta_l,l\in\IsetA$. Then 
$\tilde{R}_{{\rm sum},L}^{\rm (l)}(D^L|\Sigma_{Y^L})$
is given by 
\beqno
&  &
\tilde{R}_{{\rm sum},L}^{(\rm l)}(D^L|\Sigma_{Y^L}) 
\\
&\defeq&
\min_{\scs 
     (\delta^L,\Sigma_d):
       \atop{\scs \Sigma_d \in{\cal S}_L(D^L), 
         \atop{\scs \delta_l\in (0,\sigma_{N_l}^2],l\in \IsetA, 
           \atop{\scs
           (\Sigma_d^{-1}+B^{-1})^{-1}
           \succeq \mbox{\footnotesize Diag.}(\delta^L)
           }
       }
    }
}
\hspace*{-10mm}
\left\{
\frac{1}{2} 
\log
\frac{|\Sigma_{Y^L}+B|}{|\Sigma_d+B|}
+\sum_{l=1}^L\frac{1}{2}\log\frac{\sigma_{N_l}^2}{\delta_l}
\right\}.
\eeqno
By simple computation we can show that 
$\tilde{R}_{{\rm sum},L}^{(\rm l)}(D^L|\Sigma_{Y^L})$ 
$={R}_{{\rm sum},L}^{(\rm l)}(D^L|\Sigma_{Y^L})$. 
Although the lower bound 
$\tilde{R}_{{\rm sum},L}^{(\rm l)}($ $D^L|\Sigma_{Y^L})$ 
of Wang {\it et al.} \cite{wa} is equal to our lower bound
${R}_{{\rm sum},L}^{(\rm l)}($ $D^L|\Sigma_{Y^L})$, 
their method to derive 
$\tilde{R}_{{\rm sum},L}^{(\rm l)}($ 
$D^L|\Sigma_{Y^L})$ 
is essentially different from our method. They derived 
the lower bound by utilizing the semidefinite 
partial order of the covariance matrices associated
with 
MMSE estimation.
Unlike our method, the method of Wang {\it et al.} 
is not directly applicable to the characterization of 
the entire rate distortion region.  

When $L=2$, Wang {\it et al.} \cite{wa} solved 
the positive semidefinite programming describing 
$\tilde{R}_{{\rm sum},2}^{(\rm l)}($ $D^2|\Sigma_{Y^2})$
to obtain the following result.
\begin{lm}[Wang {\it et al.} \cite{wa}]\label{lm:wang}
For any covariance matrix $\Sigma_{Y^2}$, there exist a pair 
$(\Sigma_{X^2},$ $\Sigma_{N^2})$ of covariance and 
diagonal covariance matrices
such that $\Sigma_{Y^2}=\Sigma_{X^2}+\Sigma_{N^2}$ 
and   
$$
\tilde{R}_{{\rm sum},2}^{(\rm l)}(D^2|\Sigma_{Y^2})
=R_{{\rm sum},2}^{(\rm u)}(D^2|\Sigma_{Y^2}).
$$
\end{lm}

From Corollary \ref{co:coZ} and Lemma \ref{lm:wang}, 
we have the following corollary.
\begin{co}\label{co:coZ0}
\beqno
&&\tilde{R}_{{\rm sum},2}^{(\rm l)}(D^2|\Sigma_{Y^2})   
     ={R}_{{\rm sum},2}^{(\rm l)}(D^2|\Sigma_{Y^2})   
\\
&=&{R}_{{\rm sum},2}(D^2|\Sigma_{Y^2})
  ={R}_{{\rm sum},2}^{(\rm u)}(D^2|\Sigma_{Y^2}).
\eeqno
\end{co}

Our method to derive  
${R}_{{\rm sum},2}^{(\rm l)}(D^2|\Sigma_{Y^2})
\leq {R}_{{\rm sum},2}(D^2|\Sigma_{Y^2})$
in Corollary \ref{co:coZ} essentially differs from the method  
of Wang {\it et al.} \cite{wa} to derive 
$\tilde{R}_{{\rm sum},2}^{(\rm l)}(D^2|\Sigma_{Y^2})
\leq {R}_{{\rm sum},2}(D^2|\Sigma_{Y^2})$.
Our method to obtain Corollary \ref{co:coZ0} is also 
quite different from that of Wagner {\it et al.} \cite{wg3} 
to prove Theorem \ref{th:SumRateOptL2}.
Hence, Corollary \ref{co:coZ0} 
provides the second alternative proof of 
Theorem \ref{th:SumRateOptL2}.

\subsection{Matching Condition Analysis}

In this subsection, we derive a matching condition for ${\cal
R}_{L}^{({\rm out})}(\Gamma,D|\Sigma_{Y^L})$ to coincide with ${\cal
R}_{L}^{\rm (in)}(\Gamma,D|\Sigma_{Y^L})$. Using the derived
matching condition we derive more explicit matching condition when
$\Gamma$ is a positive semidefinite diagonal matrix. Furthermore we
apply this result to the analysis of matching condition in the case of
vector distortion criterion.

By the third equality of Proposition \ref{pro:MainTh1}, 
the determination problem of 
${\cal R}_{L}(\Gamma,D|\Sigma_{Y^L})$
can be converted into the determination problem of 
${\cal R}_{L}(\Gamma\tilde{A}^{-1},$ 
$D+{\rm tr}[\tilde{B}]|$ $\Sigma_{X^LY^L}).$
Using Theorem \ref{th:matchTh}, we derive a matching 
condition for 
$
{\cal R}_{L}^{\rm (in)}(\Gamma\tilde{A}^{-1},
$
$
D+{\rm tr}[\tilde{B}]|\Sigma_{X^LY^L}) 
$
to coincide with 
$
{\cal R}_{L}^{\rm (out)}(\Gamma\tilde{A}^{-1},
$
$
D+{\rm tr}[\tilde{B}]|\Sigma_{X^LY^L}). 
$
For simplicity of our analysis we use the second simplified 
matching condition (\ref{eqn:match102}) in Theorem \ref{th:matchTh}. 
Note that 
\beqa
& &\left[
{}^{\rm t}(\Gamma \tilde{A}^{-1})^{-1}
(\Sigma_{X^L}^{-1}+\Sigma_{N^L}^{-1})
(\Gamma \tilde{A}^{-1})^{-1}\right]^{-1}
\nonumber\\
&=&\Gamma\tilde{A}^{-1}
(\Sigma_{X^L}^{-1}+\Sigma_{N^L}^{-1})^{-1}
{}^{\rm t}(\Gamma\tilde{A}^{-1})
=\tilde{B}.
\label{eqn:zap}
\eeqa
By (\ref{eqn:zap}), the second matching condition 
in Theorem \ref{th:matchTh}, the third equality of 
Proposition \ref{pro:MainTh1}, and Property 
\ref{pr:PropZ} part c), we establish the following. 
\begin{Th}
\label{th:matchThZ}
Let $\mu_{\min}^{\ast}$ be the minimum eigenvalue of
$$
\tilde{B}
=\Gamma\left(
\Sigma_{N^L}+\Sigma_{N^L}\Sigma_{X^L}^{-1}\Sigma_{N^L}
\right){}^{\rm t}\Gamma.
$$
If we have 
\beqno
0<D&\leq (L+1)\mu_{\min}^{\ast}-
{\rm tr}\left[
\Gamma(\Sigma_{N^L}+\Sigma_{N^L}\Sigma_{X^L}^{-1}\Sigma_{N^L})
{}^{\rm t}\Gamma
\right],
\eeqno
then 
\beqno
& &{\cal R}_L^{({\rm in})}(\Gamma,D|\Sigma_{Y^L})
=\hat{\cal R}_L^{({\rm in})}(\Gamma,D|\Sigma_{Y^L})
\\
&=&{\cal R}_L(\Gamma,D|\Sigma_{Y^L})
 ={\cal R}_L^{({\rm out})}(\Gamma,D|\Sigma_{Y^L}).
\eeqno
\end{Th} 

An important feature of the multiterminal rate 
distortion problem is that the rate distortion region 
${\cal R}_L(\Gamma,D|\Sigma_{Y^L})$
remains the same for any choice of covariance matrix $\Sigma_{X^L}$
and diagonal covariance matrix $\Sigma_{N^L}$ satisfying
$\Sigma_{Y^L}$ $=\Sigma_{X^L}+\Sigma_{N^L}$. 
Using this feature and Theorem \ref{th:matchThZ}, we 
find a good pair $(\Sigma_{X^L},$
$\Sigma_{N^L})$ to provide an explicit strong 
sufficient condition for
${\cal R}_L^{({\rm in})}(\Gamma,$ $D|\Sigma_{Y^L})$ 
and ${\cal R}_L^{({\rm out})}(\Gamma,$ $D|\Sigma_{Y^L})$ 
to match.

In the following argument we consider the case where $\Gamma$ 
is the following positive definite diagonal matrix:
\beq
\Gamma=
\left(
\begin{array}{cccc}
\gamma_1 &           &        & \mbox{\huge 0}\\
          & \gamma_2 &        &          \\
          &           & \ddots &          \\
\mbox{\huge 0} &      &        & \gamma_{L}\\
\end{array}
\right)
,\quad \gamma_l\in[1,+\infty).
\label{eqn:diagZ}
\eeq
Set
$
\gamma^L \defeq (\gamma_1,\gamma_2,\cdots,\gamma_L) 
\in [1,+\infty)^L.
$
We call $\gamma^L$ the weight vector. Since $\Gamma$ 
is specified by the weight vector $\gamma^L$, 
we write ${\cal R}_L(\Gamma,D|\Sigma_{Y^L})$ 
as ${\cal R}_L(\gamma^L,D|\Sigma_{Y^L})$. 
Similar notations are adopted for other regions.

We choose $\Sigma_{N^L}$ so that 
$\Sigma_{N^L}=\delta\Gamma^{-2}$. 
Set $\tilde{\Sigma}_{X^L}\defeq$ $\Gamma{\Sigma}_{X^L}\Gamma$ and 
$\tilde{\Sigma}_{Y^L}\defeq$ $\Gamma{\Sigma}_{Y^L}\Gamma$.
Then, we have
\beq
\left.
\ba{rcl}
\tilde{B}&=&\delta I_L + \delta^2 \tilde{\Sigma}^{-1}_{X^L}, 
\\
\tilde{\Sigma}_{X^L}&=&\tilde{\Sigma}_{Y^L}-\delta I_L.
\ea
\right\}
\label{eqn:zzl}
\eeq
Let 
$
\eta_{\min}\defeq \eta_1 \leq \eta_2 \leq \cdots
\leq \eta_L\defeq \eta_{\max} 
$
be the ordered list of $L$ eigenvalues of ${\Sigma}_{Y^L}$ and let 
$
\tilde{\eta}_{\min}\defeq \tilde{\eta}_1 \leq \tilde{\eta}_2 \leq \cdots
\leq \tilde{\eta}_L\defeq \tilde{\eta}_{\max}
$
be the ordered list of $L$ eigenvalues 
of $\tilde{\Sigma}_{Y^L}$. Set $\gamma_{\max}
\defeq\max_{1\leq l \leq L}\gamma_i$. Since 
$
\eta_{\min}I_L \preceq {\Sigma}_{Y^L}\preceq \eta_{\max}I_L, 
$
we have 
$$
     \eta_{\min}I_L \preceq
\eta_{\min}\Gamma^2 \preceq
\tilde{\Sigma}_{Y^L}\preceq 
\eta_{\max}\Gamma^2 \preceq
\gamma_{\max}^2\eta_{\max}I_L,
$$
from which we obtain
\beq
\eta_{\min}\leq \tilde{\eta}_{\min}\leq 
\tilde{\eta}_{\max} \leq \gamma_{\max}^2\eta_{\max}.
\label{eqn:asoo20}
\eeq
We choose $\delta$ so that $0<\delta$ $ < \tilde{\eta}_{\min}$. 
Then, by (\ref{eqn:zzl}), we have
\beq
\left.
\ba{l}
\ds\mu_{\min}^{*}
=\delta +{\frac{\delta^2}{\tilde{\eta}_{\max}-\delta}},
\\
\ds{\rm tr}[\tilde{B}]={\rm tr}\left[\delta I_L 
+\delta^2 \tilde{\Sigma}_{X^L}^{-1}\right]
=L\delta+{\ds \sum_{l=1}^L}{ \frac{\delta^2}{\tilde{\eta}_i-\delta}}.
\ea
\right\}
\label{eqn:asoo}
\eeq
From (\ref{eqn:asoo}), we have  
\beqa
\hspace*{-3mm}& &(L+1)\mu_{\min}^{*}-{\rm tr}[\tilde{B}]
=\delta + { \frac{(L+1)\delta^2}{\tilde{\eta}_{\max}-\delta}}
            -\sum_{l=1}^L{ \frac{\delta^2}{\tilde{\eta}_{l}-\delta}}
\nonumber\\
&=&\delta + { \frac{L\delta^2}{\tilde{\eta}_{\max}-\delta}}
           -\sum_{l=1}^{L-1}{ \frac{\delta^2}{\tilde{\eta}_{l}-\delta}}
\nonumber\\
\hspace*{-3mm}
&\geq& \delta +  L\frac{\delta^2}{\tilde{\eta}_{\max}-\delta}
       -(L-1)\frac{\delta^2}{\tilde{\eta}_{\min}-\delta}
\nonumber\\
\hspace*{-3mm}
&=& L\tilde{\eta}_{\max}
    \left(\frac{\tilde{\eta}_{\max}}{\tilde{\eta}_{\max}-\delta}-1\right)
\nonumber\\
\hspace*{-3mm}
& & -(L-1)\tilde{\eta}_{\min}
    \left(\frac{\tilde{\eta}_{\min}}{\tilde{\eta}_{\min}-\delta}-1\right).
\label{eqn:ssaz2}
\eeqa
By an elementary computation we can show that 
the right member of (\ref{eqn:ssaz2}) takes the maximum value 
\beqno
& &{(\sqrt{L}-\sqrt{L-1})^2}\cdot
   \frac{\tilde{\eta}_{\max}\tilde{\eta}_{\min}}
{\tilde{\eta}_{\max}-\tilde{\eta}_{\min}}
\nonumber\\
&=&\frac{1}{(\sqrt{L}+\sqrt{L-1})^2}\cdot
    \frac{\tilde{\eta}_{\max}\tilde{\eta}_{\min}}
   {\tilde{\eta}_{\max}-\tilde{\eta}_{\min}}
\eeqno
at
$$
\delta=
\frac{(\sqrt{L}-\sqrt{L-1})\tilde{\eta}_{\max}\tilde{\eta}_{\min}}
{\sqrt{L}\tilde{\eta}_{\max}-\sqrt{L-1}\tilde{\eta}_{\min}}.
$$
Furthermore, taking (\ref{eqn:asoo20}) into account, we obtain
\beqno
\frac{\tilde{\eta}_{\max} \tilde{\eta}_{\min}}
     {\tilde{\eta}_{\max}-\tilde{\eta}_{\min}}
&=& 
\left[\tilde{\eta}_{\min}^{-1}
 -\tilde{\eta}_{\max}^{-1}\right]^{-1}
  \geq \left[{\eta}_{\min}^{-1}
-\gamma_{\max}^{-2}{\eta}_{\max}^{-1}\right]^{-1}
\nonumber\\
&=& \frac{{\eta}_{\max}{\eta}_{\min}}
   {{\eta}_{\max}-\gamma_{\max}^{-2}{\eta}_{\min}}.
\eeqno
Hence if
\beqno
0<D\leq \frac{1}{(\sqrt{L}+\sqrt{L-1})^2}\cdot
    \frac{\eta_{\max}\eta_{\min}}
   {\eta_{\max}-\gamma_{\max}^{-2}\eta_{\min}},
\eeqno
then the matching condition holds. Summarizing the above 
argument, we obtain the following corollary from 
Theorem \ref{th:matchThZ}.
\begin{co}\label{co:matchTh}
Let $\gamma^L\in [1,+\infty)^L$ be a weight vector and 
let $\gamma_{\max}$$=\max_{1\leq l\leq L}\gamma_l$. 
If
\beqno
0<D\leq \frac{1}{(\sqrt{L}+\sqrt{L-1})^2}\cdot
    \frac{\eta_{\max}\eta_{\min}}
   {\eta_{\max}-\gamma_{\max}^{-2}\eta_{\min}},
\eeqno
then we have 
\beqa
& &{\cal R}_L^{({\rm in})}(\gamma^L,D|\Sigma_{Y^L})
  =\hat{\cal R}_L^{({\rm in})}(\gamma^L,D|\Sigma_{Y^L})
\nonumber\\
&=&{\cal R}_L(\gamma^L,D|\Sigma_{Y^L}) 
  ={\cal R}_L^{({\rm out})}(\gamma^L,D|\Sigma_{Y^L}).
\label{eqn:ppa2}
\eeqa
In particular, if 
$$
0<D\leq \frac{1}{(\sqrt{L}+\sqrt{L-1})^2}\cdot \eta_{\min},
$$
then we have (\ref{eqn:ppa2}) for any weight vector 
$\gamma^L\in$ $[1,\infty)^L$. If $\gamma_{\max}=1$ and 
$$
0<D\leq \frac{1}{(\sqrt{L}+\sqrt{L-1})^2}\cdot
    \frac{\eta_{\max}\eta_{\min}}
   {\eta_{\max}-\eta_{\min}},
$$
then we have
\beqno
& &{\cal R}_L^{({\rm in})}(D|\Sigma_{Y^L})
  =\hat{\cal R}_L^{({\rm in})}(D|\Sigma_{Y^L})
\nonumber\\
&=&{\cal R}_L(D|\Sigma_{Y^L})
  ={\cal R}_L^{({\rm out})}(D|\Sigma_{Y^L}).
\eeqno
\end{co}

Fix $\gamma^L\in [1,+\infty)^L$ arbitrarily.
Consider the region ${\cal R}_L($ 
$\gamma^L|\Sigma_{Y^L})$ 
and the minimum distortion
$D_L(\gamma^L,R^L|\Sigma_{Y^L})$ induced by 
${\cal R}_L(\gamma^L,D|\Sigma_{Y^L})$. 
Those are formally defined by
\beqno
{\cal R}_L(\gamma^L |\Sigma_{Y^L}) 
\defeq \left\{(R^L,D): 
R^L\in {\cal R}_L(\gamma^L, D|\Sigma_{Y^L})\right\}, 
\\
{D}_L(\gamma^L,R^L|\Sigma_{Y^L}) 
\defeq \inf 
\left\{D:(R^L,D)\in {\cal R}_L(\gamma^L|\Sigma_{Y^L})\right\}.
\eeqno
Similarly, we define
\beqno
\lefteqn{{\cal R}_L^{\rm (in)}(\gamma^L |\Sigma_{Y^L})} 
\\
&\defeq& \left\{(R^L,D): R^L
\in {\cal R}_L^{\rm (in)}(\gamma^L, D|\Sigma_{Y^L})\right\}, 
\\
\lefteqn{{\cal R}_L^{\rm (out)}(\gamma^L |\Sigma_{Y^L})}
\\ 
&\defeq& 
\left\{(R^L,D): R^L
\in {\cal R}_L^{\rm (out)}(\gamma^L, D|\Sigma_{Y^L})\right\}, 
\\
\lefteqn{{D}_L^{\rm (u)}(\gamma^L,R^L|\Sigma_{Y^L})}
\\ 
&\defeq& \inf \left\{D:(R^L,D)
\in {\cal R}_L^{\rm (in)}(\gamma^L|\Sigma_{Y^L})\right\},
\\
\lefteqn{{D}_L^{\rm (l)}(\gamma^L,R^L|\Sigma_{Y^L})} 
\\
&\defeq&\inf 
\left\{D:(R^L,D)\in {\cal R}_L^{\rm (out)}(\gamma^L|\Sigma_{Y^L})
\right\}.
\eeqno
From Theorem \ref{th:conv4vv} and Corollary \ref{co:matchTh}, 
we obtain the following corollary.
\begin{co}\label{co:matchTh2}
For any $R^L\geq 0$ and any $\gamma^L \in [1,+\infty)^L$, 
we have 
\beqno
       {D}_L^{\rm (u)}(\gamma^L,R^L|\Sigma_{Y^L}) 
&\geq &{D}_L^{\rm}(\gamma^L,R^L|\Sigma_{Y^L})
\\
&\geq &{D}_L^{\rm (l)}(\gamma^L,R^L |\Sigma_{Y^L}).
\eeqno
For each $\gamma^L \in [1,+\infty)^L$, if we have 
$$
0<D_L^{\rm (u)}(\gamma^L,R^L |\Sigma_{Y^L})
\leq \frac{1}{(\sqrt{L}+\sqrt{L-1})^2}
\cdot\eta_{\min},
$$
then 
\beqno
{D}_L^{({\rm u})}(\gamma^L,R^L|\Sigma_{Y^L})
&=&{D}_L(\gamma^L,R^L|\Sigma_{Y^L}) 
\\
&=&{D}_L^{({\rm l})}(\gamma^L,R^L|\Sigma_{Y^L}).
\nonumber
\eeqno
\end{co}

We apply Corollary \ref{co:matchTh2}
to the derivation of matching condition in the case 
of vector distortion criterion. We consider the region
${\cal R}_L(\Sigma_{Y^L})$ and the distortion 
rate region ${\cal D}_L(R^L|\Sigma_{Y^L})$ 
induced by ${\cal R}(D^L|\Sigma_{Y^L})$. 
Those two regions are formally defined by 
\beqno
& & {\cal R}_L(\Sigma_{Y^L})
\defeq 
\left\{
(R^L,D^L): R^L\in {\cal R}_L(D^L|\Sigma_{Y^L})
\right\},
\\
& &{\cal D}_L(R^L|\Sigma_{Y^L})
\defeq 
\left\{D^L: (R^L,D^L)\in {\cal R}_L(\Sigma_{Y^L})\right\}.
\eeqno
Similarly, we define 
\beqno
& & {\cal R}_L^{\rm (in)}(\Sigma_{Y^L})
\defeq 
\left\{
(R^L,D^L): R^L\in {\cal R}_L^{\rm (in)}(D^L|\Sigma_{Y^L})
\right\},
\\
& &{\cal D}_L^{\rm (in)}(R^L|\Sigma_{Y^L})
\defeq 
\left\{D^L: (R^L,D^L)\in {\cal R}_L^{\rm (in)}(\Sigma_{Y^L})\right\}.
\eeqno
Although the distortion rate region is merely an alternative 
characterization of the rate distortion region, 
the former is more convenient than the latter 
for our analysis of matching condition.  
We examine a part of the boundary of 
${\cal D}^{(\rm in)}(R^L|\Sigma_{Y^L})$ 
which coincides with the boundary of ${\cal D}(R^L|\Sigma_{Y^L})$. 
By definition of $D_L(\gamma^L,R^L |\Sigma_{Y^L})$
and $D_L^{\rm (u)}(\gamma^L,R^L|\Sigma_{Y^L})$, we have
\beqa
D_L(\gamma^L,R^L |\Sigma_{Y^L})
&=&\min_{D^L\in {\cal D}_L(R^L|\Sigma_{Y^L})}
{\sum_{l=1}^L\gamma_i^2D_l},
\label{eqn:aas1k}\\
D_L^{\rm (u)}(\gamma^L,R^L |\Sigma_{Y^L})
&=&\min_{D^L\in {\cal D}_L^{\rm (in)}(R^L|\Sigma_{Y^L})}
{\sum_{l=1}^L \gamma_l^2D_l}.
\label{eqn:aas1k1}
\eeqa
Consider the following two hyperplanes: 
\beqno
\Pi_L(\gamma^L)&\defeq& 
\left\{D^L:
\sum_{l=1}^L\gamma_l^2D_l=D_L(\gamma^L,R^L |\Sigma_{Y^L})
\right\},  
\\
\Pi_L^{\rm (u)}(\gamma^L)&\defeq & 
\left\{D^L:
\sum_{l=1}^L\gamma_l^2D_l=D_L^{(u)}(\gamma^L,R^L |\Sigma_{Y^L})
\right\}.  
\eeqno
It can easily be verified that the region 
${\cal D}_L(R^L |\Sigma_{Y^L})$
is a closed convex set. Then by (\ref{eqn:aas1k}), 
$\Pi_L(\gamma^L)$ becomes a supporting hyperplane 
of ${\cal D}_L(R^L |\Sigma_{Y^L})$ and every $D^L\in$ 
$\Pi_L(\gamma^L)\cap$ ${\cal D}_L(R^L |\Sigma_{Y^L})$ 
is on the boundary of ${\cal D}_L(R^L |\Sigma_{Y^L})$.
On the other hand, by its definition the region 
${\cal D}_L^{\rm (in)}(R^L |\Sigma_{Y^L})$ is 
also a closed convex set. Then by (\ref{eqn:aas1k1}), 
$\Pi_L^{\rm (u)}(\gamma^L)$ becomes a supporting hyperplane of 
${\cal D}_L^{\rm (in)}(R^L |\Sigma_{Y^L})$ and every 
$D^L\in$$\Pi_L^{\rm (u)}(\gamma^L)\cap$ 
${\cal D}_L^{\rm (in)}(R^L |\Sigma_{Y^L})$ 
is on the boundary of ${\cal D}_L^{\rm (in)}(R^L |\Sigma_{Y^L})$.
Set
\beqa
\zeta_L &\defeq &\frac{1}{(\sqrt{L}+\sqrt{L-1})^2}\eta_{\min},
\nonumber\\
{\cal T}_L(\zeta_L)
\nonumber
&\defeq&
\left\{\gamma^L\in [1,+\infty)^L: 
D_L^{(u)}(\gamma^L,R^L |\Sigma_{Y^L})\leq \zeta_L
\right\}.
\nonumber
\eeqa
Then by Corollary \ref{co:matchTh2}, for any 
$\gamma^L\in{\cal T}_L(\zeta_L)$, we have 
$
\Pi_L^{\rm (u)}(\gamma^L)= \Pi_L(\gamma^L),
$
which together with ${\cal D}_L^{\rm (in)}(R^L$ $|\Sigma_{Y^L})$ 
$\subseteq {\cal D}_L(R^L$ $|\Sigma_{Y^L})$
implies that 
every $D^L$ $\in \Pi_L^{\rm (u)}(\gamma^L)\cap$ 
${\cal D}_L^{\rm (in)}(R^L |\Sigma_{Y^L})$ 
must belong to 
$\Pi_L(\gamma^L)\cap$ ${\cal D}_L(R^L |\Sigma_{Y^L})$. 
Hence this $D^L$ must be on the boundary 
of ${\cal D}_L(R^L $ $|\Sigma_{Y^L})$.
It can easily be verified that an existence of 
$\Pi_L^{\rm (u)}(\gamma^L)$ satisfying 
$\gamma^L\in {\cal T}_L(\zeta_L)$ is equivalent to 
$
\Pi_L^{\rm (u)}(\gamma^L)$ $\cap\{ D^L\geq 0\}$
$\subseteq {\cal D}_L^{\rm (+)}(\zeta_L),$
where
\beqno
{\cal D}_L^{\rm (+)}(\zeta_L)
&\defeq&\left\{ D^L: D^L\geq 0,\sum_{l=1}^LD_l
\leq \zeta_L
\right\}.
\eeqno
Summarizing the above argument, we establish the following.
\begin{Th}
The distortion rate region ${\cal D}_L(R^L |\Sigma_{Y^L})$ 
and its inner bound ${\cal D}_L^{\rm (in)}(R^L |\Sigma_{Y^L})$ 
share their boundaries at 
${\cal D}_L^*(\zeta_L)\cap {\cal D}_L^{\rm (in)}(R^L |\Sigma_{Y^L})$, 
where 
\beqno
{\cal D}_L^*(\zeta_L)&\defeq&
\bigcup_{\gamma^L\in {\cal T}_L(\zeta_L)}\Pi_L^{\rm (u)}(\gamma^L)
\\
&=& 
\bigcup_{\Pi_L^{\rm (u)}(\gamma^L)\cap \{D^L\geq 0\}
\subseteq {\cal D}_L^{(+)}(\zeta_L)}
\Pi_L^{\rm (u)}(\gamma^L).
\eeqno
\end{Th}

\begin{figure}[t]
\bc
\includegraphics[width=7.2cm]{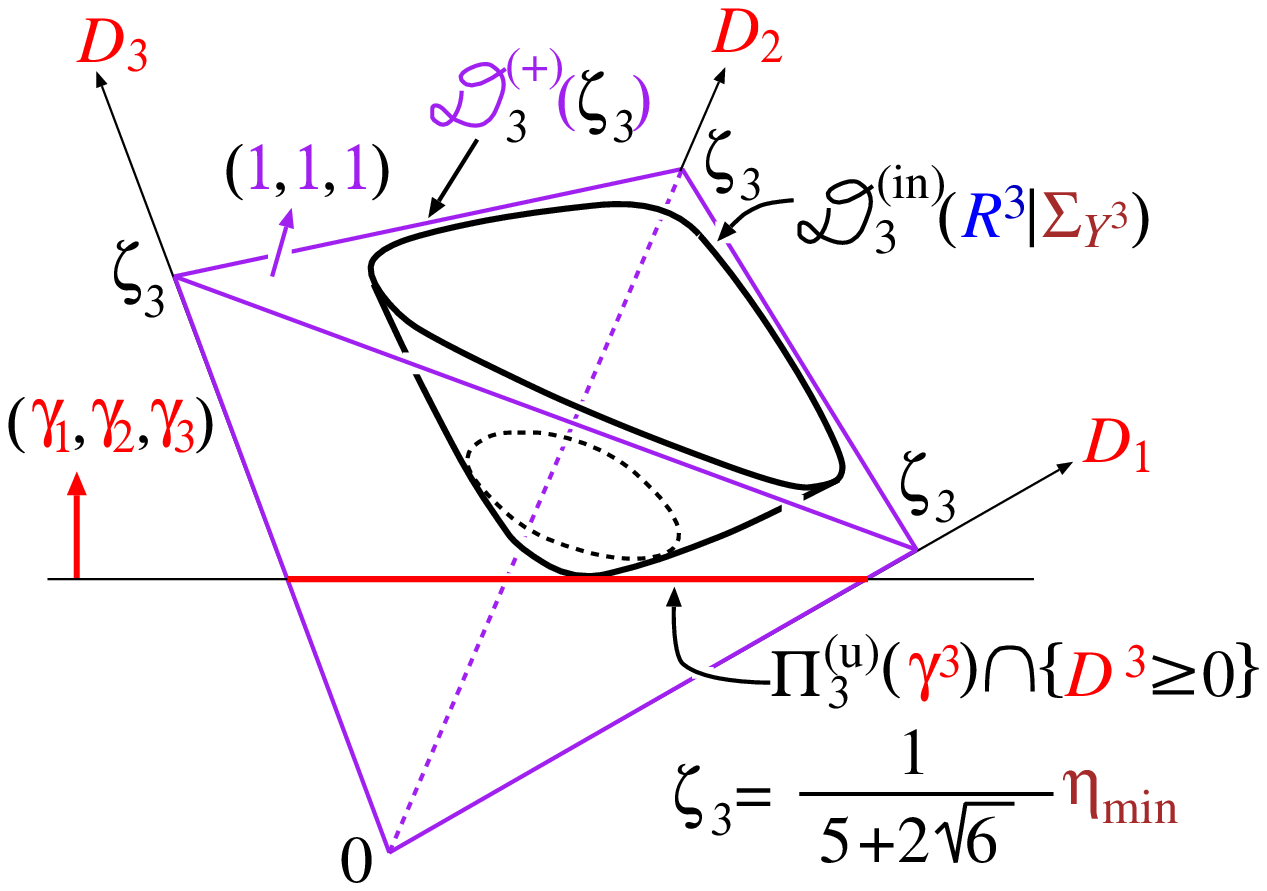}
\ec
\small
\caption{
${\cal D}_3^{\rm (in)}(R^3 |\Sigma_{Y^3})$, 
$\Pi_L^{\rm (u)}(\gamma^3)\cap \{D^3\geq 0\},$
and ${\cal D}_3^{(+)}$$({\zeta_3})$ in the case of $L=3$. 
In this figure we are in a position so that 
we can view the supporting hyperplane 
$\Pi_3^{\rm (u)}(\gamma^3)$ as a horizontal line. 
}
\label{fig:FigReg}
\end{figure}

When $L=3$, we show  
${\cal D}_3^{\rm (in)}(R^3 |\Sigma_{Y^3})$, 
${\cal D}_3^{(+)}(\zeta_3)$, and $\Pi_3^{\rm (u)}(\gamma^3)$
$\cap \{D^3\geq 0\}$ 
in Fig. \ref{fig:FigReg}.   

\subsection{Sum Rate Characterization for 
the Cyclic Shift Invariant Source}

In this subsection we further examine an explicit characterization 
of $R_{{\rm sum},L}($ $D|\Sigma_{Y^l})$ when the source has 
a certain symmetrical property. Let 
\beqno
\tau&=&
\left(
\ba{cccccc}
    1 &2&\cdots&      l &\cdots&    L\\ 
\tau(1)&\tau(2)&\cdots&\tau(l)&\cdots&\tau(L)
\ea
\right)
\eeqno
be a cyclic shift on 
$\IsetA$, that is,
$$
\tau(1)=2,\tau(2)=3,\cdots,\tau(L-1)=L,\tau(L)=1.
$$
Let
$
p_{X_{\IsetA}}(x_{\IsetA})
=p_{X_1X_2\cdots X_L}(x_1,x_2,\cdots,x_L)
$     
be a probability density function of $X^L$. 
The source ${X^L}$ is said to be cyclic 
shift invariant if we have
\beqno
p_{X_{\IsetA}}(x_{\tau(\IsetA)})
&=&p_{X_1X_2\cdots X_L}(x_2,x_3,\cdots,x_L,x_1)
\\
&=&p_{X_1X_2\cdots X_L}(x_1,x_2,\cdots,x_{L-1},x_{L})
\eeqno 
for any $(x_1,x_2,$ $\cdots,x_L)$ $\in {\cal X}^L$. 
In the following argument we assume that $X^L$ satisfies 
the cyclic shift invariant property. 
We further assume that $N_l, l\in \IsetA$ are i.i.d. Gaussian 
random variables with mean 0 and variance $\epsilon$.
Then, the observation $Y^L=X^L+N^L$ also satisfies 
the cyclic shift invariant property. 
We assume that the covariance matrix $\Sigma_{N^L}$ of $N^L$ is 
given by $\epsilon I_L$. Then $\tilde{A}$ and $B$ are given by
\beqno  
\tilde{A}&=&
\left(\epsilon\Sigma_{X^L}^{-1}+I_L\right)^{-1},\:
B=\epsilon\left(I_L+\epsilon\Sigma_{X^L}^{-1}\right).
\eeqno
Fix $r>0$, let $N_l(r),$ $l\in \IsetA$ be $L$ 
i.i.d. Gaussian random variables with mean 0 and 
variance $\epsilon/(1-{\baseN}^{-2r})$. The covariance 
matrix $\Sigma_{N^L(r)}$ for the random 
vector $N^L(r)$ is given by 
$$
\Sigma_{N^L(r)}=\frac{1-{\baseN}^{-2r}}{\epsilon}I_L. 
$$
Let $\mu_l, l\in\IsetA$ be 
$L$ eigenvalues of the matrix $\Sigma_{Y^L}$ and let 
$\beta_l=\beta_l(r), l\in\IsetA$ 
be $L$ eigenvalues of the matrix 
$$
{}^{\rm t}\tilde{A}\left(\Sigma_{X^L}^{-1}+
\frac{1-{\baseN}^{-2r}}{\epsilon}I_L 
\right)\tilde{A}.
$$
Using the eigenvalues of $\Sigma_{Y^L}$, 
$\beta_l(r), l\in\IsetA$ 
can be written as
$$
\beta_l(r)=\frac{1}{\epsilon}
\left[
1-\frac{\epsilon}{\mu_l}-
\left(1-\frac{\epsilon}{\mu_l} \right)^2{\rm e}^{-2r}
\right].
$$
Let $\xi$ be a nonnegative number that satisfies 
$$
\sum_{l=1}^L\{[\xi-\beta_l^{-1}]^{+}+\beta_l^{-1}\}
=D+{\rm tr}[B].
$$ 
Define  
\beqno
\tilde{\omega}(D,r)
&\defeq&
\prod_{l=1}^L\left\{[\xi-\beta_l^{-1} ]^{+}
+\beta_l^{-1}\right\}.
\eeqno
The function $\tilde{\omega}(D,r)$ has 
an expression of the so-called water filling 
solution to the following optimization problem:
\beqa
\tilde{\omega}(D,r)
= \max_{\scs \xi_l\beta_{l}\geq 1,l\in\IsetA, 
  \atop{\scs
          \sum_{l=1}^L\xi_l\leq D+{\rm tr}[B]
       }   
      }\prod_{l=1}^L\xi_{l}. 
\eeqa
Set
\beqno
\underline{\tilde{J}}(D,r)
&\defeq &\frac{1}{2}\log
   \left[\ts 
   \frac{\ds {\baseN}^{2Lr}\left|\Sigma_{Y^L}+B\right|}
        {\ds \tilde{\omega}(D,r)}
  \right],
\\
\pi(r)&\defeq&{\rm tr}
\left[\tilde{A}^{-1}\left(\Sigma_{X^L}^{-1}
+\frac{1-{\rm e}^{-2r}}{\epsilon}I_L
\right)^{-1}{}^{\rm t}\tilde{A}^{-1}\right].
\eeqno
By definition we have
\beq
\pi(r)  
=\sum_{l=1}^{L}\frac{1}{\beta_l(r)}.
\label{eqn:zeta1}
\eeq
Since $\pi(r)$ is a monotone decreasing function 
of $r$, there exists a unique $r$ such that 
$\pi(r)=D+{\rm tr}[B]$, we denote it by 
$r^{\ast}(D+$ ${\rm tr}[B])$.  
We can show that $\tilde{\omega}(D,r)$ satisfies the following property.
\begin{pr}\label{pr:prFz}
$\quad$
\begin{itemize}
\item[a)] For $D>0$, 
\beqno
& &(\underbrace{r,r,\cdots,r}_L) 
\in {\cal B}_L(\tilde{A}^{-1},D+{\rm tr}[B])
\\
&\Leftrightarrow &\pi(r)\leq D+{\rm tr}[B]
 \Leftrightarrow r\geq r^{\ast}(D+{\rm tr}[B]),
\\
& &
\tilde{\omega}(D,r^\ast)
=|\tilde{A}|^{-2}
\left|\Sigma_{X^L}^{-1}
 +\frac{1-{\rm e}^{-2r^{\ast}}}{\epsilon}I_L\right|^{-1}.
\eeqno
\item[b)]
The function $\tilde{\omega}(D,r)$ is a convex function of 
$r\in[r^*(D+{\rm tr}[B]),\infty)$.
\end{itemize}
\end{pr}

Proof of Property \ref{pr:prFz} part a) is easy. 
We omit the detail.   
Proof of Property \ref{pr:prFz} part b) will  
be given in Section V. 
Set 
\beqno
R_{{\rm sum},L}^{(\rm u)}(D|\Sigma_{Y^L})
&\defeq& \underline{\tilde{J}}(D,r^{\ast})
\\
&=& \ds\frac{1}{2}
\log\left[{|\Sigma_{Y^L}+B|}{\rm e}^{2Lr*}
\prod_{l=1}^L\beta_{l}(r)\right]
\vspace*{1mm}\\
&=&\ds\sum_{l=1}^L\frac{1}{2}
    \log\left\{ \frac{\mu_l}{\epsilon}\left[ {\rm e}^{2r*}-1\right]
    +1\right\}
\\
R_{{\rm sum},L}^{(\rm l)}(D|\Sigma_{Y^L})
&\defeq & \min_{r\geq r^*(D+{\rm tr}[B])}\underline{\tilde{J}}(D,r).
\eeqno
\newcommand{\dela}{
Set 
\beqno
R_{{\rm sum},L}^{(\rm l)}(D|\Sigma_{Y^L})\defeq 
\min_{r\geq r^*(D+{\rm tr}[B])}\underline{\tilde{J}}(D,r).
\eeqno
}
Then we have the following. 
\begin{Th}\label{th:sr0} Assume that the source 
$X^L$ and its noisy version $Y^L=X^L+N^L$ are 
cyclic shift invariant. Then, we have    
$$
R_{{\rm sum},L}^{(\rm l)}(D|\Sigma_{Y^L})
\leq R_{{\rm sum},L}(D|\Sigma_{Y^L})
\leq R_{{\rm sum},L}^{(\rm u)}(D|\Sigma_{Y^L}).
$$
\end{Th}

Proof of this theorem will be stated in Section V.
We next examine a necessary and sufficient condition for 
$R_{{\rm sum},L}^{(\rm l)}(D$ $|\Sigma_{Y^L})$ 
to coincide with $R_{{\rm sum},L}^{(\rm u)}($ $D|\Sigma_{Y^L})$. 
It is obvious that this condition is equivalent to the condition 
that the function 
$\underline{\tilde{J}}\left(D,r \right)$,
$r\geq r^{\ast}=$ $r^{\ast}(D+{\rm tr}[B])$, 
attains the minimum at $r=r^{\ast}$. Set
\beqno
\mu_{\min}\defeq \min_{1\leq l\leq L}\mu_{l},
\mu_{\max}\defeq \max_{1\leq l\leq L}\mu_{l}.
\eeqno
Let $l_0\in \Lambda_L$ be the largest integer such that 
$\mu_{\max}=\mu_{l_0}$ and let $l_1=l_1(r)\in \Lambda_L$ 
be the largest integer such that 
$$
\beta_{l_1}(r)=\max_{1\leq l\leq L}\beta_l(r).
$$
The following is a basic lemma to derive our 
necessary and sufficient matching condition on 
$R_{{\rm sum},L}^{(\rm l)}(D|\Sigma_{Y^L})$ 
$=R_{{\rm sum},L}^{(\rm u)}(D|\Sigma_{Y^L})$. 
\begin{lm}\label{lm:pro3bz}
The function $\underline{\tilde{J}}\left(D,r\right),$ 
$r\in [r^*(D+{\rm tr}[B]),\infty)$ 
attains the minimum at $r=r^{\ast}$ 
if and only if  
\beqa
& &\frac{1}{2}
   \left(\frac{{\rm d}}{{\rm d}r}
   \underline{\tilde{J}}\left(D,r \right)
   \right)_{r=r^{\ast}}
\nonumber\\
&=&\sum_{l=1}^L
\frac{
{\rm e}^{2r^{\ast}}
\left[{\rm e}^{2r^{\ast}}-1
+\frac{\epsilon}{\mu_l}\right]-
\left(1-\frac{\epsilon}{\mu_{l_1}}\right)
\left[{\rm e}^{2r^{\ast}}-1+
\frac{\epsilon}{\mu_{l_1}}\right]
}{
\left[{\rm e}^{2r^{\ast}}-1+
\frac{\epsilon}{\mu_{l_1}}
\right]^2
}
\nonumber\\
&\geq&0.
\label{eqn:zxx}
\eeqa
\end{lm}

Proof of Lemma \ref{lm:pro3bz} will  
be given in Section V. 
Note that for any $l\in {\Lambda_L}$, we have  
\beqa
&&
{\rm e}^{2r^{\ast}}
\left[{\rm e}^{2r^{\ast}}-1
+\frac{\epsilon}{\mu_l}\right]-
\left(1-\frac{\epsilon}{\mu_{l_1}}\right)
\left[{\rm e}^{2r^{\ast}}-1+
\frac{\epsilon}{\mu_{l_1}}\right]
\nonumber\\
&\geq&
{\rm e}^{2r^{\ast}}
\left[{\rm e}^{2r^{\ast}}-1
+\frac{\epsilon}{\mu_{l_0}}\right]-
\left(1-\frac{\epsilon}{\mu_{l_1}}\right)
\left[{\rm e}^{2r^{\ast}}-1+
\frac{\epsilon}{\mu_{l_1}}\right]
\nonumber\\
&\geq& \epsilon
\left(\frac{1}{\mu_{l_0}}-\frac{1}{\mu_{l_1}}\right).
\label{eqn:aaaza}
\eeqa
From (\ref{eqn:zxx}) in Lemma \ref{lm:pro3bz} 
and (\ref{eqn:aaaza}),  
we can see that $l_0=l_1$ is a sufficient 
matching condition for    
$R_{{\rm sum},L}^{(\rm l)}(D$ $|\Sigma_{Y^L})$ 
$=R_{{\rm sum},L}^{(\rm u)}($ $D|\Sigma_{Y^L})$.  
\begin{figure}[t]
\bc
\includegraphics[width=8.2cm]{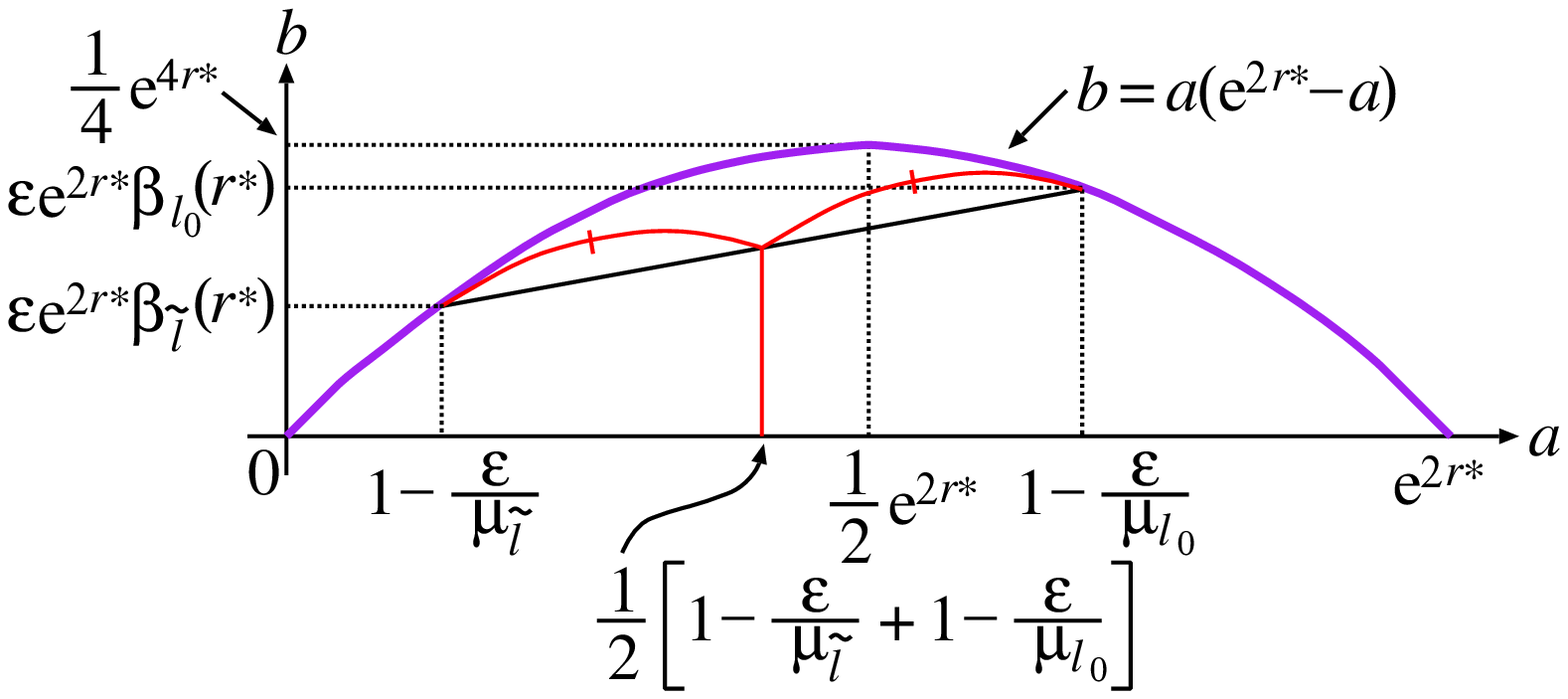}
\ec
\small
\caption{
The graph of $b=a({\rm e}^{2r^{\ast}}-a)$.
}
\label{fig:EigenDist}
\end{figure}
Let $\tilde{\mu}$ be the second largest eigenvalue of 
$\Sigma_{Y^L}$ and let $\tilde{l}\in\Lambda_L$ 
be the largest integer such that $\tilde{\mu}=\mu_{\tilde{l}}$. 
From the graph of 
$b=a({\rm e}^{2r^{\ast}}-a)$ shown in Fig. \ref{fig:EigenDist}, 
we can see that 
$$
\frac{1}{2}
\left[1-\frac{\epsilon}{\tilde{\mu}}
+1-\frac{\epsilon}{{\mu}_{\max}}\right] \leq 
\frac{1}{2}{\rm e}^{2r^{\ast}}
$$
or equivalent to 
\beq
{\rm e}^{2r^{\ast}}-1\geq
\left[1-\epsilon
\left(\frac{1}{\tilde{\mu}}+\frac{1}{ \mu_{\max} } \right)\right]
\label{eqn:mcond1a}
\eeq
is a necessary and sufficient condition for $l_0=l_1$.
Hence (\ref{eqn:mcond1a}) is 
a sufficient matching condition. 
Next, we derive another simple matching condition.
Note that 
\beqa
&&{\rm e}^{2r^{\ast}}
\left[{\rm e}^{2r^{\ast}}-1
+\frac{\epsilon}{\mu_l}\right]-
\left(1-\frac{\epsilon}{\mu_{l_1}}\right)
\left[{\rm e}^{2r^{\ast}}-1+
\frac{\epsilon}{\mu_{l_1}}\right]
\nonumber\\
&\geq& 
{\rm e}^{2r^{\ast}}\left[
{\rm e}^{2r^{\ast}}-1+\frac{\epsilon}{\mu_{\max}}
-\frac{1}{4}{\rm e}^{2r^{\ast}}
\right]
\nonumber\\
&=&
\frac{3}{4}{\rm e}^{2r^{\ast}}\left[
{\rm e}^{2r^{\ast}}-1-\frac{1}{3}
\left(1-\frac{4\epsilon}{\mu_{\max}}\right)
\right].
\nonumber
\eeqa
Hence, if we have
\beq
{\rm e}^{2r^{\ast}}-1\geq
\frac{1}{3}\left(1-\frac{4\epsilon}{\mu_{\max}}\right),
\label{eqn:mcond2}
\eeq
then the condition (\ref{eqn:zxx}) holds.
For $\epsilon\in(0, \mu_{\min})$, 
define 
\beqno
s(\epsilon)
&\defeq &
\frac{1}{2}\log\left\{1+
\min\left\{
\left[
1-\epsilon\left(\frac{1}{\tilde{\mu}}+\frac{1}{\mu_{\max}}\right)
\right]^{+}\right.\right.
,
\\
& &\qquad\qquad\qquad\qquad
\left.\left.
\frac{1}{3}\left[1-\frac{4\epsilon}{\mu_{\max}}\right]^{+}
\right\}
\right\}.
\eeqno
Then the condition (\ref{eqn:mcond1a}) or 
(\ref{eqn:mcond2}) is equivalent to 
$r^{\ast}\geq s(\epsilon)$. Furthermore, this condition 
is equivalent to $0\leq D\leq D_{\rm th}(\epsilon)$, 
where  
\beqno
D_{\rm th }(\epsilon)
&\defeq&
\sum_{l=1}^L
\frac{1}{\beta_l(s(\epsilon))}-{\rm tr}[B]
=\sum_{l=1}^L
\frac{\mu_l\epsilon}{{\mu_l}
\left[{\rm e}^{2s(\epsilon)}-1\right]+\epsilon}.
\eeqno
Summarizing the above argument we have the following.  
\begin{Th}
\label{th:matchTh2a}
We suppose that  $Y^L$ is cyclic shift invariant. 
Fix $\epsilon\in(0, \mu_{\min})$ arbitrary. 
If
$0\leq D \leq D_{\rm th }(\epsilon)$, then we have 
$$
{R}_{{\rm sum},L}^{\rm (l)}(D|\Sigma_{Y^L})
={R}_{{\rm sum},L}(D|\Sigma_{Y^L})
={R}_{{\rm sum},L}^{\rm (u)}(D|\Sigma_{Y^L}).
$$
Furthermore, the curve $R=R_{{\rm sum},L}(D|\Sigma_{Y^L})$ has 
the following parametric form:
\beq
\left.
\ba{rcl}
R
&=&\ds\sum_{l=1}^L\frac{1}{2}
    \log\left\{ \frac{\mu_l}{\epsilon}\left[ {\rm e}^{2r}-1\right]
    +1\right\},
\vspace*{2mm}\\
D&=&\ds\sum_{l=1}^L\frac{1}{\beta_l(r)}
-{\rm tr}[B]
= \ds \sum_{l=1}^L\ds
\frac{\mu_l\epsilon}{\mu_l({\rm e}^{2r}-1)+\epsilon}
\vspace*{1mm}\\
& &\mbox{for }r\in [s(\epsilon),\infty).
\ea
\right\}
\label{eqn:az}   
\eeq
\end{Th}

Since $D_{\rm th}(\epsilon)$ is a monotone increasing 
function of $\epsilon$, to choose $\epsilon$ 
arbitrary close to $\mu_{\min}$ is a choice yielding 
the best matching condition. Note here that 
we can not choose $\epsilon=\mu_{\min}$ because
$\pi(r)$ becomes infinity in this case. 
Letting $\epsilon$ arbitrary close to $\mu_{\min}$ and 
considering the continuities of $D_{\rm th}(\epsilon)$ and the functions 
in the right hand side of (\ref{eqn:az}) with respect to 
$\epsilon$, we have the following.
\begin{Th}\label{th:matchTh2b}
We suppose that  $Y^L$ is cyclic shift invariant. 
If
$0\leq D \leq D_{\rm th }(\mu_{\min})$, then we have 
$$
{R}_{{\rm sum},L}^{\rm (l)}(D|\Sigma_{Y^L})
={R}_{{\rm sum},L}(D|\Sigma_{Y^L})
={R}_{{\rm sum},L}^{\rm (u)}(D|\Sigma_{Y^L}).
$$
Furthermore, the curve $R=R_{{\rm sum},L}(D|\Sigma_{Y^L})$ has 
the following parametric form:
$$
\left.
\ba{rcl}
R
&=&\ds\sum_{l=1}^L\frac{1}{2}
    \log\left\{ \frac{\mu_l}{\mu_{\min}}\left[ {\rm e}^{2r}-1\right]
    +1\right\},
\vspace*{2mm}\\
D&=&
\ds \sum_{l=1}^L\ds
\frac{\mu_l\mu_{\min}}{{\mu_l}({\rm e}^{2r}-1)+\mu_{\min} },
\mbox{ for }r\in [s(\mu_{\min}),\infty).
\ea
\right\}
$$
\end{Th}


Let ${1}^L\defeq (1,1,\cdots,1)$ be a $L$ 
dimensional vector whose $L$ components
are all 1. We consider the characterization of 
$R_{{\rm sum},L}(D\cdot 1^L|\Sigma_{Y^L})$. 
From Theorem \ref{th:matchTh2b}, we obtain the following
corollary.
\begin{co}
Suppose that $Y^L$ is cyclic shift invariant.
If
$0\leq D \leq \frac{1}{L}D_{\rm th }(\mu_{\min})$, then we have 
\beqno
& &{R}_{{\rm sum},L}^{\rm (l)}(D\cdot 1^L|\Sigma_{Y^L})
\\
&=&{R}_{{\rm sum},L}(D\cdot 1^L|\Sigma_{Y^L})
={R}_{{\rm sum},L}^{\rm (u)}(D\cdot 1^L|\Sigma_{Y^L}).
\eeqno
Furthermore, the curve 
$R=R_{{\rm sum},L}(D\cdot 1^L|\Sigma_{Y^L})$ has 
the following parametric form:
$$
\left.
\ba{rcl}
R
&=&\ds\sum_{l=1}^L\frac{1}{2}
    \log\left\{ \frac{\mu_l}{\mu_{\min}}\left[ {\rm e}^{2r}-1\right]
    +1\right\},
\vspace*{2mm}\\
D&=&
\ds \frac{1}{L}\sum_{l=1}^L\ds
\frac{\mu_l\mu_{\min}}{{\mu_l}({\rm e}^{2r}-1)+\mu_{\min}},
\mbox{ for }r\in [s(\mu_{\min}),\infty).
\ea
\right\}
$$
\end{co}

Here we consider the case where $\Sigma_{Y^L}$ has at most two eigenvalues.
In this case we have $\tilde{\mu}=\mu_{\min}$. 
Then we have $s(\mu_{\min})=0$ and $D_{\rm th}(0)={\rm tr}[\Sigma_{Y^L}]$. 
This implies that 
$R=R_{{\rm sum},L}(D\cdot 1^L|\Sigma_{Y^L})$  is determined for all 
$0\leq D \leq $ $ \frac{1}{L}{\rm tr}[\Sigma_{Y^L}]$. 
Wagner {\it et al.} \cite{wg3} determined 
$R=R_{{\rm sum},L}(D\cdot 1^L|\Sigma_{Y^L})$ 
in a special case where $\Sigma_{Y^L}$ satisfies
$[\Sigma_{Y^L}]_{ll}=\sigma^2$ for $l\in \Lambda_L$ 
and $[\Sigma_{Y^L}]_{ll^{\prime}}=c\sigma^2,0<c<1$ for 
$l\ne l^{\prime}\in\Lambda_L$. In this special case $\Sigma_{Y^L}$ 
has two distinct eigenvaules. Hence our result includes their result 
as a special case. 


Yang and Xiong \cite{yxb} determined 
$R_{{\rm sum},L}(D\cdot 1^L|\Sigma_{Y^L})$ 
in the case where $\Sigma_{Y^L}$ has 
two distinct eigenvalues. 
Wang {\it et al.} \cite{wa} determined 
$R_{{\rm sum},L}(D\cdot 1^L|\Sigma_{Y^L})$ 
for another case of $\Sigma_{Y^L}$. 
The class of information sources satisfying 
the cyclic shift invariant property is different from the class 
of information sources investigated by Yang and Xiong \cite{yxb} and 
Wang {\it et al.} \cite{wa} although we have some overlap between them.



\section{Proofs of the Results}

\renewcommand{\irb}[1]{{\color[named]{Black}#1\normalcolor}}
\renewcommand{\irg}[1]{{\color[named]{Black}#1\normalcolor}}
\renewcommand{\irBr}[1]{{\color[named]{Black}#1\normalcolor}}
\renewcommand{\irBw}[1]{{\color[named]{Black}#1\normalcolor}}


\subsection{Derivation of the Outer Bounds
}

In this subsection we prove the results on outer bounds of 
the rate distortion region.    
We first state two important lemmas which are mathematical cores of 
the converse coding theorem. For $l\in\IsetA$, set 
\beq
\irb{W_l} \defeq \varphi_l(\irBw{\vc Y}_l), 
\irb{r_l^{(n)}}\defeq \frac{1}{n}I(\irBw{\vc Y}_l;\irb{W_l}|{\vc X}^K). 
\eeq
For ${Q}\in {\cal O}_K$, set $\irg{Z}^K \defeq QX^K$. For 
$$
{\vc X}^K=(X^K(1),X^K(2),\cdots, X^K(n))
$$ 
we set 
$$
 {\vc Z}^K \defeq Q{\vc X}^K
=(QX^K(1), QX^K(2), \cdots,QX^K(n)).
$$
Furthermore, for 
$\hat{\vc X}^K=(\hat{X}^K(1),$ 
           $\hat{X}^K(2),$ 
   $\cdots, \hat{X}^K(n))$,
we set
$$
 \hat{\vc Z}^K=Q\hat{\vc X}^K
\defeq (Q\hat{X}^K(1), 
  Q\hat{X}^K(2), \cdots, 
  Q\hat{X}^K(n)).
$$
We have the following two lemmas. 
\begin{lm}\label{lm:lm1}
For any $k\in\IsetB$ and any $Q\in{\cal O}_K$, we have 
\beqno
& &\left.h(\irg{\vc Z}_k \right| \irg{\vc Z}_{[k]}^K \irb{W}^L)
   \leq  h(\irg{{\vc Z}}_k -\hat{\irg{{\vc Z}}}_{k}
   \left.\right| \irg{{\vc Z}}_{[k]}^K -\hat{\irg{{\vc Z}}}_{[k]}^K)
\\
&\leq& 
\frac{n}{2}\log 
\left\{
(2\pi {\rm e})
\left[
Q\left({\ts \frac{1}{n}}
\Sigma_{{\lvc X}^K-\hat{\lvc X}^K}^{-1}\right)^{-1}{}^{\rm t}Q
\right]_{kk}^{-1}
\right\},
\eeqno
where $h(\cdot)$ stands for the differential entropy. 
\end{lm}
%

\begin{lm}\label{lm:lm2} 
For any $k\in\IsetB$ and any $Q\in{\cal O}_K$, we have
\beqno
& &
h(\irg{\vc Z}_k|\irg{\vc Z}_{[k]}^K\irb{W}^L)
\\
&\geq& 
\frac{n}{2}\log 
\left\{
(2\pi {\rm e})\left[
Q\left(\Sigma_{X^K}^{-1}
 +{}^{\rm t}A
\Sigma_{N_{\IsetA}(\irb{r_{\irBr{\IsetA}}^{(n)}})}^{-1}A\right){}^{\rm t}Q
\right]_{kk}^{-1}
\right\}.
\eeqno
\end{lm}

Proofs of Lemmas \ref{lm:lm1} and \ref{lm:lm2} will 
be stated in Appendixes A and B, respectively. 
The following lemma immediately follows 
from Lemmas \ref{lm:lm1} and \ref{lm:lm2}. 
\begin{lm}\label{lm:co0} For any $\Sigma_{X^KY^L}$ and for 
any 
$(\varphi_1^{(n)},$ 
$\varphi_2^{(n)}, \cdots,$ 
$\varphi_L^{(n)},$ $\psi^{(n)})$, we have 
$$
\left({\ts \frac{1}{n}}\Sigma_{{\lvc X}^K-\hat{\lvc X}^K}\right)^{-1}
\preceq
\Sigma_{X^K}^{-1}
 +{}^{\rm t}A\Sigma_{N_{\IsetA}(\irb{r_{\IsetA}^{(n)}})}^{-1}A.
$$
\end{lm}

From Lemma \ref{lm:lm2}, we obtain the following lemma.
\begin{lm}\label{lm:co1} For any $S\subseteq \IsetA $, we have  
\beq
I({\vc X}^K;W_S) \leq 
\frac{n}{2}
\log 
\left|I+\Sigma_{X^K}{}^{\rm t}A\Sigma_{N_{S}(r_S^{(n)})}^{-1}A\right|.
\eeq
\end{lm}

{\it Proof:} For each $l\in \IsetA-S$, we choose 
$W_l$ so that it takes a constant value. In this case
we have $r_l^{(n)}=0$ for $l\in \IsetA-S$. 
Then by Lemma \ref{lm:lm2}, 
for any $k\in\IsetB$, we have
\beqa
\hspace*{-6mm}& &
h(\irg{\vc Z}_k|\irg{\vc Z}_{[k]}^K\irb{W}_{\irBr{S}})
\nonumber\\
\hspace*{-6mm}&\geq& 
\frac{n}{2}\log 
\left\{
(2\pi {\rm e})\left[
Q\left(\Sigma_{X^K}^{-1}
 +{}^{\rm t}A
\Sigma_{N_{S}(\irb{r_{\irBr{S}}^{(n)}})}^{-1}A\right){}^{\rm t}Q
\right]_{kk}^{-1}
\right\}.
\label{eqn:conv999}
\eeqa
We choose an orthogonal matrix $Q$ $\in{\cal O}_K$ so that 
$$
Q\left(\Sigma_{X^K}^{-1}
+{}^{\rm t}A\Sigma_{N_S(r_S^{(n)})}^{-1}A\right){}^{\rm t}Q
$$ 
becomes the following diagonal matrix: 
\beq
Q\left(\Sigma_{X^K}^{-1}
+{}^{\rm t}A\Sigma_{N_S(r_S^{(n)})}^{-1}A\right){}^{\rm t}Q
=\left(
\begin{array}{cccc}
\lambda_1 &           &        & \mbox{\huge 0}\\
          & \lambda_2 &        &          \\
          &           & \ddots &          \\
\mbox{\huge 0} &      &        & \lambda_{K}\\
\end{array}
\right).
\label{eqn:diag}
\eeq
Then we have the following chain of inequalities:  
\beqa
\lefteqn{   
I({\vc X}^K;W_S)
=h({\vc X}^K)-h({\vc X}^K|W_S )}
\nonumber\\ 
&\MEq{a}& h({\vc X}^K) 
         -h({\vc Z}^K|W_S )
\leq  h({\vc X}^K) 
         -\sum_{k=1}^{K} 
         h({\vc Z}_k|{\vc Z}_{[k]}^KW_S)
\nonumber\\
&\MLeq{b} &\frac{n}{2} 
         \log \left[ (2\pi {\rm e})^{K} 
\left|\Sigma_{X^K}\right| \right]
\nonumber\\
&     &\hspace*{-3mm}
             +\sum_{k=1}^{K} \frac{n}{2} \log 
             \left\{
             \frac{1}{2\pi {\rm e}}\left[
             Q\hspace*{-1mm}\left(\Sigma_{X^K}^{-1}
             +{}^{\rm t}A\Sigma_{N_S(r_S^{(n)})}^{-1}A\right){}^{\rm t}Q
             \right]_{kk}
             \right\}
\nonumber\\
&\MEq{c}& \frac{n}{2} \log \left| \Sigma_{X^K} \right| 
         +\sum_{k=1}^{K}\frac{n}{2}\log \lambda_l 
\nonumber\\
&=&\frac{n}{2} \log \left|\Sigma_{X^K} \right| 
   +\frac{n}{2}\log \left|\Sigma_{X^K}^{-1}
   +{}^{\rm t}A\Sigma_{N_S(r_S^{(n)})}^{-1}A\right| 
\nonumber\\
&=& \frac{n}{2}\log \left|I+\Sigma_{X^K}
{}^{\rm t}A\Sigma_{N_{S}(r_S^{(n)})}^{-1}A\right|. 
\nonumber
\eeqa
Step (a) follows from the rotation invariant property 
of the (conditional) differential entropy. 
Step (b) follows from (\ref{eqn:conv999}).
Step (c) follows from (\ref{eqn:diag}).
\hfill \IEEEQED 

We first prove the inclusion 
$
{\cal R}_{L}(\DisT|$ $\Sigma_{X^KY^L})
\subseteq 
{\cal R}_{L}^{({\rm out})}(\DisT$ $|\Sigma_{X^KY^L})
$
stated in Theorem \ref{th:conv2}. 
Using Lemmas \ref{lm:lm1}, \ref{lm:lm2}, \ref{lm:co1} 
and a standard argument on the proof of converse coding theorems, 
we can prove the above inclusion. 

{\it Proof of 
${\cal R}_L(\DisT|\Sigma_{X^KY^L})
\subseteq {\cal R}_L^{ ({\rm out})}(\DisT|\Sigma_{X^KY^L})$:}
We first observe that 
\beq
 W_S\to {\vc Y}_S\to {\vc X}^K \to {\vc Y}_{\coS}\to W_{\coS}
\label{eqn:MkPr1}
\eeq
hold for any subset $S$ of $\IsetA$. 
Assume $(R_1, R_2,$ $\!\cdots, R_L) 
\in {\cal R}_{L}(\DisT|\Sigma_{X^KY^L})$. 
Then, there exists a sequence
$\{(\varphi_1^{(n)},\varphi_2^{(n)},$ 
$\cdots,\varphi_L^{(n)},\psi^{(n)}\}_{n=1}^{\infty}$
such that 
\beq
\left.
\ba{l}
\ds\limsup_{n\to\infty}R_l^{(n)}\leq R_l, l\in \IsetA,
\vspace{1mm}\\
\ds \limsup_{n\to\infty}
\frac{1}{n}\Sigma_{{\lvc X}^K-\hat{{\lvc X}}^K}
\preceq \DisT.
\ea
\right\}
\label{eqn:goddz1}
\eeq
We set
\beq
r_l\defeq \limsup_{n\to\infty}r_l^{(n)}
=\limsup_{n\to\infty}
\frac{1}{n}I({\vc Y}_l;W_S|{\vc X}^K).
\eeq
For any subset $S \subseteq \IsetA$, we have 
the following chain of inequalities:
\beqa
& &
\sum_{l\in S}nR_l^{(n)}
\geq  
\sum_{l\in S}\log M_l
\geq  \sum_{l\in S}H(W_l)
\geq H(W_S|W_{\coS})
\nonumber\\
&=&I({\vc X}^K;W_S|W_{\coS}) + H(W_S|W_{\coS}{\vc X}^K)
\nonumber\\
&{\stackrel{({\rm a})}{=}}&I({\vc X}^K;W_S|W_{\coS}) 
+\sum_{l\in S}H(W_l|{\vc X}^K) 
\nonumber\\
&{\stackrel{({\rm b})}{=}}&I({\vc X}^K;W_S|W_{\coS}) 
+\sum_{l\in S}H(W_l|{\vc X}^K) 
\nonumber\\
&{\stackrel{({\rm c})}{=}}&I({\vc X}^K;W_S|W_{\coS})
+n\sum_{l\in S}r_l^{(n)}, 
\label{eqn:conv1zf}
\eeqa
where steps (a),(b) and (c) follow from (\ref{eqn:MkPr1}). 
We estimate a lower bound of $I({\vc X}^K;W_S|W_{\coS})$. 
Observe that
\beqa
I({\vc X}^K;W_S|W_{S^{\rm c}})
& =& I({\vc X}^K;W^L)-I({\vc X}^K;W_{\coS}). 
\label{eqn:prthc0}
\eeqa
Since an upper bound of $I({\vc X}_{\coS};W_{\coS})$ 
is derived by Lemma \ref{lm:co1}, it suffices 
to estimate a lower bound of $I({\vc X}^K;$ $W^L)$.
We have the following chain of inequalities:
\beqa
& &
I({\vc X}^K;W^L)
=h({\vc X}^K)-h({\vc X}^K|W^L)
\nonumber\\
&\geq &h({\vc X}^K)-h({\vc X}^K|\hat{{\vc X}}^K)
\geq h({\vc X}^K)-h({\vc X}^K-\hat{{\vc X}}^K)
\nonumber\\
&\geq &\frac{n}{2}\log\left[(2\pi {\rm e})^K
 \left|\Sigma_{{X}^K}\right|\right] 
-\frac{n}{2}\log \left[(2\pi {\rm e})^K
\left| {\ts\frac{1}{n}}\Sigma_{{\lvc X}^K-\hat{{\lvc X}}^K}
\right|\right]
\nonumber\\
&=&\frac{n}{2}\log
\frac{\left|\Sigma_{{X}^K}\right|}
{\left| {\ts\frac{1}{n}}\Sigma_{{\lvc X}^K-\hat{{\lvc X}}^K}\right|}
.
\label{eqn:prthz120}
\eeqa
Combining (\ref{eqn:prthc0}), (\ref{eqn:prthz120}), and  
Lemma \ref{lm:co1}, we have 
\beqno
& & 
I({\vc X}^K;W_S|W_{S^{\rm c}}) +n\sum_{l\in S}r_l^{(n)}
\nonumber\\
&\geq & \frac{n}{2}\log 
\left[
\frac{\prod_{l\in S}{\baseN}^{2r_l^{(n)}} \left|\Sigma_{X^K}\right|}
{
\left|I+ \Sigma_{X^K}{}^{\rm t}A
\Sigma_{N_{S^{\rm c}}(r_{S^{\rm c}}^{(n)})}^{-1}A
\right|
\left|{\ts \frac{1}{n}}\Sigma_{{\lvc X}^K-\hat{{\lvc X}}^K}\right|
}
\right]
\\
&=& \frac{n}{2}\log 
\left[
\frac{\prod_{l\in S}{\baseN}^{2r_l^{(n)}} }
{\left|\Sigma_{X^K}^{-1} 
+ {}^{\rm t}A\Sigma_{N_{S^{\rm c}}(r_{S^{\rm c}}^{(n)}) }^{-1}A\right|
\left|{\ts \frac{1}{n}}\Sigma_{{\lvc X}^K-\hat{{\lvc X}}^K}\right|}
\right].
\eeqno
Note here that
$
I({\vc X}^K;W_S|W_{\coS})+n\sum_{i\in S}r_i^{(n)}
$
is nonnegative. Hence, we have 
\beqa
& &
I({\vc X}^K;W_S|W_{\coS})+n\sum_{i\in S}r_i^{(n)}
\nonumber\\
&\geq & n\underline{J}_{S}\left(
\left.
\left|
{\ts\frac{1}{n}}\Sigma_{{\lvc X}^K-\hat{{\lvc X}}^K}
\right|,r_S^{(n)}\right|r_{\coS}^{(n)}\right).
\label{eqn:prthdzq}
\eeqa
Combining (\ref{eqn:conv1zf}) and (\ref{eqn:prthdzq}), we obtain
\beqa
\sum_{l\in S}R_l^{(n)} 
&\geq & \underline{J}_{S}\left(
\left.
\left|
{\ts\frac{1}{n}}\Sigma_{{\lvc X}^K-\hat{{\lvc X}}^K}
\right|,r_S^{(n)}\right|r_{\coS}^{(n)}\right)
\label{eqn:prth99}
\eeqa
for $S\subseteq \IsetA$. On the other hand, 
by Lemma \ref{lm:co0}, we have 
\beqa
\Sigma_{X^K}^{-1}
+{}^{\rm t}A\irBr{\Sigma_{N_\IsetA(r_{\IsetA}^{(n)})}^{-1}}A
&\succeq&
{\ts \frac{1}{n}}\Sigma_{{\lvc X}^K-\hat{\lvc X}^K}^{-1}.
\label{eqn:prth101}
\eeqa
By letting $n\to\infty$ 
in (\ref{eqn:prth99}) and (\ref{eqn:prth101}) 
and taking (\ref{eqn:goddz1}) into account, 
we have for any $S\subseteq \IsetA$
\beqa
\sum_{l\in S}R_l 
&\geq &\underline{J}_{S}(\left|\DisT\right|,r_S|r_{\coS}),
\label{eqn:prth104}
\eeqa
and 
\beq
\Sigma_{X^K}^{-1}+{}^{\rm t}A\irBr{\Sigma_{N^L(r^L)}^{-1}}A
\succeq \DisT^{-1}.
\label{eqn:prth103} 
\eeq
From (\ref{eqn:prth104}) and (\ref{eqn:prth103}),
${\cal R}_{\Iset}(\DisT|\Sigma_{X^KY^L})
\subseteq {\cal R}_{\Iset}^{({\rm out})}(\DisT|$$\Sigma_{X^KY^L})$
is concluded. \hfill \IEEEQED

{\it Proof of Theorem \ref{th:conv2a}:} 
We choose an orthogonal matrix $Q$$\in {\cal O}_K$ so that 
\beqno
& &Q\Gamma^{-1}\left(\Sigma_{X^K}^{-1}
+{}^{\rm t}A\Sigma_{N^L(r^L)}^{-1}A\right){}^{\rm t}\Gamma^{-1} {}^{\rm t}Q
\\
&=&\left[
\begin{array}{cccc}
\alpha_1 &           &        & \mbox{\huge 0}\\
          & \alpha_2 &        &               \\
          &           & \ddots &              \\
\mbox{\huge 0} &      &        & \alpha_{K}   \\
\end{array}
\right].
\eeqno
Then we have
\beqa
&&Q\Gamma\left(\Sigma_{X^K}^{-1}
+{}^{\rm t}A\Sigma_{N^L(r^L)}^{-1}A\right)^{-1}{}^{\rm t}\Gamma {}^{\rm t}Q
\nonumber\\
&=&\left[
\begin{array}{cccc}
\alpha_1^{-1} &           &        & \mbox{\huge 0}\\
          & \alpha_2^{-1} &        &           \\
          &           & \ddots &               \\
\mbox{\huge 0} &      &        & \alpha^{-1}_{K}\\
\end{array}
\right].
\label{eqn:diagaa}
\eeqa
For $\DisT \in {\cal A}(r^L)$, set 
$$
\tilde{\Sigma}_d 
\defeq Q\Gamma\DisT {}^{\rm t}\Gamma {}^{\rm t}Q,\quad 
\xi_k\defeq \left[\tilde{\Sigma}_d\right]_{kk}.
$$ 
Since 
$$
\Gamma \DisT {}^{\rm t}\Gamma \succeq \Gamma(\Sigma_{X^L}^{-1}
+{}^{\rm t}A\Sigma_{N^L(r^L)}^{-1}A)^{-1}{}^{\rm t}\Gamma,
$$
$(\ref{eqn:diagaa})$, 
and ${\rm tr}[\Gamma\Sigma_d{}^{\rm t}\Gamma]\leq D$, 
we have
\beq
\left.
\ba{l}
\xi_k\geq \alpha_k^{-1},\mbox{ for }k\in\IsetB,
\vspace{1mm}\\
\ds \sum_{k=1}^K\xi_k={\rm tr}\left[\tilde{\Sigma}_d\right]
={\rm tr}\left[\Gamma\Sigma_d{}^{\rm t}\Gamma\right]\leq D.
\ea
\right\}
\label{eqn:aa2}
\eeq
Furthermore, by Hadamard's inequality we have 
\beqa
|\Sigma_d|=|\Gamma|^{-2}|\tilde{\Sigma}_d| \leq 
|\Gamma|^{-2}\prod_{k=1}^K[\tilde{\Sigma}_d]_{kk}
=|\Gamma|^{-2}\prod_{k=1}^K\xi_{k}. 
\label{eqn:aa3}
\eeqa
Combining (\ref{eqn:aa2}) and (\ref{eqn:aa3}), we obtain 
\beqno
& &\theta(\Gamma,D,r^L)
=\max_{\scs \DisT: \DisT \in {\cal A}_L({r^L}),
      \atop{\scs 
       {\rm tr}[\Gamma\DisT {}^{\rm t}\Gamma]\leq D}
      }
    \left|\DisT \right|
\\
&\leq& |\Gamma|^{-2}\max_{\scs \xi_k\alpha_{k}\geq 1,k\in\IsetB, 
          \atop{\scs
          \sum_{k=1}^K\xi_k\leq D
          }   
      }\prod_{k=1}^K\xi_{k} 
={\omega}(\Gamma,D,r^L).
\eeqno
The equality holds when 
$\tilde{\Sigma}_d$ is a diagonal matrix.
\hfill\IEEEQED

{\it Proof of Theorem \ref{th:sr0}:}
Assume that
$(R_1, R_2,$ $\!\cdots, R_L) \in {\cal R}_{L}(D|\Sigma_{Y^L})$. 
Then, there exists a sequence 
$\{(\varphi_1^{(n)},\varphi_2^{(n)},$ 
$\cdots,\varphi_L^{(n)},\phi^{(n)}\}_{n=1}^{\infty}$
such that 
\beq
\left.
\ba{l}
\ds\limsup_{n\to\infty}R_l^{(n)}\leq R_l, l\in \IsetA
\vspace{1mm}\\
\ds \limsup_{n\to\infty}
\frac{1}{n}\Sigma_{{\lvc Y}_{\IsetA}-\hat{{\lvc Y}}_{\IsetA}}
\preceq \DisT,\:
{\rm tr }[\DisT]\leq D
\vspace{1mm}\\
\mbox{ for some }\DisT.
\ea
\right\}
\label{eqn:godd}
\eeq
For each $j=0,1,\cdots,L-1$, we use 
$(\varphi_{\tau^j{(1)}}^{(n)}, \varphi_{\tau^j(2)}^{(n)}, \cdots,$ 
 $\varphi_{\tau^j(L)}^{(n)})$ 
for the encoding of $({\vc Y}_1, {\vc Y}_2, \cdots, {\vc Y}_L)$. 
For $l\in {\IsetA}$ and for $j=0,1,\cdots,L-1$, set 
\beqno
W_{j,l}&\defeq & \varphi_{\tau^j(l)}^{(n)}({\vc Y}_l), 
r_{j,l}^{(n)}\defeq \frac{1}{n}I({\vc Y}_l;W_{j,l}|{\vc X}^L).
\eeqno
In particular,
$$
r_{0,l}^{(n)}=r_l^{(n)}=\frac{1}{n}I({\vc Y}_l;W_{l}|{\vc X}_i), 
\quad\mbox{for }l\in\IsetA. 
$$
Furthermore, set 
\beqno
r_{\tau^j(\IsetA)}^{(n)}
&\defeq&(r^{(n)}_{j,1},r_{j,2}^{(n)},\cdots,r_{j,L}^{(n)}),
\mbox{ for }j=0,1,\cdots,L-1, 
\\
r^{(n)}&\defeq& \frac{1}{L}\sum_{l=1}^{L}r_l^{(n)}.
\eeqno 
By the cyclic shift invariant property of 
${\vc X}_{\IsetA}$ and ${\vc Y}_{\IsetA}$, we have
for $j=0,1,\cdots,L-1$,
\beq
 \frac{1}{L}\sum_{l=1}^Lr^{(n)}_{j,l}
=\frac{1}{L}\sum_{l=1}^Lr^{(n)}_{0,i}=r^{(n)}.
\label{eqn:godd0}
\eeq
For $j=0,1,\cdots,L-1$ and for $l\in \IsetA$, set 
\beqno
&&\hat{\vc Y}_{j,l} \defeq
\phi_{\tau^j(l)}(
\varphi_{\tau^j(1)}({\vc Y}_1),
\varphi_{\tau^j(2)}({\vc Y}_2), \cdots,
\varphi_{\tau^j(L)}({\vc Y}_L)), 
\\
& &\hat{\vc Y}_{\tau^j(\IsetA)} 
\defeq \left[
\ba{l}
\hat{\vc Y}_{j,1}\\ 
\hat{\vc Y}_{j,2}\\
\vdots \\
\hat{\vc Y}_{j,L}
\ea 
\right]. 
\eeqno
By the cyclic shift invariant property of ${\vc Y}_{\IsetA}$, 
we have 
\beqa
\lefteqn{{\rm E} \langle {\vc Y}_l-\hat{\vc Y}_{j,l},
        {\vc Y}_{l^{\prime}}-\hat{\vc Y}_{j,l^{\prime}}\rangle}
\nonumber\\
&=&{\rm E} \langle {\vc Y}_{\tau(l)}-\hat{\vc Y}_{j,l},
        {\vc Y}_{\tau(l^{\prime})}-\hat{\vc Y}_{j,l^{\prime}}\rangle
\label{eqn:aaaz}
\eeqa
for $(l,l^{\prime}) \in \Lambda_L^2$ and for 
$j=0,1,\cdots,L-1.$ For $\DisT=[d_{ll^{\prime}}]$, set 
\beqno
& &\tau^j(\DisT)\defeq [d_{\tau^j(l)\tau^j(l^{\prime})}],\:
\overline{\DisT}\defeq
\frac{1}{L}\sum_{j=0}^{L-1}\tau^j(\DisT).
\eeqno
Then, we have 
\beqa
& &\limsup_{n\to\infty}
\frac{1}{L}\sum_{j=0}^{L-1}
{\ts \frac{1}{n}}\Sigma_{{\lvc Y}_{\IsetA}-\hat{{\lvc Y}}_{\tau^j(\IsetA)}}
\nonumber\\
&\MEq{a}&\limsup_{n\to\infty}
\frac{1}{L}\sum_{j=0}^{L-1}
{\ts \frac{1}{n}}\Sigma_{{\lvc Y}_{\tau^j(\IsetA)}-\hat{{\lvc Y}}_{\tau^j(\IsetA)}}
\nonumber\\
&\MPreq{b}&
\frac{1}{L}\sum_{j=0}^{L-1}\tau^j(\DisT)\MEq{c}\overline{\DisT}.
\label{eqn:godd2}
\eeqa
Step (a) follows from (\ref{eqn:aaaz}). 
Step (b) follows from (\ref{eqn:godd}). 
Step (c) follows from the definition of $\overline{\DisT}$.
From ${\vc Y}_{\IsetA}$, we construct an 
estimation $\hat{\vc X}_{\IsetA}$ of 
${\vc X}_{\IsetA}$ by 
$\hat{\vc X}_{\IsetA}=\tilde{A}\hat{\vc Y}_{\IsetA}.$
Then for $j=0,1,\cdots,L-1$, we have the following:
\beqa
& &\Sigma_{X_{\IsetA}}^{-1}
    +\Sigma_{N_{\tau^j(\IsetA)}(r_{\tau^j(\IsetA)}^{(n)})}^{-1}
\nonumber\\
&\MEq{a}&\Sigma_{X_{\tau^j(\IsetA)}}^{-1}
    +\Sigma_{N_{\tau^j(\IsetA)}(r_{\tau^j(\IsetA)}^{(n)})}^{-1}
\nonumber\\
&\MSueq{b}&
{\ts \frac{1}{n}}
\Sigma_{{\lvc X}_{\tau^j(\IsetA)}-\hat{\lvc X}_{\tau^j(\IsetA)}}^{-1}
 \MEq{c}{\ts \frac{1}{n}}
\Sigma_{{\lvc X}_{\IsetA}-\hat{\lvc X}_{\tau^j(\IsetA)}}^{-1}
\nonumber\\
&=&\left[\tilde{A}
\left(
{\ts \frac{1}{n}}\Sigma_{{\lvc Y}_{\IsetA}-\hat{\lvc Y}_{\tau^j(\IsetA)}}
\right)
{}^{\rm t}\tilde{A}+\Sigma_{X_{\IsetA}|Y_{\IsetA}}
\right]^{-1}.
\label{eqn:zaaa0}
\eeqa
Steps (a) and (c) follow from the cyclic shift invariant 
property of ${X}_{\IsetA}$ and ${\vc X}_{\IsetA}$, 
respectively. 
Step (b) follows from Lemma \ref{lm:co0}. From (\ref{eqn:zaaa0}), we have
\beqa
& &\frac{1}{L}\sum_{j=0}^{L-1}
\left[\Sigma_{X_{\IsetA}}^{-1}
     +\Sigma_{N_{\tau^j(\IsetA)}(r_{\tau^m(\IsetA)}^{(n)})}^{-1}\right]
\nonumber\\
&\succeq&
\frac{1}{L}\sum_{j=0}^{L-1}
\left[\tilde{A}
\left(
{\ts \frac{1}{n}}\Sigma_{{\lvc Y}_{\IsetA}-\hat{\lvc Y}_{\tau^j(\IsetA)}}
\right)
{}^{\rm t}\tilde{A}+\Sigma_{X_{\IsetA}|Y_{\IsetA}}
\right]^{-1}
\nonumber\\
&\MSueq{a}&
\left[\tilde{A}
\left(
\frac{1}{L}\sum_{j=0}^{L-1}
{\ts \frac{1}{n}}\Sigma_{{\lvc Y}_{\IsetA}-\hat{\lvc Y}_{\tau^j(\IsetA)}}
\right)
{}^{\rm t}\tilde{A}+\Sigma_{X_{\IsetA}|Y_{\IsetA}}
\right]^{-1}
\nonumber\\
&=&
\left[\tilde{A}
\left(
\frac{1}{L}\sum_{j=0}^{L-1}
{\ts \frac{1}{n}}\Sigma_{{\lvc Y}_{\IsetA}-\hat{\lvc Y}_{\tau^j(\IsetA)}}
+B\right)
{}^{\rm t}\tilde{A}
\right]^{-1}.
\label{eqn:zaaa1}
\eeqa
Step (a) follows form that
$
(\tilde{A}\Sigma{}^{\rm t}\tilde{A}
+\Sigma_{X_{\IsetA}|Y_{\IsetA}})^{-1}
$
is convex with respect to $\Sigma$. On the 
other hand, we have 
\beqa
& &\frac{1}{L}\sum_{j=0}^{L-1}
   \left[\Sigma_{X_{\IsetA}}^{-1}
         +\Sigma_{N_{\tau^j(\IsetA)} (r_{\tau^j(\IsetA)}^{(n)}) }^{-1}
   \right]
\nonumber\\
&=&\Sigma_{X_{\IsetA}}^{-1}
    +\left(\frac{1}{L}\sum_{l=1}^{L}
    \frac{1-{\rm e}^{-2r_l^{(n)}}}{\epsilon}\right)I_L
\nonumber\\   
&\MPreq{a}&
 \Sigma_{X_{\IsetA}}^{-1}
    +\left(
    \frac{1-{\rm e}^{-2\frac{1}{L}\sum_{l=1}^{L}r_l^{(n)}}}
    {\epsilon}
    \right)I_L
\nonumber\\
&=&\Sigma_{X_{\IsetA}}^{-1}
     +\left(
     \frac{1-{\rm e}^{-2r^{(n)}}}
    {\epsilon}
    \right)I_L.
\label{eqn:zaaa2}
\eeqa
Step (a) follows from that $1-{\rm e}^{-2a}$ is a concave
function of $a$. Combining (\ref{eqn:zaaa1}) 
and (\ref{eqn:zaaa2}), we obtain
\beqno
& &\Sigma_{X_{\IsetA}}^{-1}
    +\left(
    \frac{1-{\rm e}^{-2r^{(n)}}}{\epsilon}
    \right)I_L
\\
&\succeq &
\left[
\tilde{A}
\left(
\frac{1}{L}\sum_{j=0}^{L-1}
{\ts \frac{1}{n}}\Sigma_{{\lvc Y}_{\IsetA}-\hat{\lvc Y}_{\tau^j(\IsetA)}}
+B\right)
{}^{\rm t}\tilde{A}
\right]^{-1},
\eeqno
from which we obtain
\beqa
& &\frac{1}{L}\sum_{j=0}^{L-1}
{\ts \frac{1}{n}}\Sigma_{{\lvc Y}_{\IsetA}-\hat{\lvc Y}_{\tau^j(\IsetA)}}+B
\nonumber\\
&\succeq &
\left[
{}^{\rm t}\tilde{A}\left\{
\Sigma_{X_{\IsetA}}^{-1}
    +\left(
    \frac{1-{\rm e}^{-2r^{(n)}}}{\epsilon}
     \right)I_L
\right\}\tilde{A}
\right]^{-1}.
\label{eqn:coverse1001}
\eeqa
Next we derive a lower bound of the sum rate part. 
For each $j=0,1,\cdots,L-1$, we have the following 
chain of inequalities:
\beqa
& &
\sum_{l\in \IsetA}nR_l^{(n)}\geq \sum_{l\in \IsetA}\log M_l 
\geq \sum_{l\in \IsetA}H(W_{j,l})
\nonumber\\
&\geq&H(W_{\tau^j(\IsetA)})=I({\vc X}_\IsetA;W_{\tau^j(\IsetA)}) 
   +H(W_{\tau^l(\IsetA)}|{\vc X}_{\IsetA}) 
\nonumber\\
&{\stackrel{({\rm a})}{=}}&I({\vc X}_\IsetA;W_{\tau^j(\IsetA)}) 
+\sum_{l\in \IsetA}H(W_{j,l}|{\vc X}_{\IsetA}) 
\nonumber\\
&=&I({\vc X}_\IsetA;W_{\tau^j(\IsetA)}) 
+\sum_{l\in \IsetA}I({\vc Y}_{\IsetA};W_{j,l}|{\vc X}_{\IsetA}) 
\nonumber\\
&\MEq{b}& I({\vc X}_\IsetA;W_{\tau^j(\IsetA)})+nLr^{(n)}
\nonumber\\ 
&\MGeq{c}&
\frac{n}{2}\log\left[\frac{\left|\Sigma_{{X}_{\IsetA}}\right|}
{\left|{\ts\frac{1}{n}}
\Sigma_{{\lvc X}_{\IsetA}-\hat{{\lvc X}}_{\tau^j(\IsetA)}}\right|}\right]
+nLr^{(n)}  
\nonumber\\
&=&\frac{n}{2}\log
     \frac{\left|\tilde{A}\Sigma_{{Y}_{\IsetA}}{}^{\rm t}\tilde{A}
      +\Sigma_{X_{\IsetA}|Y_{\IsetA}}\right|}
      {\left|
      \tilde{A}
      \left(
      {\ts\frac{1}{n}}\Sigma_{{\lvc Y}_{\IsetA}-\hat{{\lvc Y}}_{\tau^j(\IsetA)}}
      \right)
      {}^{\rm t}\tilde{A}+\Sigma_{X_{\IsetA}|Y_{\IsetA}}
      \right|}
+nLr^{(n)} 
\nonumber\\
&=&\frac{n}{2}\log
     \frac{\left|\Sigma_{{Y}_{\IsetA}}+B\right|}
      {\left|
      {\ts\frac{1}{n}}\Sigma_{{\lvc Y}_{\IsetA}-\hat{{\lvc Y}}_{\tau^j(\IsetA)}}+B
      \right|}
+nLr^{(n)}. 
\label{eqn:prthz0zz}
\eeqa
Step (a) follows from (\ref{eqn:MkPr1}). 
Step (b) follows from (\ref{eqn:godd2}). 
Step (c) follows from (\ref{eqn:prthz120}). 
From (\ref{eqn:prthz0zz}), we have
\beqa
& &\sum_{l\in \IsetA}R_l^{(n)}
=\frac{1}{L}\sum_{j=0}^{L-1}\sum_{l\in \IsetA}R_l^{(n)}
\nonumber\\
&\geq&
{\ds \frac{1}{L}\sum_{j=0}^{L-1}}
\frac{1}{2}
      \log 
      \frac{\left|\Sigma_{{Y}_{\IsetA}}+B\right|}
      {\left|
      {\ts\frac{1}{n}}\Sigma_{{\lvc Y}_{\IsetA}-\hat{{\lvc Y}}_{\tau^j(\IsetA)}}+B
      \right|}
+Lr^{(n)}
\nonumber\\
&\MGeq{a}&\frac{1}{2}
\log 
      \frac{\left|\Sigma_{{Y}_{\IsetA}}+B\right|}
      {\left|{\ds \frac{1}{L}\sum_{j=0}^{L-1}}
      {\ts\frac{1}{n}}\Sigma_{{\lvc Y}_{\IsetA}-\hat{{\lvc Y}}_{\tau^j(\IsetA)}}+B
      \right|}
+Lr^{(n)}. 
\label{eqn:prthz20}
\eeqa
Step (a) follows from that $-\log|\Sigma+B|$ is convex 
with respect to $\Sigma$. 
Letting $n\to\infty$ in (\ref{eqn:coverse1001}) 
and (\ref{eqn:prthz20}) and taking (\ref{eqn:godd2}) 
into account, we have
\beqa
\hspace*{-2mm}\ds\sum_{l\in \IsetA}R_l
&\geq&
\ds \frac{1}{2}
      \log 
      \frac{\left|\Sigma_{{Y}_{\IsetA}}+B\right|}
      {\left|
      \overline{\DisT}+B
      \right|}
+Lr,
\label{eqn:prthz17}
\\
\hspace*{-2mm}\overline{\DisT}+B
&\succeq&
\ds \left[
  {}^{\rm t}\tilde{A}\left\{
  \Sigma_{X_{\IsetA}}^{-1}
    +\left(
    \frac{1-{\rm e}^{-2r}}{\epsilon}
     \right)I_L\right\}\tilde{A}\right]^{-1},
\label{eqn:prthz18}
\\
\hspace*{-2mm}
{\rm tr}[\overline{\DisT}+B]&=&{\rm tr}[\DisT]+{\rm tr}[B]
\leq D+{\rm tr}[B].
\label{eqn:prthz21}
\eeqa
Now we choose an orthogonal matrix $Q\in {\cal O}_L$ 
so that 
\beqno
Q
{}^{\rm t}\tilde{A}\left\{\Sigma_{X_{\IsetA}}^{-1}
+\left(\frac{1-{\rm e}^{-2r}}{\epsilon}\right)I_L\right\}
\tilde{A}{}^{\rm t}Q
&=&\left[
\begin{array}{cccc}
\beta_1 &           &        & \mbox{\huge 0}\\
          & \beta_2 &        &               \\
          &           & \ddots &             \\
\mbox{\huge 0} &      &        & \beta_{L}   \\
\end{array}
\right].
\eeqno
Set 
\beqno
\hat{\Sigma}_d &\defeq&Q\DisT {}^{\rm t}Q,
\hat{B}_d \defeq QB{}^{\rm t}Q,
\xi_l\defeq \left[\hat{\Sigma}_d+\hat{B}\right]_{ll}.
\eeqno 
From (\ref{eqn:prthz18}) and (\ref{eqn:prthz21}), 
we have
\beq
\left.
\ba{l}
\xi_l\geq \beta_l^{-1}(r), l\in\IsetA,
\vspace{1mm}\\
\ds \sum_{l=1}^L\xi_l
={\rm tr}\left[\hat{\Sigma}_d+\hat{B}\right]
={\rm tr}\left[\Sigma_d+B\right]\leq D+{\rm tr}[B].
\ea
\right\}
\label{eqn:aa2z}
\eeq
From (\ref{eqn:aa2z}), we have
\beqa
\sum_{l=1}^L\frac{1}{\beta_l(r)}
    &\leq&\sum_{l=1}^L\xi_l
    ={\rm tr}[\hat{\Sigma_d}+\hat{B}]
\leq D+{\rm tr}[B]
\nonumber\\
&\Leftrightarrow& r\geq r^*(D+{\rm tr}[B]).
\label{eqn:aa03zz}
\eeqa
Furthermore, by Hadamard's inequality we have 
\beqa
|\Sigma_d+B|=|\hat{\Sigma}_d +\hat{B}|
\leq \prod_{l=1}^L[\hat{\Sigma}_d +\hat{B}]_{ll}
=\prod_{l=1}^L\xi_{l}. 
\label{eqn:aa3z}
\eeqa
Combining (\ref{eqn:aa2z}) and (\ref{eqn:aa3z}), we obtain 
\beq
|\Sigma_d+B|
\leq\max_{\scs \xi_l\beta_{l}\geq 1,l\in{\IsetA}, 
          \atop{\scs
          \sum_{l=1}^L\xi_l\leq D+{\rm tr}[B]
          }   
      }\prod_{l=1}^L\xi_{l} 
=\tilde{\omega}(D,r).
\label{eqn:prthz22}
\eeq
Hence, from (\ref{eqn:prthz17}), (\ref{eqn:aa03zz}), 
and (\ref{eqn:prthz22}), we have 
\beqno
\sum_{l=1}^L R_l
&\geq& 
\min_{r\geq r^*(D+{\rm tr}[B])}
\frac{1}{2}
\log
\left[
\frac{{\rm e}^{Lr}|\Sigma_Y+B|}{\tilde{\omega}(D,r)}
\right]
\\
&=&\min_{r\geq r^*(D+{\rm tr}[B])}\underline{\tilde{J}}(D,r)
=R_{{\rm sum},L}(D|\Sigma_{Y^L}),
\eeqno
completing the proof. 
\hfill \IEEEQED

\subsection{Derivation of the Inner Bound}

In this subsection we prove 
${\cal R}_L^{({\rm in})}(\DisT$ $|\Sigma_{X^KY^L})$
$\subseteq$ ${\cal R}_L(\DisT$ $|\Sigma_{X^KY^L})$ 
stated in Theorem \ref{th:conv2}. 

{\it Proof of ${\cal R}_L^{({\rm in})}(\DisT|\Sigma_{X^KY^L})$
$\subseteq$ ${\cal R}_L(\DisT|\Sigma_{X^KY^L})$:} 
Since $\hat{\cal R}_L^{({\rm in})}($ $\DisT|\Sigma_{X^KY^L})$ $\subseteq$ 
${\cal R}_L(\DisT|\Sigma_{X^KY^L})$ is proved by Theorem \ref{th:direct}, 
it suffices to show 
${\cal R}_L^{({\rm in})}(\DisT|\Sigma_{X^KY^L})$
$=$ 
$\hat{\cal R}_L^{({\rm in})}(\DisT|\Sigma_{X^KY^L})$
to prove ${\cal R}_L^{({\rm in})}(\DisT|\Sigma_{X^KY^L})$
$\subseteq$ ${\cal R}_L(\DisT|\Sigma_{X^KY^L})$. 
We assume that 
$R^L\in {\cal R}_L^{({\rm in})}$ $(\DisT|\Sigma_{X^KY^L})$. 
Then, there exists nonnegative vector $r^L$ such that
\beqno
& &\left(\Sigma_{X^K}^{-1}
  +{}^{\rm t}{A}\Sigma_{N^L(r^L)}^{-1}A\right)^{-1}\preceq \DisT
\eeqno
and 
\beq
\sum_{i\in S} R_l\geq J_S(r_S|r_{\coS})
\mbox{ for any }S \subseteq \IsetA. 
\label{eqn:zsa1}
\eeq
Let $V_l, l\in \IsetA$ be $L$ independent zero mean Gaussian 
random variables with variance $\sigma_{V_l}^2$.
Define Gaussian random variables $U_i, l\in \IsetA$ by 
$
U_l=X_l+N_l+V_l. 
$
By definition it is obvious that
\beq
\left.
\ba{l}
U^L\to Y^L \to X^K \\
U_S\to Y_S \to X^K \to Y_{\coS}\to U_{\coS}\\
\mbox{ for any } S\subseteq \IsetA.  
\ea
\right\}
\label{eqn:gau00sz} 
\eeq
For given $r_l \geq 0, l\in \IsetA$, choose 
$\sigma_{V_l}^2$ so that 
$\sigma_{V_l}^2=\sigma_{N_l}^2/({\baseN}^{2r_l}-1)$
when $r_l>0$. When $r_l=0$, we choose $U_l$ so that $U_l$
takes constant value zero. In the above choice the covariance 
matrix of $N^L+V^L$ becomes 
$\Sigma_{N^L(r^L)}$. Define 
the linear function ${\psi}$ of $U^L$ by 
$$
{\psi}\left(U^L\right) = 
(\Sigma_{X^K}^{-1} +{}^{\rm t}A\Sigma_{N^L(r^L)}^{-1}A)^{-1}
{}^{\rm t}A\Sigma_{N^L(r^L)}^{-1}U^L
.
$$
Set $\hat{X}^L={\psi}\left(U^L\right)$ and 
\beqa 
d_{kk}
& \defeq & {\rm E}\left[||{X}_k-\hat{X}_k||^2\right],
1 \leq k \leq K,
\nonumber\\
d_{kk^{\prime}}
& \defeq &
{\rm E}\left[
       \left({X}_k-\hat{X}_k\right)
       \left({X}_{k^{\prime}}-\hat{X}_{k^{\prime}}\right)
       \right],
1 \leq k\ne k^{\prime}\leq K.
\nonumber
\eeqa
Let $\Sigma_{{X}^K-\hat{X}^K}$ be a covariance matrix 
with $d_{kk^{\prime}}$ in its $(k,k^{\prime})$ element. By simple 
computations we can show that
\beq
\Sigma_{X^K-\hat{X}^K}
=(\Sigma_{X^K}^{-1} +{}^{\rm t}A\Sigma_{N^L(r^L)}^{-1}A)^{-1}\preceq \DisT
\label{eqn:gau2zz} 
\eeq
and that for any $S\subseteq \IsetA$, 
\beqa
& &J_S(r_S|r_{\coS})=I(Y_S;U_S|U_{\coS}).
\label{eqn:gau1z} 
\eeqa
From (\ref{eqn:gau00sz}) and (\ref{eqn:gau2zz}), we have 
$U^L\in {\cal G}(\DisT)$. Thus, from (\ref{eqn:gau1z}) 
$
{\cal R}_L^{({\rm in})}
(\DisT|\Sigma_{X^KY^L})\subseteq 
\hat{\cal R}_L^{({\rm in})}(\DisT|\Sigma_{X^KY^L})
$
is concluded. \hfill \IEEEQED 

\subsection{Proofs of the Results on Matching Conditions}

We first observe that the condition
$$
{\rm tr}\left[
\Gamma
\left(\Sigma_{X^K}^{-1}+{}^{\rm t}
A\Sigma_{N^L(r^L)}^{-1}A\right)^{-1}{}^{\rm t}\Gamma
\right]\leq D
$$
is equivalent to 
\beq
\sum_{k=1}^K\frac{1}{\alpha_{k}(r^L)}\leq D.
\label{eqn:aa00z}
\eeq
 
{\it Proof of Lemma \ref{lm:pro3}:}
Let $\IsetB=\{1,2,\cdots,K\}$ and let $S \subseteq \IsetB$ 
be a set of integers that satisfies $\alpha_i^{-1}\geq \xi$ 
in the definition of $\theta(\Gamma,D,r^L)$. 
Then, $\theta(\Gamma,D,r^L)$ is computed as 
\beqno
& &\theta(\Gamma,D,r^L)
\\
&=&{\ts \frac{1}{(K-|S|)^{K-|S|}}}
   \left(\prod_{j\in S}\frac{1}{\alpha_j}\right)
   \left(D-\sum_{k\in S} \frac{1}{\alpha_k}\right)^{K-|S|}.
\eeqno
Fix $l\in\IsetA$ arbitrarily and set 
$
\Psi_l\defeq 2r_l-$ $\log \theta(\Gamma,D,r^L).
$
Computing the partial derivative of $\Psi_l$ by $r_l$, we obtain
\beqa
\lefteqn{
\frac{\partial \Psi_l}{\partial r_l}
=\sum_{j\in S}
\left(
\frac{\partial \alpha_j}{\partial r_l}
\right)
\left[\frac{1}{\alpha_j}
-\frac{K-|S|}
{D-{\ds \sum_{k\in S}}\frac{1}{\alpha_k}}\frac{1}{\alpha_j^2}
\right]
+2}
\nonumber\\
&\MGeq{a}&\sum_{j\in S}
\left(
\frac{\partial \alpha_j}{\partial r_l}
\right)
\left[\frac{1}{\alpha_j}
-\frac{K-|S|}
{{\ds \sum_{k\in \IsetB-S}}\frac{1}{\alpha_k}}\frac{1}{\alpha_j^2}
\right]+2
\nonumber\\
&\geq&
\sum_{j\in S}
\left(
\frac{\partial \alpha_j}{\partial r_l}
\right)
\left[\frac{1}{\alpha_j}
-
\frac{\alpha_{\max}}{\alpha_j^2}
\right]+2
\nonumber\\
&\MEq{b}&
\sum_{j\in S}
\left(
\frac{\partial \alpha_j}{\partial r_l}
\right)
\left[\frac{\alpha_j-\alpha_{\max}}{\alpha_j^2}
\right]+
\frac{\sigma_{N_l}^2{\baseN}^{2r_l}}{||\hat{\vc a}_l||^2}
\sum_{j=1}^L
\left(
\frac{\partial \alpha_j}{\partial r_l}
\right)
\nonumber\\
&\geq&
\sum_{j\in S}
\left(
\frac{\partial \alpha_j}{\partial r_l}
\right)
\left[
\frac{\sigma_{N_l}^2{\baseN}^{2r_l}}{||\hat{\vc a}_l||^2}
-\frac{\alpha_{\max}}{\alpha_j}
\left(\frac{1}{\alpha_{\max}}-\frac{1}{\alpha_j}\right)
\right]
\nonumber\\
&\geq&
\left[
\frac{\sigma_{N_l}^2{\baseN}^{2r_l}}{||\hat{\vc a}_l||^2}
-\frac{\alpha_{\max}}{\alpha_{\min}}
\left(\frac{1}{\alpha_{\max}}-\frac{1}{\alpha_{\min}}\right)
\right]
\sum_{j\in S}
\left(
\frac{\partial \alpha_j}{\partial r_l}
\right).\quad
\label{zsz0}
\eeqa
Step (a) follows from the following inequality 
which is equivalent to (\ref{eqn:aa00z}):
\beqno
D-\sum_{k\in S}\frac{1}{\alpha_{k}(r^L)}
\geq 
\sum_{k\in \Lambda_K-S}\frac{1}{\alpha_{k}(r^L)}.
\label{eqn:aa00zpp}
\eeqno
Step (b) follows from Lemma \ref{lm:Egn1}.  
Hence, by (\ref{zsz0}) and Lemma \ref{lm:Egn1}, 
$\frac{\partial \Psi_l}{\partial r_l}$ is nonnegative if
\beqno
\frac{\sigma_{N_l}^2{\baseN}^{2r_l}}{||\hat{\vc a}_l||^2}
    -\frac{\alpha_{\max}}{\alpha_{\min}}
   \left(\frac{1}{\alpha_{\min}}-\frac{1}{\alpha_{\max}}\right)
\geq 0,
\eeqno
completing the proof. 
\hfill\IEEEQED

{\it Proof of Lemma \ref{lm:pro4}: }
Without loss of generality we 
may assume $k=1$. For $T$$\in {\cal O}_K(\hat{\vc a}_l,k)$,
the matrix $C^*(\Gamma^{-1}T,r_l)$ has the form: 
\beqno
C^*(\Gamma^{-1}T,r_l)
=\left[
\ba{c|c}
c_{11}^*(\Gamma^{-1}T,r_l)& {\vc c}^*_{1[1]}(\Gamma^{-1}T)
\\\hline
{}^{\rm t}{\vc c}^*_{1[1]}(\Gamma^{-1}T)&
C_{22}^*(\Gamma^{-1}T)
\ea\right],
\eeqno
where $C_{22}^*(\Gamma^{-1}T)$ is a $(K-1)\times (K-1)$
matrix with $c_{kk^{\prime}}^*(\Gamma^{-1}T),$ $(k,k^{\prime})$
$\in (\Lambda_K-\{1\})^2$ in its $(k,k^{\prime})$ element. 
Since $C^*(\Gamma^{-1}T,r_l)\preceq$
$\alpha_{\max}^{*}(r_l)I_K$, we must have
$C_{22}^*(\Gamma^{-1}T)\preceq$
$\alpha_{\max}^{*}(r_l)I_{K-1}$.
%
Then we have
\beqa
C^*(\Gamma^{-1}T,r_l)
\preceq 
\left[
\ba{c|c}
c_{11}^*(\Gamma^{-1}T,r_l)& {\vc c}^*_{1[1]}(\Gamma^{-1}T)
\\\hline
{}^{\rm t}{\vc c}^*_{1[1]}(\Gamma^{-1}T)&
\alpha_{\max}^*(r_l)I_{K-1}
\ea\right].
\label{eqn:match300}
\eeqa
Let $\lambda$ be the minimum eigenvalue of 
the matrix in the right hand side of (\ref{eqn:match300}).
Then, by (\ref{eqn:match300}), we have 
$\lambda \geq \alpha_{\min}^*(r_l)$ and 
$\lambda$ satisfies the following:
\beqa
\lefteqn{(\lambda-c_{11}^{*}(\Gamma^{-1}T,r_l))
(\lambda-\alpha_{\max}^*(r_l))}
\nonumber\\
&&-||{\vc c}^*_{1[1]}(\Gamma^{-1}T)||^2=0.
\label{eqn:math301}
\eeqa
From (\ref{eqn:math301}), we have
\beqno
c_{11}^{*}(\Gamma^{-1}T,r_l)
&=&\lambda +\frac{||{\vc c}^*_{1[1]}(\Gamma^{-1}T)||^2}
{\alpha_{\max}^*(r_l)-\lambda}
\\
&\geq & 
\alpha_{\min}^*(r_l)
+\frac{||{\vc c}^*_{1[1]}(\Gamma^{-1}T)||^2}
{\alpha_{\max}^*(r_l)-\alpha_{\min}^*(r_l)}
\\
&\geq & \alpha_{\min}(r^L)
+\frac{||{\vc c}^*_{1[1]}(\Gamma^{-1}T)||^2}
{\alpha_{\max}^*-\alpha_{\min}(r^L)},
\eeqno
completing the proof. 
\hfill\IEEEQED

Next we prove Theorems \ref{th:matchTh} 
and \ref{th:matchTh2z}. For simplicity of notation 
we set
\beqno
&&a(r^L)\defeq\frac{1}{\alpha_{\min}(r^L)},
b(r^L)\defeq \frac{1}{\alpha_{\max}(r^L)},
b^*\defeq\frac{1}{\alpha_{\max}^*}.
\eeqno
Then the condition (\ref{eqn:match600}) in 
Lemma \ref{lm:pro3} 
is rewritten as
\beq
a(r^L)\left[\frac{a(r^L)}{b(r^L)}-1\right]\leq 
\frac{\sigma_{N_l}^2{\baseN}^{2r_l}}{||\hat{\vc a}_l||^2}.
\label{eqn:match601aaa}
\eeq

{\it Proof of Theorem \ref{th:matchTh}: }For 
$(l,k)\in \Lambda_L\times \Lambda_K$, 
we choose $T\in{\cal O}_K(\hat{\vc a}_l,k)$. 
By Lemma \ref{lm:pro4}, we have  
\beq
\frac{\sigma_{N_l}^2{\baseN}^{2r_l}}{||\hat{\vc a}_l||^2}
\geq 
\left[
\chi_k^{\ast}-\frac{1}{a(r^L)} -\frac{a(r^L)b^*||
{\vc c}^{\ast}_{k[k]}||^2}{a(r^L)-b^*}
\right]^{-1}.
\label{eqn:match602aaa}
\eeq
It follows from (\ref{eqn:match601aaa}), 
(\ref{eqn:match602aaa}), and Lemma \ref{lm:pro3} 
that if for any $l\in \Lambda_L$, 
there exist $k\in \Lambda_K$ and 
$T\in{\cal O}_K(\hat{\vc a}_{l},k)$ such that    
\beq
a(r^L)\left[\frac{a(r^L)}{b(r^L)}-1\right]\leq 
\left[\chi_k^*-\frac{1}{a(r^L)}
-\frac{a(r^L)b^*||{\vc c}^{\ast}_{k[k]}||^2}{a(r^L)-b^*}\right]^{-1}
\label{eqn:match703}
\eeq
holds for $r^L\in{\cal B}(\Gamma,D)$, then 
$\theta(\Gamma,D,r^L)$ satisfies the MD condition on 
${\cal B}_L($ $\Gamma,D)$.  
Since the left hand side of (\ref{eqn:match703}) 
is a monotone decreasing function 
of $b(r^L)$ and $b(r^L)\geq b^*$, 
\beq
a(r^L)\left[\frac{a(r^L)}{b^*}-1\right]\leq 
\left[\chi_k^*-\frac{1}{a(r^L)}
-\frac{a(r^L)b^*||{\vc c}^{\ast}_{k[k]}||^2}{a(r^L)-b^*}\right]^{-1}
\label{eqn:match604z}
\eeq
implies (\ref{eqn:match703}). Observe that 
(\ref{eqn:match604z}) is equivalent to  
\beqa
\hspace*{-15mm}
\lefteqn{
a(r^L) \left[\frac{a(r^L)}{b^*}-1 \right] \cdot 
   \left[\chi_k^*-\frac{1}{a(r^L)}
   -\frac{a(r^L)b^*||{\vc c}^{\ast}_{k[k]}||}{a(r^L)-b^*}\right]\leq 1
}
\nonumber\\
\hspace*{-15mm}
&\Leftrightarrow&
\left(\frac{a(r^L)}{b^*}-1\right)\chi_k^{\ast}
-\frac{1}{b^*} - a(r^L)||{\vc c}^{\ast}_{k[k]}||^2\leq 0.
\label{eqn:match705}
\eeqa
Solving (\ref{eqn:match705}) with respect to
$a(r^L)$, we have
\beqa
a(r^L)&\leq& \ds
\frac{\chi_k^{\ast}+\frac{1}{b^*}}
{\frac{1}{b^*}\chi_k^{\ast}-||{\vc c}^{\ast}_{k[k]}||^2}
=\frac{b^*\chi_k^{\ast}+1}{\chi_k^{\ast}-b^*||{\vc c}^{\ast}_{k[k]}||^2}
\nonumber\\
&=&b^*+\frac{1+(b^*)^2||{\vc c}^{\ast}_{k[k]}||^2}
   {\chi_k^*-b^*||{\vc c}^{\ast}_{k[k]}||^2}.
\label{eqn:match777}
\eeqa
On the other hand, by (\ref{eqn:aa00z}), we have 
\beq
a(r^L)\leq D-(K-1)b(r^L)\leq D-(K-1)b^*.
\label{eqn:match778}
\eeq
Then we have the following.
\beqa
&&D\leq Kb^*+\frac{1+(b^*)^2||{\vc c}^{\ast}_{k[k]}||^2}
   {\chi_k^*-b^*||{\vc c}^{\ast}_{k[k]}||^2}.
\nonumber\\
&\Leftrightarrow&
D-(K-1)b^*\leq b^*+\frac{1+(b^*)^2||{\vc c}^{\ast}_{k[k]}||^2}
   {\chi_k^*-b^*||{\vc c}^{\ast}_{k[k]}||^2}.
\nonumber\\
&\Rightarrow& (\ref{eqn:match777}) 
\mbox{ holds under }(\ref{eqn:match778}).
\nonumber\\
&\Rightarrow& (\ref{eqn:match777}) 
\mbox{ holds for }r^L\in {\cal B}(\Gamma,D).
\nonumber\\
&\Leftrightarrow& 
(\ref{eqn:match604z})
\mbox{ holds for }r^L\in {\cal B}(\Gamma,D).
\nonumber\\
&\Rightarrow& 
(\ref{eqn:match703})
\mbox{ holds for }r^L\in {\cal B}(\Gamma,D).
\nonumber
\eeqa
Hence, if for any $l\in\IsetA$, there exist 
$k\in \Lambda_K$ and 
$T\in {\cal O}_K($ $\hat{\vc a}_l,k)$ such that    
$$
D\leq 
\frac{K}{\alpha_{\max}^*}
+\frac{1+\frac{||{\vc c}^{\ast}_{k[k]}(\Gamma^{-1}T)||^2}
{(\alpha_{\max}^*)^2}}
{\chi_k^*(\Gamma^{-1}T)-\frac{||{\vc c}^{\ast}_{k[k]}(\Gamma^{-1}T)||^2}
{\alpha_{\max}^*}},
$$
then $\theta(\Gamma,D,r^L)$ satisfies the MD condition on 
${\cal B}_L($ $\Gamma,D)$. Thus, by Lemma \ref{lm:lem1}, 
\beqno
\lefteqn{
D\leq\frac{K}{\alpha_{\max}^*}}
\\
& & 
+\min_{l\in \IsetA}
 \max_{
  \scs k\in \IsetB
  \atop{ 
  \scs T\in {\cal O}_K(\hat{\lvc a}_l,k)
            }
      }
\frac{1+\frac{||{\vc c}^{\ast}_{k[k]}(\Gamma^{-1}T)||^2}
{(\alpha_{\max}^*)^2}}{\chi_k^*(\Gamma^{-1}T)-
\frac{||{\vc c}^{\ast}_{k[k]}(\Gamma^{-1}T)||^2}
{\alpha_{\max}^*}}
\eeqno
is a sufficient matching condition. 
\hfill\IEEEQED

{\it Proof of Theorem \ref{th:matchTh2z}: } 
The inequality (\ref{eqn:match600}) in 
Lemma \ref{lm:pro3} 
is rewritten as
\beq
[a(r^L)-b(r^L)]\frac{a(r^L)}{b(r^L)}\leq \tau_l{\baseN}^{2r_l}.
\label{eqn:match601}
\eeq
From (\ref{eqn:match601}), we can see that if we have     
\beq
[a(r^L)-b(r^L)]\frac{a(r^L)}{b(r^L)}\leq \tau^*
\label{eqn:match602}
\eeq
on ${\cal B}_L(\Gamma,D)$, then $\theta(\Gamma, D,r^L)$ 
satisfies the MD condition on ${\cal B}_L(\Gamma,D)$.
On the other hand, from (\ref{eqn:aa00z}), we obtain 
\beq
a(r^L)\leq D-(K-1)b(r^L).
\label{eqn:match603} 
\eeq
Under (\ref{eqn:match603}), we have 
\beqno
&    &[a(r^L)-b(r^L)]\frac{a(r^L)}{b(r^L)}
\nonumber\\ 
&\leq&\left[D-Kb(r^L)\right]
\frac{D-(K-1)b(r^L)}{b(r^L)}. 
\eeqno
Hence the following is a sufficient condition 
for (\ref{eqn:match602}) to hold: 
\beq
\left[D-Kb(r^L)\right]
\frac{D-(K-1)b(r^L)}{b(r^L)}\leq \tau^*.
\label{eqn:match604} 
\eeq
Solving (\ref{eqn:match604}) with respect to 
$D$, we obtain
\beq
D \leq K b(r^L)
+\frac{1}{2}\left[\sqrt{b^2(r^L)+4\tau^*b(r^L)}-b(r^L)\right].
\label{eqn:match605}  
\eeq
Since the right hand side of (\ref{eqn:match605}) is 
a monotone increasing function of $b(r^L)$ and 
$b(r^L)\geq 1/\alpha_{\max}^{*}$ by Lemma \ref{lm:Egn1}, 
the condition    
$$
D \leq {\frac{K}{\alpha_{\max}^{*}}}
  +\frac{1}{2\alpha_{\max}^{*}}
 \left\{\sqrt{1+4\alpha_{\max}^{*}\tau^*}-1\right\} 
$$
is a sufficient condition for (\ref{eqn:match602}) 
to hold. \hfill\IEEEQED 

Next, we prove Lemma \ref{lm:pro3bz}. To prove this lemma 
we prepare a lemma shown below. 
\begin{lm}\label{lm:lm8b}{\rm 
A necessary and sufficient condition for 
$
\underline{\tilde{J}}($ $D,r)$ 
to take the maximum at $r=r^*$ is 
$$
\left(\frac{{\rm d}}{{\rm d}r}\underline{\tilde{J}}(D,r)
\right)_{r=r^{\ast}}\geq 0. 
$$
}
\end{lm}

{\it Proof:} For simplicity of notation we set 
$
\underline{\tilde{J}}(r)\defeq \underline{\tilde{J}}(D$ $,r).$
Suppose that 
\beq
\left(
\frac{{\rm d}\underline{\tilde{J}}(r)}{{\rm d}r}
\right)_{r=r^{\ast}}\geq 0.
\label{eqn:assum1}
\eeq
Under (\ref{eqn:assum1}), we assume that  
$\underline{\tilde{J}}(r)$ does not take the minimum
at $r=r^*$. Then there exists $\epsilon>0$ and 
$\tilde{r} > r^{\ast}$ such that 
$\underline{\tilde{J}}(\tilde{r})\leq 
\underline{\tilde{J}}({r}^{\ast})-\epsilon$.
Since 
$\underline{\tilde{J}}(r)$ is a convex function 
of $r\geq r^*$, we have
\beqa
& &\underline{\tilde{J}}(\tau \tilde{r}+(1-\tau)r^{\ast})
\leq \tau\underline{\tilde{J}}(\tilde{r})+(1-\tau)
\underline{\tilde{J}}(r^{\ast})
\nonumber\\
&\leq & \tau(\underline{\tilde{J}}(r^{\ast})-\epsilon)
         +(1-\tau)\underline{\tilde{J}}(r^{\ast})
= \underline{\tilde{J}}(r^{\ast})-\tau\epsilon
\label{eqn:assz2}
\eeqa
for any $\tau\in (0,1]$. From (\ref{eqn:assz2}), we obtain 
\beq
\frac{
\underline{\tilde{J}}(r^{\ast}+\tau(\tilde{r}-r^{\ast}))
-\underline{\tilde{J}}(r^{\ast})}{\tau(\tilde{r}-r^{\ast})}\leq 
-\frac{\epsilon}{\tilde{r}-r^{\ast}}
\label{eqn:aa00a}
\eeq 
for any $\tau\in (0,1]$. By letting $\tau \to 0$ 
in (\ref{eqn:aa00a}), we have
$$
\left(
\frac{{\rm d}\underline{\tilde{J}}(r)}{{\rm d}r}
\right)_{r=r^{\ast}}\leq -\frac{\epsilon}{\tilde{r}-r^{\ast}}<0, 
$$
which contradicts (\ref{eqn:assum1}).  
Hence under (\ref{eqn:assum1}), $\underline{\tilde{J}}(r)$ 
takes the minimum at $r=r^{\ast}$. 
It is obvious that when 
$\left(\frac{{\rm d}\underline{\tilde{J}}(r)}{{\rm d}r}\right)_{r=r^{\ast}}<0$, 
$\underline{\tilde{J}}(r)$ does not take the minimum at $r=r^{\ast}$. 
\hfill
\IEEEQED

{\it Proof of Lemma \ref{lm:pro3bz}:}
We first derive expression of $\tilde{\omega}(D,r)$ 
using $\beta_l=\beta_l(r),$ $l\in \IsetA$ in 
a neighborhood of $r=r^{\ast}$. 
Let $S(r)=\{l:\beta_l(r)<\beta_{l_1}(r)\}.$
By definition, $L-|S(r)|$ is equal to the multiplicity of the 
$\beta_{l_1}(r)$.  In particular, for $r=r^{\ast}$, we have
\beq
\frac{1}{\beta_{l_1}(r^{\ast})}
=\frac{1}{L-|S(r^{\ast})|}
\left(D+{\rm tr}[B]-\sum_{l\in S(r^{*})} \frac{1} {\beta_{l}(r^{*}) }\right).
\label{eqn:asss}
\eeq
Since $\beta_l(r),l\in \Lambda_L$ are strictly monotone increasing 
functions of $r$, there exists small positive number $\delta$ 
such that for any $r \in [r^{*},r^{*}+\delta)$, we have 
\beqno
S(r)&=&S(r^{\ast}),
\\
\frac{1}{\beta_{l_1}(r)}
&<&\frac{1}{L-|S(r)|}
\left(D+{\rm tr}[B]-\sum_{l\in S(r)}\frac{1}{\beta_l(r)} \right)
\\
&<&\frac{1}{\beta_{k}(r)}\quad\mbox{ for } {k}\notin S(r^{\ast}).
\eeqno
The function $\tilde{\omega}(D,r), $$r \in [r^{*},r^{*}+\delta)$ is computed as 
\beqno
\tilde{\omega}(D,r)
&=&{\ts \frac{1}{(L-|S(r^{\ast})|)^{L-|S(r^{\ast})|}}}
   \left(\prod_{l\in S(r^{\ast})}\frac{1}{\beta_l(r)}\right)
\\
& & \times \left(D+{\rm tr}[B]
-\sum_{l\in S(r^{\ast})} \frac{1}{\beta_l(r)}\right)^{L-|S(r^{\ast})|}.
\eeqno
In the following we use the simple notations 
$\beta_l$ and $S$ for $\beta_l(r^{\ast})$ and $S(r^{\ast})$, 
respectively. Computing the derivative of 
$\underline{\tilde{J}}\left(D,r \right)$
at $r=r^{\ast}$, we obtain
\beqa
& &\frac{1}{2}
   \left(\frac{{\rm d}}{{\rm d}r}
   \underline{\tilde{J}}\left(D,r \right)
   \right)_{r=r^{\ast}}
 \nonumber\\
&=&
\frac{1}{\epsilon{\rm e}^{2r^{\ast}}}
\sum_{l\in S}
\left(1-\frac{\epsilon}{\mu_l}\right)^2
\hspace*{-1mm}
\left[\frac{1}{\beta_l}
-\frac{L-|S|}{D+{\rm tr}[B]-{\ds \sum_{l\in S}}
\frac{1}{\beta_l}}\frac{1}{\beta_l^2}
\right]\hspace*{-1mm}
+L
\nonumber\\
&\MEq{a}&
\frac{1}{\epsilon{\rm e}^{2r^{\ast}}}
\sum_{l\in S}
\left(1-\frac{\epsilon}{\mu_l}\right)^2
 \left[\frac{1}{\beta_l}-\frac{\beta_{l_1}}{\beta_l^2}
\right]
+L
\nonumber\\
&=&
\frac{1}{\epsilon{\rm e}^{2r^{\ast}}}
\sum_{l=1}^L
\left(1-\frac{\epsilon}{\mu_l}\right)^2
 \left[\frac{1}{\beta_l}-\frac{\beta_{l_1}}{\beta_l^2}
\right]
+L
\nonumber\\
&=&
\sum_{l=1}^L
\left\{
\frac{
\left(1-\frac{\epsilon}{\mu_l}\right)
\left[ {\rm e}^{2r^{\ast}}-1+ \frac{\epsilon}{\mu_{l}} \right]
}
{\left[{\rm e}^{2r^{\ast}}-1+\frac{\epsilon}{\mu_{l}}\right]^2}
\right.
\nonumber\\
& &\qquad \quad -\left.
\frac{
\left(1-\frac{\epsilon}{\mu_{l_1}}\right)
\left[{\rm e}^{2r^{\ast}}-1+\frac{\epsilon}{\mu_{l_1}}\right]
}
{\left[{\rm e}^{2r^{\ast}}-1+\frac{\epsilon}{\mu_{l}}\right]^2}
\right\}+L
\nonumber\\
&=&\sum_{l=1}^L
\frac{
{\rm e}^{2r^{\ast}}
\hspace*{-1mm}
\left[{\rm e}^{2r^{\ast}}
\hspace*{-1mm}-1
+\frac{\epsilon}{\mu_l}\right]-
\left(1-\frac{\epsilon}{\mu_{l_1}}\right)
\left[{\rm e}^{2r^{\ast}}
\hspace*{-1mm}-1+
\frac{\epsilon}{\mu_{l_1}}\right]
}{
\left[{\rm e}^{2r^{\ast}}-1+
\frac{\epsilon}{\mu_{l}}
\right]^2
}
\nonumber\\
&\geq &0.
\nonumber
\eeqa
Step (a) follows from (\ref{eqn:asss}). 
\hfill\IEEEQED

\section{Conclusion}

We have considered the distributed source coding of correlated
Gaussian sources $Y_l,l\in \Lambda_L$ which are $L$ 
observations of $K$ remote sources $X_k, k\in \Lambda_K$. 
We have studied the remote source coding 
problem where the decoder wish to reconstruct $X^K$ 
and have derived explicit outer bounds 
${\cal R}_L^{\rm (out)}(\Gamma, D^L|\Sigma_{X^KY^L})$
and  
${\cal R}_L^{\rm (out)}(\Gamma, D|\Sigma_{X^KY^L})$ 
of 
${\cal R}_L($ $\Gamma,D^L|\Sigma_{X^KY^L})$ and 
${\cal R}_L(\Gamma,D|\Sigma_{X^KY^L})$, respectively. 
Those outer bounds are described in 
a form of positive semi definite programming.  
On the outer bound ${\cal R}_L^{\rm (out)}(\Gamma, D|\Sigma_{X^KY^L})$, 
we have shown that it has a form of the water 
filling solution. Using this form, we have derived 
two different matching conditions for 
${\cal R}_L^{\rm (out)}($ $\Gamma, D|\Sigma_{X^KY^L})$
to coincide with      
${\cal R}_L($ $\Gamma, D|\Sigma_{X^KY^L})$.

In the case of $K=L,A=I_L$, we have considered 
the multiterminal \RateDist problem where 
the decoder wishes to reconstruct 
$Y^L=X^L+N^L$. Using the strong relation between
the remote source coding problem and the multiterminal 
\RateDist problem, we have obtained the outer bounds 
${\cal R}_L^{\rm (out)}(\Gamma,D^L|\Sigma_{Y^L})$  
and 
${\cal R}_L^{\rm (out)}(\Gamma,D|\Sigma_{Y^L})$, 
of 
${\cal R}_L(\Gamma,D^L|\Sigma_{Y^L})$  
and 
${\cal R}_L(\Gamma,D|\Sigma_{Y^L})$, 
respectively. Furthermore, using this relation, 
we have obtained the matching condition 
for ${\cal R}_L^{\rm (out)}(\Gamma,D|\Sigma_{Y^L})$
to coincide with ${\cal R}_L($ $\Gamma,D|\Sigma_{Y^L})$. 

In the remote source coding problem, finding 
an explicit condition for 
${\cal R}_L^{\rm (out)}(\Gamma,D^L|\Sigma_{X^KY^L})$ 
to be tight is left to us as a future work.  
Similarly, in the multiterminal source coding 
problem, finding an explicit condition for   
${\cal R}_L^{\rm (out)}(\Gamma,D^L|\Sigma_{Y^L})$ 
to be tight is also left to us as a future work. 
To investigate those problems we must examine the solutions to the
problems of positive semi definite programming describing those
two outer bounds. Those analysis are rather mathematical problems 
in the field of convex optimization. 

\section*{\empty}
\appendix

{\it Proof of Property \ref{pr:prFz} part b):}
Since 
$$
\underline{\tilde{J}}(D,r)
=Lr-\log \tilde{\omega}(D,r)+\frac{1}{2}\log |\Sigma_{Y^L}+B|,
$$
it suffices to prove the concavity of 
$
\log \tilde{\omega}(D,r)
$
with respect to $r\geq r^{\ast}$. We first observe that
$\log \tilde{\omega}(D,r)$ has the following expression: 
$$
\log \tilde{\omega}(D,r)=\max_{
        \scs \sum_{l=1}^L\xi_l\leq D+{\rm tr}[B], 
        \atop{\scs \xi_l \beta_l(r)\geq 1}} \sum_{l=1}^L \log \xi_l
$$
For each $j\in \{1,2\}$, let $\xi_l^{(j)},l=1,2,\cdot,L$ be 
$L$ positive numbers that attain 
$\log \tilde{\omega}(D,r^{(j)})$.
Let $t_1,t_2$ be a pair of nonnegative 
numbers such that $t_1+t_2=1$.
Then we have 
\beqa
& &t_1\log \tilde{\omega}(D,r^{(1)})+
   t_2\log \tilde{\omega}(D,r^{(2)})
\nonumber\\
&=&\sum_{i=1}^L\left(t_1\log\xi_i^{(1)}+t_2\log\xi_i^{(2)}\right)
\nonumber\\
&\MLeq{a}&\sum_{i=1}^L\log
\left(t_1\xi_i^{(1)}+t_2\xi_i^{(2)}\right). 
\label{eqn:ssz0z}
\eeqa
Step (a) follows from the concavity of the logarithm functions. Since 
\beqno
\{\beta_{l}(r)\}^{-1}
&=&\frac{\mu_l\epsilon}{\mu_l-\epsilon}
\frac{{\rm e}^{2r}}{\mu_l[{\rm e}^{2r}-1]+\epsilon}
\\
&=&\frac{\mu_l\epsilon}{\mu_l-\epsilon}
+\frac{\mu_l\epsilon}{\mu_l[{\rm e}^{2r}-1]+\epsilon}
\eeqno
$\{\beta_{l}(r)\}^{-1}$ is a convex function of $r\geq r^{\ast}$.
Then we have 
\beqa
t_1\xi_i^{(1)}+t_2\xi_i^{(2)}&\geq &
t_1\{\beta_{i}(r^{(1)})\}^{-1} +t_2\{\beta_{i}(r^{(2)})\}^{-1}
\nonumber\\
&\geq& \{\beta_{i}(t_1r^{(1)}+t_2r^{(2)})\}^{-1},
\label{eqn:ss0}
\eeqa
for $l=1,2,\cdots,L$. Furthermore, we have 
\beq
\sum_{l=1}^L \left( t_1\xi_l^{(1)}+t_2\xi_l^{(2)}\right)
=t_1\sum_{l=1}^L \xi_l^{(1)}+t_2\sum_{l=1}^L\xi_l^{(2)}\leq D\,.
\label{eqn:ss1}
\eeq
From (\ref{eqn:ss0}), (\ref{eqn:ss1}), and the definition of 
$\log \tilde{\omega}(D,r)$, we have 
\beq
\sum_{l=1}^L \log \left( t_1\xi_l^{(1)}+t_2\xi_l^{(2)}\right)
\leq \log \tilde{\omega}\left(D,t_1r_1^{(1)}+t_2r_2^{(2)}\right).
\label{eqn:ss2}
\eeq
From (\ref{eqn:ssz0z}) and (\ref{eqn:ss2}), we have 
\beqno
& &t_1\log \tilde{\omega}(D,r^{(1)})+
   t_2\log \tilde{\omega}(D,r^{(2)})
\nonumber\\
&\leq &\log \tilde{\omega}\left(D,t_1r_1^{(1)}+t_2r_2^{(2)}\right),
\eeqno
completing the proof.
\hfill \IEEEQED

\subsection{
Proof of Lemma \ref{lm:lm1}
}

In this appendix we prove Lemma \ref{lm:lm1}. 
To prove this lemma we need some preparations.
For $k\in\IsetB$ and for $Q$$\in{\cal O}_K$, set
$$
F_k(\Sigma|Q)\defeq 
\sup_{\scs p_{\hat{X}^K|X^K}:
\atop{\scs \Sigma_{X^K-\hat{X}^K}\preceq \Sigma }}
h(Z_k-\hat{Z}_k|Z_{[k]}^K-\hat{Z}_{[k]}^K).
$$
To compute $F_k(\Sigma|Q)$, define two random variables 
by 
$$
\tilde{X}^K\defeq X^K-\hat{X}^K, \tilde{Z}^K\defeq Z^K-\hat{Z}^K.
$$
Note that by definition we have $\tilde{Z}^K=Q\tilde{X}^K$. 
Let $p_{X^K\tilde{X}^K}$ $(x^K,\tilde{x}^K)$ be a density 
function of $(X^K,\tilde{X}^K)$. 
Let $q_{Z^K\tilde{Z}^K}$ $(z^K,\tilde{z}^K)$ be a density 
function of $(Z^K,\tilde{Z}^K)$ induced by the orthogonal 
matrix $Q$, that is,
$$
q_{Z^K\tilde{Z}^K}(z^K,\tilde{z}^K)
\defeq 
p_{{}^{\rm t}QZ^K {}^{\rm t}Q\tilde{Z}^K}({}^{\rm t}Qz^K,{}^{\rm t}Q\tilde{z}^K).
$$
Expression of $F_k(\Sigma|Q)$ using the above density 
functions is the following.
\beqno
&  &F_k(\Sigma|Q)
=\sup_{\scs p_{\tilde{X}^K|X^K}:
\atop{\scs \Sigma_{\tilde{X}^K}\preceq \Sigma }}
h(\tilde{Z}_k|\tilde{Z}_{[k]}^K)
\nonumber\\
&=&\sup_{\scs p_{\tilde{X}^K|X^K}:
\atop{\scs \Sigma_{\tilde{X}^K}\preceq \Sigma }}-\int
q_{\tilde{Z}^K}(z^K)
\log q_{\tilde{Z}_k|\tilde{Z}_{[k]}^K}(z_k|z_{[k]}^K){\rm d}z^K
\nonumber\\
&=&\sup_{\scs p_{\tilde{X}^K|X^K}:
\atop{\scs \Sigma_{\tilde{X}^K}\preceq \Sigma }}-\int
q_{\tilde{Z}^K}(z^K)
\log \frac{ q_{\tilde{Z}^K}(z^K)}
          { q_{\tilde{Z}_{[k]}^K}(z_{[k]}^K)}{\rm d}z^K.
\eeqno
The following two properties on $F_k(\Sigma|Q)$ are 
useful for the proof of Lemma \ref{lm:lm1}.
\begin{lm}\label{lm:lm7}
$F_k(\Sigma|Q)$ is concave with respect to $\Sigma$.
\end{lm}
\begin{lm}\label{lm:lm8}
$$
F_k(\Sigma|Q)=\frac{1}{2}
\log
 \left\{
 {(2\pi{\rm e})}\left[Q\Sigma^{-1} {}^{\rm t}Q\right]_{kk}^{-1}
\right\}.
$$ 
\end{lm}

We first prove Lemma \ref{lm:lm1} using those two lemmas and 
next prove Lemmas \ref{lm:lm7} and \ref{lm:lm8}.

{\it Proof of Lemma \ref{lm:lm1}:} 
We have the following chain of inequalities:
\beqno
& &\left. h(\irg{\vc Z}_k \right| \irg{\vc Z}_{[k]}^K \irb{W}^K)
\leq
   h( \irg{{\vc Z}}_k -\hat{\irg{{\vc Z}}}_k
      \left. \right| \irg{{\vc Z}}_{[k]}^K -\hat{\irg{{\vc Z}}}_{[k]}^K)
\nonumber\\
&\leq&\sum_{t=1}^n
   h(\irg{Z}_k(t) -\hat{\irg{Z}}_k(t)
     \left. \right|
     \irg{Z}_{[k]}^K(t) -\hat{\irg{Z}}_{[k]}^K(t))
\nonumber\\
&\MLeq{a}&\sum_{t=1}^nF_k
\left.\left(
\Sigma_{{X}^K(t)-\hat{\irg{X}}^K(t)}\right|Q 
\right)
\nonumber\\
&\MLeq{b}& nF_k
\left.\left({\ts\frac{1}{n}}\sum_{t=1}^n
\Sigma_{{X}^K(t)-\hat{\irg{X}}^K(t)}\right|Q 
\right)
\nonumber\\
&=&nF_k\left.\left(
     \ts{\frac{1}{n}}\Sigma_{{\lvc X}^K-\hat{\irg{\lvc X}}^K}
        \right|Q\right)
\nonumber\\
&\MEq{c}&\frac{n}{2}\log 
\left\{
(2\pi {\rm e})
\left[
Q
\left(
{\ts \frac{1}{n}}\Sigma_{{\lvc X}^K-\hat{\lvc X}^K}
\right)^{-1} {}^{\rm t}Q 
\right]_{kk}^{-1}
\right\}.
\eeqno
Step (a) follows from the definition of $F_k(\Sigma|Q)$.
Step (b) follows from Lemma \ref{lm:lm7}.
Step (c) follows from Lemma \ref{lm:lm8}.
\hfill \IEEEQED

{\it Proof of Lemma \ref{lm:lm7}:}
For given covariance matrices 
$\Sigma^{(0)}$ and $\Sigma^{(1)}$, let 
$p_{\tilde{X}^K|X^K}^{(0)}$ and 
$p_{\tilde{X}^K|X^K}^{(1)}$
be conditional densities achieving 
$F_k(\Sigma^{(0)}|Q)$ and $F_k(\Sigma^{(1)}|Q)$, respectively.
For $0\leq \alpha \leq 1$, define 
a conditional density parameterized with $\alpha$ by 
$$
p_{\tilde{X}^K|X^K}^{(\alpha)}
= (1-\alpha) p_{\tilde{X}^K|X^K}^{(0)}
+     \alpha p_{\tilde{X}^K|X^K}^{(1)}.
$$
Let $p_{X^K\tilde{X}^K}^{(\alpha)}$ 
be a density function of $(X^K,\tilde{X}^K)$ 
defined by 
$(p_{\tilde{X}^K|X^K}^{(\alpha)},$
$p_{X^K}^{(\alpha)})$. Let 
$\Sigma_{\tilde{X}}^{(\alpha)}$ be a covariance matrix 
computed from the density $p_{\tilde{X}^K}^{(\alpha)}$. 
Since 
$$
p_{\tilde{X}^K}^{(\alpha)}
= (1-\alpha)p_{\tilde{X}^K}^{(0)}
+    \alpha p_{\tilde{X}^K}^{(1)},
$$
we have 
\beqa
\Sigma_{\tilde{X}}^{(\alpha)}
&=&
 (1-\alpha)\Sigma_{\tilde{X}}^{(0)}
   +\alpha \Sigma_{\tilde{X}}^{(1)}
\nonumber
\\
&\preceq &
 (1-\alpha)\Sigma^{(0)}
   +\alpha \Sigma^{(1)}.
\label{eqn:cov0}
\eeqa
Let $q_{Z^K\tilde{Z}^K}^{(\alpha)}$ be a density function 
of $(Z^K,\tilde{Z}^K)$ induced by the orthogonal matrix $Q$, that is, 
$$
q_{Z^K\tilde{Z}^K}^{(\alpha)}(z^K,\tilde{z}^K)
\defeq p_{{}^{\rm t}QZ^K {}^{\rm t}Q\tilde{Z}^K}^{(\alpha)}
({}^{\rm t}Qz^K,{}^{\rm t}Q\tilde{z}^K).
$$
By definition it is obvious that
$$
            q_{\tilde{Z}^K}^{(\alpha)}
= (1-\alpha)q_{\tilde{Z}^K}^{(0)}
+    \alpha q_{\tilde{Z}^K}^{(1)}.
$$
Then we have 
\beqno
&  & (1-\alpha)F_k(\Sigma^{(0)}|Q) 
      + \alpha F_k(\Sigma^{(1)}|Q)
\\
&=& -(1-\alpha)\int q_{\tilde{Z}^K}^{(0)}(z^K)
\log \frac{q_{\tilde{Z}^K}^{(0)}(z^K)}
          {q_{\tilde{Z}_{[k]}^K}^{(0)}(z_{[k]}^K)}{\rm d}z^K
\\
& &
 \quad \quad \:\:- \alpha\int q_{\tilde{Z}^K}^{(1)}(z^K)
\log \frac{q_{\tilde{Z}^K}^{(1)}(z^K)}
          {q_{\tilde{Z}_{[k]}^K}^{(1)}(z_{[k]}^K)}{\rm d}z^K
\\
&\MLeq{a}& -\int q_{\tilde{Z}^K}^{(\alpha)}(z^K)
\log \frac{q_{\tilde{Z}^K}^{(\alpha)}(z^K)}
          {q_{\tilde{Z}_{[k]}^K}^{(\alpha)}(z_{[k]}^K)}{\rm d}z^K
\\ 
&=&-\int q_{\tilde{Z}^K}^{(\alpha)}(z^K)
    \log q_{\tilde{Z}_k|\tilde{Z}_{[k]}^K}^{(\alpha)}(z_k|z_{[k]}^K){\rm d}z^K
\\
&\MLeq{b}& 
F_k \left(\left.(1-\alpha)\Sigma^{(0)}+\alpha \Sigma^{(1)}
\right|Q\right).
\eeqno
Step (a) follows from log sum inequality. 
Step (b) follows from the definition of 
$F_k(\Sigma|Q)$ and (\ref{eqn:cov0}).
\hfill \IEEEQED

{\it Proof of Lemma \ref{lm:lm8}:}
Let 
\beqno
q_{\tilde{Z}^K}^{(\rm G)}(z^K)
&\defeq&
\frac{1}{(2\pi{\rm e})^{\frac{K}{2}}
\left|\Sigma_{\tilde{Z}^K}\right|^{\frac{1}{2}}}
{\rm e}^{\scs 
-\frac{1}{2}
{}^{\rm t}[z^K]
\Sigma_{\tilde{Z}^K }^{-1}
\mbox{\scriptsize $[z^K]$ }} 
\eeqno
and let 
$$
{q_{\tilde{Z}_k|\tilde{Z}_{[k]}^K}^{({\rm G})}(z_k|z_{[k]}^K)}
=\frac{q_{\tilde{Z}^K}^{(\rm G)}(z^K)}
{q_{\tilde{Z}_{[k]}^K}^{({\rm G})}(z_{[k]}^K)}
$$
be a conditional density function induced by 
$q_{\tilde{Z}^K}^{(\rm G)}(\cdot)$.
We first observe that
\beq
\int q_{\tilde{Z}^K}(z^K)
    \log 
\frac
{q_{\tilde{Z}_k|\tilde{Z}_{[k]}^K}(z_k|z_{[k]}^K)}
{q_{\tilde{Z}_k|\tilde{Z}_{[k]}^K}^{({\rm G})}(z_k|z_{[k]}^K)}
{\rm d}z^K
\geq 0 . \label{eqn:div}
\eeq
From (\ref{eqn:div}), we have the following chain of inequalities:
\beqa
\lefteqn{
h(\tilde{Z}_k|\tilde{Z}_{[k]}^K)
=-\int q_{\tilde{Z}^K}(z^K)
    \log q_{\tilde{Z}_k|\tilde{Z}_{[k]}^K}(z_k|z_{[k]}^K)
    {\rm d}z^K}
\nonumber\\
&\leq &-\int q_{\tilde{Z}^K}(z^K)
    \log q_{\tilde{Z}_k|\tilde{Z}_{[k]}^K}^{({\rm G})}(z_k|z_{[k]}^K)
    {\rm d}z^K
\nonumber\\
&=& -\int q_{\tilde{Z}^K}(z^K)
    \log \frac{q_{\tilde{Z}^K}^{({\rm G})}(z^K)}
              {q_{\tilde{Z}_{[k]}^K}^{({\rm G})}(z_{[k]}^K)}
     {\rm d}z^K
\nonumber\\
&=& - \int q_{\tilde{Z}^K}(z^K)
      \log q_{\tilde{Z}^K}^{({\rm G})}(z^K)
      {\rm d}z^K
\nonumber\\
& & + \int q_{\tilde{Z}^K}(z^K)
      \log q_{\tilde{Z}_{[k]}^K}^{({\rm G})}(z_{[k]}^K)
      {\rm d}z^K
\nonumber\\
&\MEq{a}&
   - \int  q_{\tilde{Z}^K}^{({\rm G})}(z^K)
      \log q_{\tilde{Z}^K}^{({\rm G})}(z^K)
      {\rm d}z^K
\nonumber\\
& & + \int q_{\tilde{Z}^K}^{({\rm G})}(z^K)
      \log q_{\tilde{Z}_{[k]}^K}^{({\rm G})}(z_{[k]}^K)
      {\rm d}z^K
\nonumber\\
&=& \frac{1}{2}
    \log
    \left\{
     {(2\pi{\rm e})}
     \frac{|\Sigma_{\tilde{Z}^K     }|}
          {|\Sigma_{\tilde{Z}_{[k]}^K}|}
   \right\}
\MEq{b}\frac{1}{2}
    \log
    \left\{
     {(2\pi{\rm e})}
     \left[\Sigma_{\tilde{Z}^K}^{-1}\right]_{kk}^{-1}
    \right\}
\nonumber\\
&=&\frac{1}{2}
   \log
   \left\{
   {(2\pi{\rm e})}
   \left[Q\Sigma_{\tilde{X}^K}^{-1}{}^{\rm t}Q\right]_{kk}^{-1}
   \right\}
\nonumber\\
&\MLeq{c}&\frac{1}{2}
   \log
   \left\{
   {(2\pi{\rm e})}\left[Q\Sigma^{-1}{}^{\rm t}Q \right]_{kk}^{-1}
   \right\}.
\nonumber
\eeqa
Step (a) follows from the fact that 
$q_{\tilde{Z}^L}$ and 
$q_{\tilde{Z}^L}^{({\rm G})}$ 
yield the same moments of the quadratic form 
$\log q_{\tilde{Z}^L}^{({\rm G})}$.  
Step (b) is a well known formula on the determinant of matrix.
Step (c) follows from $\Sigma_{\tilde{X}^L} \preceq \Sigma$.
Thus
$$
F_k(\Sigma|Q) \leq \frac{1}{2}
\log
 \left\{
 {(2\pi{\rm e})}\left[Q\Sigma^{-1}{}^{\rm t}Q\right]_{kk}^{-1}
\right\}
$$ 
is concluded. Reverse inequality holds by letting 
$p_{\tilde{X}^K|{X}^K}$ be Gaussian with 
covariance matrix $\Sigma$. 
\hfill\IEEEQED

\subsection{
Proof of Lemma \ref{lm:lm2}
} 

In this appendix we prove Lemma \ref{lm:lm2}.
We write an orthogonal matrix $Q$ $\in {\cal O}_K$ 
as $Q=[q_{kk^{\prime}}]$, where $q_{kk^{\prime}}$ 
stands for the $(k,k^{\prime})$ element of $Q$. 
The orthogonal matrix ${Q}$ transforms 
$X^K$ into $\irg{Z}^K$$=QX^K$.
Set $\tilde{Q}=Q{}^{\rm t}A$ and let 
$\tilde{q}_{kl}$ be the $(k,l)$ element of $Q{}^{\rm t}A$.
The following lemma states an important 
property on the distribution of Gaussian random vector 
$Z^K$. This lemma is a basis of the proof 
of Lemma \ref{lm:lm2}. 

\begin{lm}\label{lm:LmO}
For any $k\in\IsetB$, we have the following.
\beq 
{Z}_k=-\frac{1}{g_{kk}}\sum_{k^{\prime}\ne k}\nu_{kk^{\prime}}{Z}_{k^{\prime}}
+\frac{1}{g_{kk}}\sum_{l=1}^{L}
\frac {\tilde{q}_{kl}}{\sigma_{N_l}^2}{Y}_l + \hat{N}_k,
\label{eqn:prlmaa00}
\eeq
where 
\beq
g_{kk}= \left[Q\Sigma_{X^K}^{-1}{}^{\rm t}Q\right]_{kk} 
+\sum_{l=1}^{L}\frac{\tilde{q}_{kl}^2}{\sigma_{N_l}^2},
\label{eqn:defajj}
\eeq
$\nu_{kk^{\prime}},$ 
$k^{\prime}\in \Lambda_K - \{k\}$ are suitable constants 
and $\hat{N}_k$ is a zero mean Gaussian random 
variables with variance $\frac{1}{g_{kk}}$. 
For each $k\in \IsetB$, $\hat{N}_k$ is independent 
of ${Z}_{k^{\prime}},k^{\prime}\in \IsetB - \{k\}$ 
and ${Y}_l,l\in \IsetA$. 
\end{lm}

{\it Proof:} Without loss of generality we may assume $k=1$.
Since $Y^L=AX^K+N^L$, we have 
\beqno
\Sigma_{X^KY^L}=
\left[
\ba{cc} 
\Sigma_{X^K} &\Sigma_{X^K}{}^{\rm t}A\\
A\Sigma_{X^K} & A\Sigma_{X^K}{}^{\rm t}A+\Sigma_{N^L}
\ea
\right].
\eeqno
Since $Z^K=QX^K$, we have
\beqno
\Sigma_{Z^KY^L}=
\left[
\ba{cc}
Q\Sigma_{X^K}{}^{\rm t}Q &   Q\Sigma_{X^K}{}^{\rm t}A\\
{}^{\rm t}A\Sigma_{X^K}{}^{\rm t}Q  & A\Sigma_{X^K}{}^{\rm t}A+\Sigma_{N^L}
\ea
\right].
\eeqno
The density function $p_{Z^KY^L}(z^K,y^L)$ of  
$(Z^K,Y^L)$ is given by 
\beqno
& &p_{Z^KY^L}(z^K,y^L)
\\
&=&
\frac{1}{(2\pi{\rm e})^{\frac{K+L}{2}}\left|\Sigma_{Z^KY^L}\right|^{\frac{1}{2}}}
{\rm e}^{\scs 
-\frac{1}{2}
{}^{\rm t}[z^K y^L]
\Sigma_{Z^KY^L}^{-1}
\mbox{\scriptsize 
$\left[\ba{c}z^K\\
y^L\ea\right]$}},
\eeqno
where $\Sigma_{Z^KY^L}^{-1}$ has the following form:
\beqno
\Sigma_{Z^KY^L}^{-1}=
\left[
\ba{cc}
   Q(\Sigma_{X^K}^{-1}+ {}^{\rm t}A\Sigma_{N^L}^{-1}A ){}^{\rm t}Q 
   & -Q{}^{\rm t}A\Sigma_{N^L}^{-1}\\
    -\Sigma_{N^L}^{-1}A{}^{\rm t}Q & \Sigma_{N^L}^{-1}
\ea
\right].
\eeqno
For $(k,k^{\prime})\in \Lambda_K^2$ and $l\in \IsetB$, set
\beq
\left.
\ba{rcl}
\nu_{kk^{\prime}}  &\defeq &\ds 
   \left[Q(\Sigma_{X^K}^{-1}+{}^{\rm t}A\Sigma_{N^L}^{-1}A)
{}^{\rm t}Q\right]_{kk^{\prime}}
\vspace*{1mm}\\
&=&\ds\left[Q\Sigma_{X^K}^{-1}{}^{\rm t}Q\right]_{kk^{\prime}}
+\sum_{l=1}^L\frac{\tilde{q}_{kl}
                       \tilde{q}_{k^{\prime}l}}{\sigma_{N_l}^2},
\vspace*{1mm}\\
\beta_{kl}&\defeq &\ds 
-\left[Q{}^{\rm t}A\Sigma_{N^L}^{-1}\right]_{kl}=
-\frac{\tilde{q}_{kl}}{\sigma_{N_l}^2}.
\ea
\right\}
\label{eqn:prlmz2z} 
\eeq
Now, we consider the following partition of 
$\Sigma_{Z^KY^L}^{-1}$: 
\beqno
\Sigma_{Z^KY^L}^{-1}
&=&
\left[
\ba{cc}
   Q(\Sigma_{X^K}^{-1}+{}^{\rm t}A\Sigma_{N^L}^{-1}A){}^{\rm t}Q 
& -Q{}^{\rm t}A\Sigma_{N^L}^{-1}\\
 -\Sigma_{N^L}^{-1}A{}^{\rm t}Q & \Sigma_{N^L}^{-1}
\ea
\right]
\nonumber\\
&=&
\left[
\ba{c|c}    
      g_{11} & {}^{\rm t}g_{12} \\\hline
      g_{12} & G_{22}    
\ea
\right],
\label{eqn:prlmz3} 
\eeqno 
where $g_{11}$, $g_{12}$, and $G_{22}$ are scalar, 
$K+L-1$ dimensional column vector, and $(K+L-1)$
$\times(K+L-1)$ 
matrix, respectively. It is obvious from 
the above partition of $\Sigma_{Z^KY^L}^{-1}$ 
that we have 
\beq
\left.
\ba{rcl}
g_{11}&=&\ds\nu_{11}=
   \left[Q\Sigma_{X^K}^{-1}{}^{\rm t}Q\right]_{11}
   +\sum_{l=1}^L\frac{\tilde{q}_{1l}^2}{\sigma_{N_l}^2},
\vspace{1mm}\\
g_{12}&=&{}^{\rm t}\left[\nu_{12}\cdots\nu_{1K}
               \beta_{11}\beta_{12}\cdots\beta_{1L}\right].
\ea
\right\}
\label{eqn:prlmz5} 
\eeq
It is well known that 
$\Sigma_{Z^KY^L}^{-1}$ has the following expression:
\beqno
\Sigma_{Z^KY^L}^{-1}
&=&\left[\ba{c|c}    
      g_{11} & {}^{\rm t}g_{12} \\\hline
      g_{12} & G_{22}    
      \ea\right]
\nonumber\\
&=&\left[
\ba{c|c}    
      1 & {}^{\rm t}0_{12} \\\hline
      \frac{1}{g_{11}}g_{12} & I_{L-1}    
\ea
\right]
\left[
\ba{c|c}    
      g_{11} & {}^{\rm t}0_{12} \\\hline
      0_{12}&G_{22}-\frac{1}{g_{11}}{}^{\rm t}g_{12}g_{12}    
\ea
\right]
\nonumber\\
& &\qquad\qquad \qquad \times
\left[
\ba{c|c}    
      1 & \frac{1}{g_{11}}{}^{\rm t}g_{12} \\\hline
      0_{12} & I_{L-1}    
\ea
\right].
\eeqno
Set
\beq
\hat{n}_1
\defeq 
[z_1|{}^{\rm t}z^K_{[1]}{}^{\rm t}y^L]
\left[
      \ba{c}
      1\\
      \hline 
      \frac{1}{g_{11}}g_{12}
      \ea
\right]
=z_1+\frac{1}{g_{11}}\left[{}^{\rm t}z_{[1]}^K{}^{\rm t}y^L\right]g_{12}.
\label{eqn:prlmz64} 
\eeq
Then, we have
\beqa
\lefteqn{
[{}^{\rm t}z^K{}^{\rm t}y^L]
\Sigma_{Z^KY^L}
    \left[\ba{c}
          z^K\\
          y^L
          \ea
    \right]
=[z_1|{}^{\rm t}z_{[1]}^K
           {}^{\rm t}y^L]
  \left[\ba{c|c}    
      g_{11} & {}^{\rm t}g_{12} \\\hline
      g_{12} & G_{22}    
      \ea\right]
 \left[\ba{c}
        z_1\\
       \hline
       \vspace*{-3.5mm}\\
       z_{[1]}^K\\
        y^L
        \ea
\right] } 
\nonumber\\
&=&[\hat{n}_1|{}^{\rm t}z^K_{[1]}{}^{\rm t}y^L]
\left[
\ba{c|c}    
      g_{11} & {}^{\rm t}0_{12}\\\hline
      0_{12}&G_{22}-\frac{1}{g_{11}}g_{12}{}^{\rm t}g_{12}    
\ea
\right]
\left[\ba{c}\hat{n}_1\\
      \hline 
      \vspace*{-3.5mm}\\
      z_{[1]}^K\\
      y^L
      \ea\right]. 
\qquad  
\label{eqn:zasds}
\eeqa
From (\ref{eqn:prlmz2z})-(\ref{eqn:prlmz64}), we have 
\beqa 
\hat{n}_1&=&z_1
+\frac{1}{g_{11}}\sum_{j=2}^L\nu_{1j}z_j
+\frac{1}{g_{11}}\sum_{l=1}^L\beta_{1l}y_l
\nonumber\\
&=&z_1
+\frac{1}{g_{11}}\sum_{j=2}^L\nu_{1j}z_j
-\frac{1}{g_{11}}\sum_{l=1}^L
\frac{\tilde{q}_{1l}}{\sigma_{N_l}^2}y_l.
\label{eqn:prlmz66} 
\eeqa
It can be seen from (\ref{eqn:zasds}) and (\ref{eqn:prlmz66}) 
that the random variable $\hat{N}_1$ defined by
$$
\hat{N}_1\defeq 
Z_1+\frac{1}{g_{11}}\sum_{j=2}^L\nu_{1j}Z_j
   -\frac{1}{g_{11}}\sum_{l=1}^L\frac{\tilde{q}_{1l}}{\sigma_{N_l}^2}Y_l
$$
is a zero mean Gaussian random variable with variance 
$\frac{1}{{g}_{11}}$ and is independent 
of $Z_{[1]}^K$ and $Y^L$. This completes the proof 
of Lemma \ref{lm:LmO}.
\hfill\IEEEQED

The followings are two variants of the entropy power 
inequality.

\begin{lm}\label{lm:lm5zz} Let ${\vc U}_i,i=1,2,3$ 
be $n$ dimensional 
random vectors with densities and let 
$T$ be a random variable taking values in a finite set. 
We assume that ${\vc U}_3$ is independent of ${\vc U}_1$, 
${\vc U}_2$, and $T$. Then,  we have 
\beqno
\EP{{\lvc U}_2+{\lvc U}_3|{\lvc U}_1T}
\geq 
\EP{{\lvc U}_2|{\lvc U}_1T}+\EP{{\lvc U}_3}.
\eeqno
\end{lm}
\begin{lm}\label{lm:lm5zzb} Let ${\vc U}_i$, $i=1,2,3$ be 
$n$ random vectors with densities. Let $T_1, T_2$ 
be random variables taking values in finite sets. 
We assume that those five random variables
form a Markov chain 
$
(T_1,{\vc U}_1) \to {\vc U}_3 \to (T_2,{\vc U}_2) 
$
in this order. Then, we have    
\beqno
& &\EP{{\lvc U}_1+{\lvc U}_2|{\lvc U}_3T_1T_2}
\\
&\geq& \EP{{\lvc U}_1|{\lvc U}_3T_1}
      +\EP{{\lvc U}_2|{\lvc U}_3T_2}.
\eeqno
\end{lm}

{\it Proof of Lemma \ref{lm:lm2}:}
By Lemma \ref{lm:LmO}, we have
\beq 
{\vc Z}_k=-\frac{1}{g_{kk}}\sum_{k^{\prime}\ne k}\nu_{kk^{\prime}}{\vc Z}_{k^{\prime}}
+ \frac{1}{g_{kk}}\sum_{l=1}^{L}
\frac {\tilde{q}_{kl}}{\sigma_{N_l}^2}{\vc Y}_l + \hat{\vc N}_k,
\label{eqn:prlmaa}
\eeq
where 
$\hat{\vc N}_k$ is a vector of $n$ independent 
copies of zero mean Gaussian random 
variables with variance $\frac{1}{g_{kk}}$. 
For each $k\in \IsetB$, $\hat{\vc N}_k$ is independent of 
${\vc Z}_{k^{\prime}}, k^{\prime}\in \IsetB$ $-\{k\}$
and ${\vc Y}_l, l \in \IsetA$. 
Set 
\beqno
h^{(n)}&\defeq& \frac{1}{n}h({\vc Z}_k|{\vc Z}_{[k]}^K,W^L).
\eeqno
Furthermore, for $l\in \IsetA$, define
\beqno
&&S_l \defeq \{l,l+1,\cdots,L\},
\Psi_l=\Psi_l({\vc Y}_{S_l}) 
\defeq \sum_{j=l}^{L} \frac {\tilde{q}_{kj}}{\sigma_{N_j}^2}{\vc Y}_j.
\eeqno
Applying Lemma \ref{lm:lm5zz} to (\ref{eqn:prlmaa}), we have
\beq  
\frac{{\baseN}^{2h^{(n)}}}{2\pi{\rm e}}
\geq \frac{1}{(g_{kk})^2}
\frac{1}{2\pi{\rm e}}{\baseN}^{\frac{2}{n}h(\Psi_1|{\svc Z}_{[k]}^K,W^L)} 
+\frac{1}{g_{kk}}.
\label{eqn:PrConvLm1}
\eeq
On the quantity $h(\Psi_1|{\svc Z}_{[k]}^K,W^L)$ 
in the right member of (\ref{eqn:PrConvLm1}), we have 
the following chain of equalities:
\beqa
\hspace*{-3mm}& &h(\Psi_1|{\vc Z}_{[k]}^K,W^L)
\nonumber\\
\hspace*{-3mm}&=&I(\Psi_1;{\vc X}^K|{\vc Z}_{[k]}^K,W^L)
   +h(\Psi_1|{\vc X}^K,{\vc Z}_{[k]}^K,W^L)
 \nonumber\\
\hspace*{-3mm}&\MEq{a}&
   I(\Psi_1;{\vc Z}^K|{\vc Z}_{[k]}^K,W^L)  
   +h(\Psi_1|{\vc X}^K,W^L)
\nonumber\\
\hspace*{-3mm}&=& I(\Psi_1;{\vc Z}_k|{\vc Z}_{[k]}^K,W^L) 
   +h(\Psi_1|{\vc X}^K,W^L)
\nonumber\\
\hspace*{-3mm}&=&h({\vc Z}_k|{\vc Z}_{[k]}^K,W^L)
    -h({\vc Z}_k|\Psi_1,{\vc Z}_{[k]}^K,W^L) 
\nonumber\\
\hspace*{-3mm}& &+h(\Psi_1|{\vc X}^K,W^L)
\nonumber\\
\hspace*{-3mm}&\MEq{b}& nh^{(n)}-h({\vc Z}_k|\Psi_1,{\vc Z}_{[k]}^K) 
           +h(\Psi_1|{\vc X}^K,W^L)
\nonumber\\
\hspace*{-3mm}&=&
nh^{(n)}-\frac{n}{2}\log\left[{2\pi{\rm e}}(g_{kk})^{-1}\right]
   +h(\Psi_1|{\vc X}^K,W^L). 
\label{eqn:PrConvLm2}
\eeqa
Step (a) follows from that ${\vc Z}^K$ can be obtained from 
${\vc X}^K$ by the invertible matrix $Q$. 
Step (b) follows from the Markov chain 
$${\vc Z}_k\to (\Psi_1,{\vc Z}_{[k]}^K)\to {\vc Y}^L\to W^L.$$ 
From (\ref{eqn:PrConvLm2}), we have
\beq
\frac{1}{2\pi{\rm e}}{\baseN}^{\frac{2}{n}h(\Psi_1|{\svc Z}_{[k]}^K,W^L)}
=\frac{{\baseN}^{2h^{(n)}}}{2\pi{\rm e}}g_{kk}\cdot
 \frac{1}{2\pi{\rm e}}{\baseN}^{\frac{2}{n}h(\Psi_1|{\svc X}^K,W^L)}.
\label{eqn:PrConvLm3}
\eeq
Substituting (\ref{eqn:PrConvLm3}) into (\ref{eqn:PrConvLm1}), 
we obtain  
\beq
\frac{{\baseN}^{2h^{(n)}}}{2\pi{\rm e}}
\geq \frac{{\baseN}^{2h^{(n)}}}{2\pi{\rm e}}\frac{1}{g_{kk}}
     \cdot\frac{1}{2\pi{\rm e}}{\baseN}^{\frac{2}{n}h(\Psi_1|{\svc X}^K,W^L)} 
     +\frac{1}{g_{kk}}.
\label{eqn:PrConvLm4}
\eeq
Solving (\ref{eqn:PrConvLm4}) with respect to 
$\frac{{\baseN}^{2h^{(n)}}}{2\pi{\rm e}}$, we obtain
\beq
\frac{{\baseN}^{2h^{(n)}}}{2\pi{\rm e}}
\geq
  \left[g_{kk} 
 -\frac{1}{2\pi{\rm e}}{\baseN}^{\frac{2}{n}h(\Psi_1|{\svc X}^K,W^L)} 
  \right]^{-1}.
\label{eqn:prlmz}
\eeq
Next, we evaluate a lower bound of  
$
{\baseN}^{\frac{2}{n}h(\Psi_1|{\svc X}^K,W^L)}.
$ 
Note that for $l=1,2,\cdots,L-1$ we have 
the following Markov chain:
\beq
\left(W_{S_{l+1}},\Psi_{l+1}({\vc Y}_{S_{l+1}})\right)\to {\vc X}^K 
\to \left(W_l,\ts \frac{\tilde{q}_{kl}}{\sigma_{N_l}^2}{\vc Y}_l\right). 
\label{eqn:Markov}
\eeq
Based on (\ref{eqn:Markov}), we apply 
Lemma \ref{lm:lm5zzb} to 
$
\frac{1}{2\pi{\rm e}}  
{\baseN}^{\frac{2}{n}h(\Psi_l|{\svc X}^K,W^L)}
$
for $l=1,2,\cdots,L-1$. Then, for $l=1,2,$ $\cdots,L-1$, 
we have the following chains of inequalities : 
\beqa
& &\frac{1}{2\pi{\rm e}} 
  {\baseN}^{\frac{2}{n}h(\Psi_l|{\svc X}^K,W^L)}
\nonumber\\
&=&\frac{1}{2\pi{\rm e}} 
  {\baseN}^{\frac{2}{n}
  h\left(\left.\Psi_{l+1}+\frac{\tilde{q}_{kl}}{\sigma_{N_1}^2}{\svc Y}_l
         \right|{\svc X}^K,W_{S_{l+1}},W_l\right)}
\nonumber\\
&\geq & \frac{1}{2\pi{\rm e}} 
        {\baseN}^{\frac{2}{n}
        h\left(\left.\Psi_{l+1}\right|{\svc X}^K,W_{S_{l+1}}\right)}
        +\frac{1}{2\pi{\rm e}} 
         {\baseN}^{
         \frac{2}{n}
         h\left(\left.\frac{\tilde{q}_{kl}}{\sigma_{N_l}^2}{\svc Y}_l\right|
         {\svc X}^K,W_l\right)
         }
\nonumber\\
&=& \frac{1}{2\pi{\rm e}} 
        {\baseN}^{\frac{2}{n}
         h\left(\left.
         \Psi_{l+1}\right|{\svc X}^K,W_{S_{l+1}}\right)}
        +\tilde{q}_{kl}^2\frac{{\baseN}^{-2r_l^{(n)}}}{\sigma_{N_l}^2}.
\label{eqn:prlmz1}
\eeqa
Using (\ref{eqn:prlmz1}) iteratively for
$l=1,2,\cdots, L-1$, we have
\beq
\frac{1}{2\pi{\rm e}} 
  {\baseN}^{\frac{2}{n}h(\Psi_1|{\svc X}^K,W^L)}
\nonumber\\
\geq 
\sum_{l=1}^{L}\tilde{q}_{kl}^2\frac{{\baseN}^{-2r_l^{(n)}}}{\sigma_{N_l}^2}.
\label{eqn:prlmz2}
\eeq
Combining (\ref{eqn:defajj}), (\ref{eqn:prlmz}), 
and (\ref{eqn:prlmz2}), we have 
\beqa
\frac{{\baseN}^{2h^{(n)}}}{2\pi{\rm e}}
&\geq&
  \left\{
  \left[Q\Sigma_{X^K}^{-1}{}^{\rm t}Q\right]_{kk} 
  +\sum_{l=1}^{L}\tilde{q}_{kl}^2
   \frac{1-{\baseN}^{-2r_l^{(n)}}}{\sigma_{N_l}^2}
   \right\}^{-1}
\nonumber\\
&=&\left[Q
         \left(
         \Sigma_{X^K}^{-1}
         +{}^{\rm t}A\Sigma_{N_{\IsetA}(r_{\IsetA}^{(n)})}^{-1}A
         \right)
         {}^{\rm t}Q\right]_{kk}^{-1},
\nonumber
\eeqa
completing the proof. 
\hfill \IEEEQED 

\section*{Acknowledgment}
The author would like to thank Dr. Yang Yang and 
Prof. Zixiang Xiong for pointing out an eariler 
mistake in the sum rate characterization 
of the rate disitortion region for 
the cyclic shift invariant sources.

\newcommand{\Skip}{}
{

}

\end{document}